\documentclass[epsf,epsfig,aps,floatfix,preprint,nofootinbib]{revtex4}
\usepackage{graphicx}
\usepackage{amssymb}
\usepackage{amsmath}
\usepackage{times}
\usepackage{latexsym,subfigure}
\usepackage{hyperref}
\usepackage{bbold}

\def\beq{\begin{equation}}
\def\eeq{\end{equation}}
\def\bey{\begin{eqnarray}}
\def\eey{\end{eqnarray}}

\def\lsim{\mathrel{\raise.3ex\hbox{$<$\kern-.75em\lower1ex\hbox{$\sim$}}}}
\def\gsim{\mathrel{\raise.3ex\hbox{$>$\kern-.75em\lower1ex\hbox{$\sim$}}}}

\newcommand{\be}{\begin{equation}}
\newcommand{\ee}{\end{equation}}

\def\lag {{\mathcal{L}}}

\newcommand{\AFBL}{A_{FB}^{m_{t \bar{t}} < 450}}
\newcommand{\AFBH}{A_{FB}^{m_{t \bar{t}} > 450}}

\def\afb {{A_{FB}}}
\def\tr {\text{Tr}}

\begin{document}

\preprint{MCTP/11-26}

\title{Tevatron Top $A_{FB}$ Versus LHC Top Physics}
\author{Moira I. Gresham, Ian-Woo Kim, Kathryn M. Zurek}
\address{Michigan Center for Theoretical Physics, University of Michigan, Ann Arbor, MI 48109  }

\date{\today}

\begin{abstract}

We carry out a comprehensive analysis of models for top $A_{FB}$ at CDF in light of new top data arriving from the LHC. We begin with a careful Tevatron analysis, considering in general which sets of effective vertices give rise to a large forward-backward asymmetry while suppressing the contribution to the total $t\bar{t}$ cross-section.  We show on general grounds that scalar models struggle to produce sufficient asymmetries consistent with CDF observations, while vector models can produce a large asymmetry with a less significant tension in the total cross-section and $t\bar{t}$ invariant mass distribution at the Tevatron.  We examine the essential observables of these models for top physics at LHC7 with $1 \mbox{ fb}^{-1}$ of data, including the total cross-section, invariant mass distribution and number of additional jets in $t\bar{t}$ events.  In the case of $t$-channel mediators, the LHC total cross-section places a strong constraint on light mediators, while the Tevatron invariant mass distributions place strong constraints on heavy mediators that are able to produce the asymmetry.  Heavy axigluons are becoming increasingly squeezed by LHC7 $t\bar{t}$ and dijet resonance searches.   We conclude that LHC7 top analyses are rapidly closing the window for viable models of the CDF top $A_{FB}$.

\end{abstract}
\maketitle

\tableofcontents

\section{Introduction}

The Large Hadron Collider (LHC) is providing an unprecedented probe of top quark properties.
While the Tevatron has to date collected on the order of a thousand tops, the LHC, with $1 \mbox{ fb}^{-1}$ of data has already nearly an order of magnitude more tops.  The improvement is due both to a larger production cross-section, and to improved rapidity coverage for leptons in semi-leptonic and fully leptonic top analyses.  In terms of the percentage error on the total cross-section, the 7 TeV LHC (LHC7) results are already competitive with the Tevatron \cite{cdftotalxsection} with only $35 \mbox{ pb}^{-1}$ \cite{CrossSection}, while the invariant mass distribution with just $200 \mbox{ pb}^{-1}$ extends to a higher $m_{t\bar{t}}$ of 2.5 TeV \cite{InvariantMass} as compared to the Tevatron reach of 1.8 TeV \cite{Aaltonen:2009iz}.  At $1 \mbox{ fb}^{-1}$, the top quark properties will be far better measured than at the Tevatron. 

At the same time, the Tevatron as a $p\bar{p}$ machine is better able at the outset to measure a forward-backward asymmetry.\footnote{Though see \cite{Hewett:2011wz,Bai:2011uk} for efforts to make a measurement of the forward-backward asymmetry at the LHC.}  The asymmetry in a particular invariant mass bin, $m_{t\bar{t},i}$, is defined by
\begin{equation}
A^{t\bar{t}}(m_{t\bar{t},i}) = \frac{N(\Delta y>0,m_{t\bar{t},i})-N(\Delta y<0,m_{t\bar{t},i})}{N(\Delta y>0,m_{t\bar{t},i})+N(\Delta y<0,m_{t\bar{t},i})},
\label{AFBdef}
\end{equation}    
with $\Delta y$ the rapidity difference between a top and an anti-top. The recent CDF anlaysis shows  $\afb = 0.475 \pm 0.114$ for $m_{t\bar{t}} > 450$ GeV~\cite{Aaltonen:2011kc} at the parton level (or $\afb = 0.266 \pm 0.062$ at the signal level)\footnote{Throughout this paper we use ``signal level'' to refer to background subtracted, raw measured quantities in the detector, and ``parton level'' to refer to unfolded results which attempt to subtract detector effects from the results.}, while the Next-to-Leading Order (NLO) Standard Model (SM) predicts much lower values $0.088 \pm 0.013$ (or $0.043 \pm 0.009$ at the signal level)~\cite{Kuhn:1998jr,Kuhn:1998kw,Bowen:2005ap,Almeida:2008ug,Ahrens:2011uf}, corresponding to a 3.4$\sigma$ deviation (3.6$\sigma$ at signal level). A measurement of the asymmetry with fully leptonic tops has also been made which is roughly consistent with the measurement in the semileptonic channel \cite{CDFLeptons}.  The D0 collaboration also observes a larger than predicted asymmetry~\cite{Abazov:2007qb}.   

Because this asymmetry is so large, any new physics (NP) that could generate such an asymmetry must have large couplings to the top as well as to the light quarks in the initial state.  A very large number of models have been proposed in the literature.  However, from a phenomenological point of view, these models mainly fall into only two categories according to the nature of the new particle exchange: (i) $s$-channel exchange of vector mediators ({\it e.g.} axigluon models)~\cite{Sehgal:1987wi,Bagger:1987fz,Djouadi:2009nb,Ferrario:2009bz,Frampton:2009rk,Chivukula:2010fk,Bauer:2010iq,Chen:2010hm,Alvarez:2010js,Delaunay:2011vv,Bai:2011ed,Barreto:2011au,Foot:2011xu,Zerwekh:2011wf,Shu:2011au,Haisch:2011up,Tavares:2011zg,Alvarez:2011hi,Barcelo:2011fw,Barcelo:2011vk,AguilarSaavedra:2011ci} or (ii) $t$-channel exchange of flavor-violating mediators~\cite{Jung:2009jz,Cheung:2009ch,Shu:2009xf,Dorsner:2009mq,Arhrib:2009hu,Barger:2010mw,Xiao:2010hm,Cheung:2011qa,Shelton:2011hq,Berger:2011ua,Grinstein:2011yv,Patel:2011eh,Craig:2011an,Ligeti:2011vt,Jung:2011zv,Nelson:2011us,Duraisamy:2011pt}\footnote{There is another class of models that can create effective axial QCD coupling from NP \cite{Gabrielli:2011jf}.  }.  Comparative studies of these models have also been carried out \cite{Jung:2009pi,Cao:2009uz,Cao:2010zb,Jung:2010yn,Choudhury:2010cd,Jung:2010ri,Delaunay:2011gv,Gresham:2011pa}, and their implications for top observables at the Tevatron observed \cite{Gresham:2011pa,Krohn:2011tw}.  The $s$-channel mediators often have maximally axial couplings (though there are exceptions such as \cite{AguilarSaavedra:2011ci}), while the $t$-channel mediators connect a light quark to the top quark in a way that appears to maximally violate flavor.  A number of studies on the implications of these models for LHC physics have also been carried out \cite{Berger:2011ua,Gresham:2011dg,Blum:2011up,AguilarSaavedra:2011vw,AguilarSaavedra:2011hz}.

There are a large number of possibilities for the spin, color, flavor and electroweak representation of a new field that fits into the two categories mentioned above.
In the literature these have been mostly built and studied one by one.  Here, by contrast, we are motivated to extract {\em general} features to determine which effective vertices are able to generate the large $\afb$ while contributing a small amount to the total cross-section.   We find that the form of the matrix element itself allows one to make general conclusions about which classes of models are successful in generating a significant asymmetry.   

We find on general grounds that perturbative\footnote{Scalar models with larger couplings can achieve larger asymmetries, though at the expense of a larger contribution to the total $t\bar{t}$ cross-section.} scalar models typically can produce no more than a $10-20\%$ ``parton level'' asymmetry for $m_{t\bar{t}} > 450 \mbox{ GeV}$, which is only somewhat larger than the asymmetry produced in the SM (at $\sim 9\%$) and well below CDF's parton level central value of $\sim 48\%$.\footnote{The data-level asymmetry yields a result about a factor of two lower than the parton level result, which has been confirmed by the theoretical study of \cite{Gresham:2011pa}.  A comparison of a parton level theoretical result to the signal level asymmetry is not valid, and will underproduce by more than $2\sigma$ the observed asymmetry.}  The reason is simply the combination of the Mandelstam variables that enters into $t$- and $u$-channel processes for scalars; the statement is independent of the color (singlet, triplet, sextet or octet) or flavor representation of the state.  By contrast, $t$-channel vectors have a matrix element that is conducive to producing a large asymmetry with a relatively small contribution to the total cross-section.  

We systematically enumerate the possibilities for the quantum numbers of $t$- and $s$- channel mediators that can produce an asymmetry and show that classes of models are strongly disfavored based on a small contribution to the total asymmetry or large contribution to the total $t\bar{t}$ cross-section at the Tevatron.  This paper is intended to be a companion to our earlier paper on $\afb$ \cite{Gresham:2011pa}, which carried out a systematic comparison of NP models to the data.  This was the only theory paper to carry out the full top reconstruction in order to compare results at the signal level.  We found that there were large acceptance effects which changed the extracted parton level comparison between the SM and the NP models.\footnote{See also \cite{Jung:2011zv} for a discussion of the acceptance effect at the parton level.}  

With LHC data quickly arriving, however, the source of strong constraints is rapidly changing, and we are particularly compelled by the fact that the LHC collaborations are now analyzing unprecedented amounts of top data that will clearly rule out a large swath of models.  We examine observables from the LHC, such as total cross-section, $t\bar{t}$ invariant mass distribution, and the number of additional jets in $t\bar{t}$ events.  In order to carry out our analysis, we have done a systematic scan in mass and coupling space for a broad class of models, described in appendix \ref{app: madgraph}.  For a subset of the models that give the best fit to the data we generate 5 million events, applying cuts and $m_{t \bar{t}}$ reconstruction mirroring the ATLAS analysis \cite{InvariantMass} in order to compare to LHC $t \bar{t}$ distributions. Many models will be strongly constrained by these analyses with just $1 \mbox{ fb}^{-1}$.\footnote{These statements must take into consideration, however, uncertainties in next-to-leading-order (NLO) corrections to the NP contributing to the total cross-section and invariant mass distribution.}  In our earlier paper on searching for flavor-violating resonances at the LHC, we proposed top-jet resonances as a means to search for $t$-channel mediators \cite{Gresham:2011dg}.  Such a search is complementary to the analysis here.  Many $t$-channels models will be constrained by existing analyses, but the models that survive can have an imprint in top-jet resonances.

The outline of this paper is as follows.  In the next section we discuss the classes of models that could generate the forward-backward asymmetry at the parton level, examining the asymmetries that are generated by the possible effective vertices, and drawing conclusions about which classes of models are viable.   The reader who is interested only in the numerical results can skip this section, and move on to Sec.~\ref{sec:tevatron}, referring to Sec.~(\ref{Sec: effective}) solely for a discussion of our conventions.  In Sec.~(\ref{sec:tevatron}), we carry out a systematic scan of models at the Tevatron, choosing a set of models as benchmarks for simulation of the large data sets necessary for invariant mass distributions.  In Sec.~(\ref{Sec: LHC}), we then examine the expected top properties at the LHC for the classes of models we consider.  In the appendices, parton level asymmetries, as well as a detailed discussion of our analysis pipeline, can be found.

\section{Effective Vertices and Top $A_{FB}$}
\label{Sec: effective}

\begin{figure}
\centering
\subfigure[~s-channel $q\bar{q}$]{\includegraphics[width=0.25\textwidth]{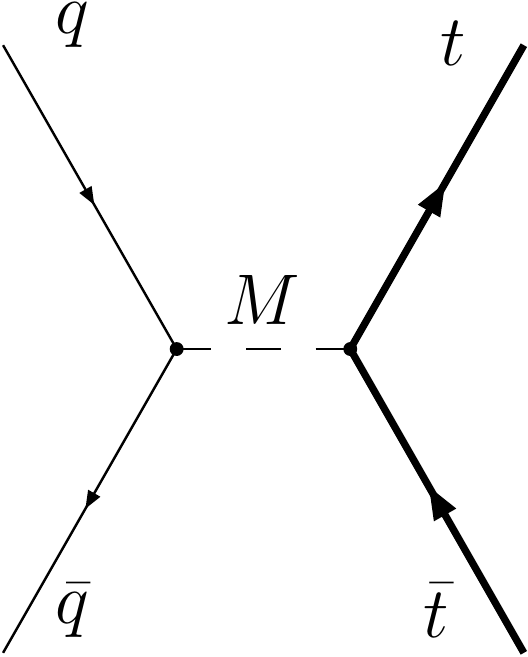}}\qquad
\subfigure[~s-channel $gg$]{\includegraphics[width=0.25\textwidth]{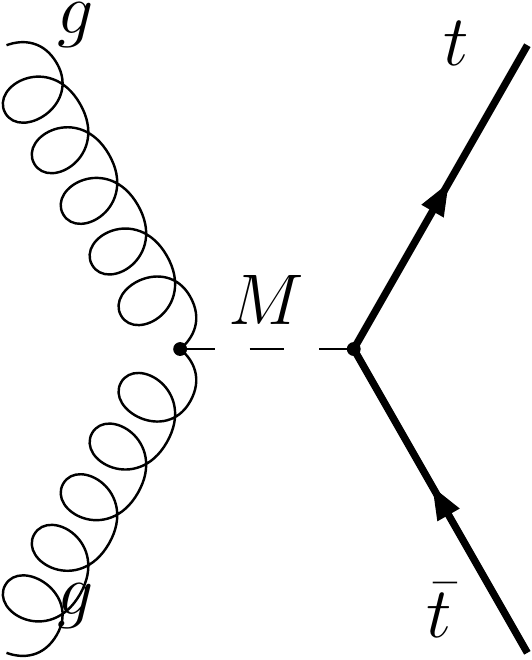}}\qquad
\subfigure[~$t\bar{t}$ production]{\includegraphics[width=0.25\textwidth]{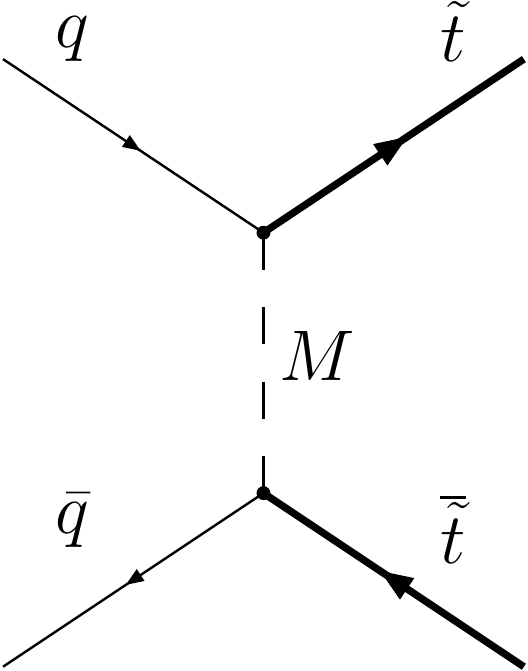}}\qquad
\subfigure[~t channel single mediator]{\includegraphics[width=0.25\textwidth]{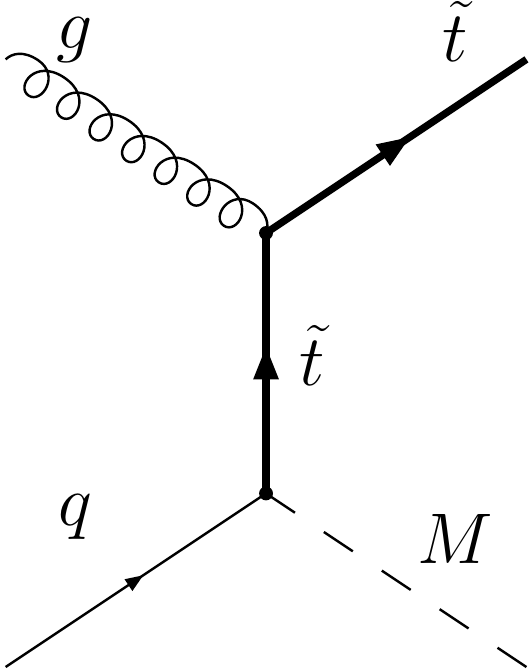}}\qquad
\subfigure[~s channel single mediator]{\includegraphics[width=0.25\textwidth]{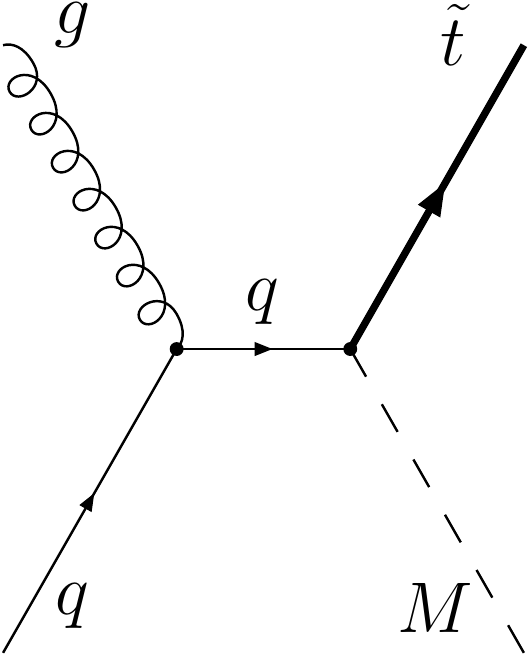}} \qquad
\subfigure[~u channel single mediator]{\includegraphics[width=0.25\textwidth]{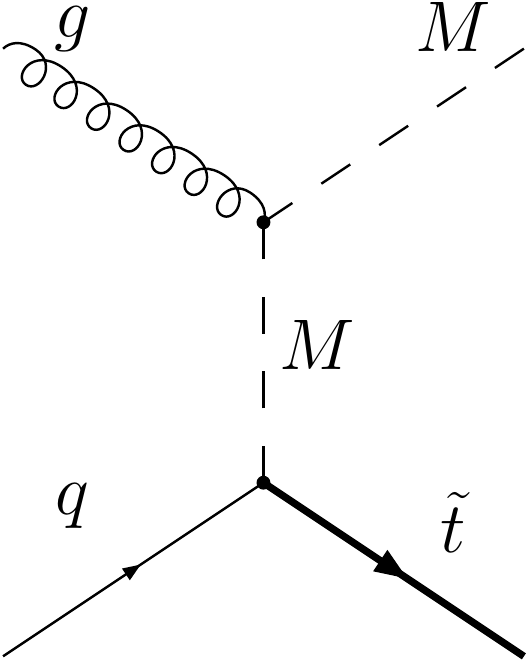}} 
\caption{Tree level production diagram involving the mediator $M$ and the coupling $g_M$.}
\label{phiproduction}
\end{figure}

Broadly speaking, either $s$-channel or $t$- (or $u$)-channel resonances can generate the top forward-backward asymmetries at tree level.  We show the diagrams that contribute both to the Tevatron $A_{FB}$ and $t\bar{t}$ production at LHC in Fig.~\ref{phiproduction}.   The structure of the differential cross-section for models that produce the asymmetry through $t$-channel exchange of a top-flavor-carrying mediator takes the same basic form according to whether the mediator is spin-0 or spin-1. Let the effective Lagrangian involving top and up quarks take the form
\beq\label{eq: NP lagrangian}
\lag_\text{NP} =  \bigg\lbrace 
	\begin{array}{l l} 	\bar{\tilde{t}}(g_L P_L + g_R P_R) t^a_r u M^a	+ \text{H.c.}   					& \text{spin-0} \\
						\bar{\tilde{t}}\gamma^\mu(g_L P_L + g_R P_R) t^a_r u M^a_\mu	+ \text{H.c.} 	& \text{spin-1}
	\end{array} 
\eeq 
where $\tilde{t} = t$ for singlets and octets and $\tilde{t}=t^c$ for anti-triplets and sextets, and $t^a_r$ are the color generators of a representation $r$---$3 \times 3$ Hermitian matrices that contain Clebsch-Gordon coefficients connecting two (anti-)quarks, normalized so that $\tr(t^a_r t^b_r)={1 \over 2}\delta^{ab}$. For singlets, we take $t^a_r=\mathbb{1}$.  Note that we are restricting ourselves to couplings to top and up quarks, though our results should not qualitatively change given couplings to down-type quarks.  We also only consider single mediator production; double mediator production is only important for light colored states, which are not present for the models we consider.

A large number of models of NP can generate these effective vertices, involving, in addition to the color group, potentially $SU(2)$ and flavor representations.  There are a limited number of flavor symmetric models that can generate the top $A_{FB}$ in the $t$ channel while satisfying existing constraints.  We show in Table~\ref{quantumnumbers} the possibilities.  There are also interactions that connect $Q_L$ to $Q_L$, but these models with flavor symmetries are typically highly constrained by light quark observables since they mix with SM CKM physics.  We do not consider them further. Interactions connecting, {\em e.g.}, $Q_L$ to $u_R$ through a spin-1 color triplet or sextet ``diquark'' are also possible, but as we will soon see (see Fig.~\ref{fig: failures}), the dominant $t$-channel interaction for top $\afb$ does not give rise to a significant positive asymmetry.    We refer the reader to \cite{Arnold:2009ay,Giudice:2011ak} for a complete tabulation in the scalar mediator case of the possible flavor symmetries and to \cite{Grinstein:2011yv,GrinsteinToAppear} for discussion of $\afb$ in the context of Minimal Flavor Violation.  In any case, the general observations that we make on the basis of the effective vertices in Eq.~(\ref{eq: NP lagrangian}) will be relatively independent of the flavor representation, and we make the appropriate qualifications where necessary.    For example, in flavor symmetric models, states in both the $t$ channel and the $s$ channel can contribute to the total asymmetry.  Scalars in the $s$ channel don't contribute to the forward-backward asymmetry, but can have an impact through their interference with $t$-channel scalars that do generate the asymmetry. 

\begin{table}
\begin{tabular}{|c|c|c|c|c|}
\hline
Interaction & $SU(3)_c$ & $SU(2)$ & $U(1)_Y$ & Flavor ($u_R,d_R,Q_L$) \\ \hline
$\bar{u}_R Q_L$ & 1,~8 & 2 & $\pm 1/2$  & $(3,1,\bar{3})$ \\ \hline
$u_R u_R$ & $3,~\bar{6}$ & 1 & -4/3 & $(3,1,1)$ \\ \hline
$d_R u_R$ & $3,~\bar{6}$ & 1 & -1/3 & $(3,1,1)$ \\ \hline
$\bar{u}_R \gamma^\mu u_R$ & $1,~8$& 1 &0 &  $(1,1,1)$ \\ \hline
$\bar{u}_R \gamma^\mu u_R$ & $1,~8$& 1 &0 &  $(8,1,1)$ \\ \hline
$\bar{d}_R \gamma^\mu u_R$ & $1,~8$ & 1 &-1 & $(\bar{3},3,1)$ \\ \hline
\end{tabular} 
\caption{Flavor symmetric interactions (in schematic notation) involving at least one $u_R$ quark that can mediate a significant positive top forward-backward asymmetry in the $t$-channel. (See also \cite{GrinsteinToAppear}.)}
\label{quantumnumbers}
\end{table}

Given the large number of possible combinations of $s$- and $t$-channel resonances from the flavor symmetric models, one despairs of ever being able to derive the characteristics of the state that can generate the asymmetry.  However, we will find that in the $t$ (or $u$) channel, the amplitudes have very distinctive shapes dependent on whether the state is a vector or scalar mediator particle.  We examine these characteristic features, and use it to draw conclusions about the nature of the mediator from the invariant mass dependence of $A_{FB}$.  These conclusions are robust independent of the particular flavor symmetric model that one employs, and allows one to make general statements on the types of characteristics that are necessary for generating a large top $A_{FB}$.

 The cross-sections arising from the NP interactions \eqref{eq: NP lagrangian} and SM interactions are given by 
  \beq
 \frac{d \sigma}{d\cos\theta} =\frac{\beta}{32 \pi \hat{s}} \left({\cal A}_{\rm SM} + {\cal A}_{\rm int} + {\cal A}_{\rm sq}\right),
 \eeq
 where \cite{Shu:2009xf, Arhrib:2009hu, Cao:2010zb}
 \beq
 {\cal A}_{SM} = \frac{2 g_s^4}{9}(1+c_\theta^2 + \frac{4 m_t^2}{\hat{s}}),
 \eeq
\beq
\mathcal{A}_\text{int} = {g_s^2 C_{(0)}^r \over 9} \bigg\lbrace 
	\begin{array}{l l} 	 { (g_L^2 + g_R^2) } {2 ( \hat{u}_t^2 +  \hat{s} m_t^2 ) +{m_t^2 \over m_{M}^2}(\hat{t}_t^2+\hat{s} m_t^2) \over \hat{s} \hat{t}_{M}} & \text{spin-1} \\
	{(g_L^2 + g_R^2)} {\hat{t}_t^2 + \hat{s} m_t^2 \over \hat{s} \hat{t}_M} & \text{spin-0}
	\end{array} 
	~~~(\hat{t} \leftrightarrow \hat{u} \text{ for diquarks}),
\eeq
and
\beq
\mathcal{A}_\text{sq} = {C_{(2)}^r \over 9}\bigg\lbrace 
	\begin{array}{l l} 	 {(g_L^4+g_R^4)\hat{u}_t^2+2 g_L^2 g_R^2 \hat{s} (\hat{s}-2 m_t^2) +\frac{m_t^4}{4 m_{M}^4}(g_L^2+g_R^2)^2(\hat{t}_{M}^2+4 \hat{s} m_{M}^2) \over \hat{t}_{M}^2} & \text{spin-1} \\
	{(g_L^2 + g_R^2)^2 \over 4}{\hat{t}_t^2 \over \hat{t}_M^2} & \text{spin-0}
	\end{array} 
	~~~(\hat{t} \leftrightarrow \hat{u} \text{ for diquarks}).
\eeq
Here $C_{(0)}^r$ and $C_{(2)}^r$ are color factors depending on the color rep of the mediator.\footnote{Specifically, $C_{(0)}^r = - \xi \tr\left( t^a_r T^A t^a_r \tilde{T^{A}} \right)$ and $C_{(2)}^r = \tr\left( t^a_r t^b_r\right) \tr\left( t^a_r t^b_r \right)$ where $\xi = -1(1)$ and $\tilde{T^A}=T^A ({T^{A}}^T)$ for octets and singlets (anti-triplets and sextets).}   
We have also defined
\beq 
 c_\theta = \beta \cos \theta \qquad  \beta = \sqrt{1-4 m_t^2/\hat{s}},
 \eeq
\beq 
\hat{t}_i \equiv \hat{t} - m_i^2 \qquad \hat{u}_i \equiv \hat{u} - m_i^2.
\eeq
The Mandelstam variables are related to the scattering angle via 
\beq
\hat{t} = -\hat{s}(1-c_\theta)/2 + m_t^2 \qquad \text{and} \qquad \hat{u} = -\hat{s}(1+c_\theta)/2 + m_t^2.
\eeq
  Note that we have not taken into account interference between NP contributions which can arise in flavor symmetric models.  For example, $s$-channel flavor conserving and $t$-channel flavor changing diagrams may interfere.  These new contributions do not give rise to any new {\em types} of terms (modulo mass terms in propagators) in the interference amplitude for the vector states, but do give rise to new contributions for the scalar states.  We discuss these terms later, but suffice for now to comment that the new terms will not change our qualitative conclusions.
 
 \begin{table}
 \centering
 \begin{tabular}{r c c c c}
 \hline
 Color rep: & \bf{1} & \bf{8} & $\bf{3}$ & \bf{6} \\
 \hline
$ C_{(0)}$ & 4 &-2/3 & 1 & -1\\
$ C_{(2)}$ & 9 & 2 & 3/4 & 3/2 \\
 \hline
 \end{tabular}
 \caption{Color factors for color representations of flavor-changing mediators.}
\end{table}

A flavor-conserving vector can give rise to an asymmetry at tree level if couplings to top and up have nonzero axial parts. Given the NP interaction Lagrangian
\begin{align}\label{eq:ExoticGluonLagrangian}
{\cal L}_\text{NP} = 
				\left( \bar{q} \, T^A \gamma^\mu(g_L^q P_L+g_R^q P_R) q + \bar{t} \, T^A \gamma^\mu(g_L^t P_L+g_R^t P_R) t \right) {G'}^A_\mu,
\end{align}
the scattering cross-sections calculated through these interactions are \cite{Cao:2010zb}:
 \beq
 {\cal{A}}_\text{int} = {2 g_s^2 \over 9}{ \hat{s}_G \over \hat{s} ( \hat{s}_G^2 + m_G^2 \Gamma_G^2)} \left( g^+ (\hat{u}_t^2 + \hat{t}_t^2 + 2 m_t^2  \hat{s} ) + g^- (\hat{u}_t^2 - \hat{t}_t^2) \right),
 \eeq
 and
  \beq
 {\cal{A}}_\text{sq} = {1 \over 9}{1  \over ( \hat{s}_G^2 + m_G^2 \Gamma_G^2)}  \left( ({g_L^q}^2 + {g_R^q}^2)\left(({g_L^t}^2 + {g_R^t}^2) (\hat{u}_t^2 + \hat{t}_t^2) +   2 g_L^t g_R^t  2 m_t^2 \hat{s} \right) + g^- g^+ (\hat{u}_t^2 - \hat{t}_t^2) \right),
 \eeq
 where
 \beq
 g^{\pm} \equiv (g^q_L \pm g^q_R)(g^t_L \pm g^t_R) \qquad \text{and} \qquad \hat{s}_G \equiv \hat{s} - m_G^2.
 \eeq
For a color singlet rather than a color octet, the interference term vanishes and the squared term is scaled by a factor $C^\mathbf{1}_{(2)} / C^\mathbf{8}_{(2)} = 9/2$. 

We now assemble these results using the parton distribution functions to gain a strong quantitative understanding of which types of interactions  can give rise to the observed forward-backward asymmetry.
The  cross-section for the process $p \bar{p} \rightarrow t \bar{t}$ is given by:
\begin{equation}
\sigma (s) = \Sigma_{i, j} \int d\hat{s} \int_{\hat{s} / s}^1 dx  {1 \over s x} \, \int d\cos\theta \; f_i(x) f_j\left( {\hat{s} \over s x} \right) \; \hat{\sigma}_{i,j}(\cos \theta, \hat{s}).
\end{equation}
We define
\begin{equation}\label{}
F_{i j}(\hat{s},s) = \int_{\hat{s} / s}^1 dx  {1 \over x} f_i(x) f_j\left( {\hat{s} \over s x} \right).
\end{equation}
Then the differential cross-section as a function of parton energy $\hat{s}$ can be expressed as
\begin{equation}\label{eq: differential cross-section}
{d \sigma (s) \over d\hat{s} \, d\cos\theta} = {1 \over s} \Sigma_{i, j} F_{i j}(\hat{s},s) \hat{\sigma}_{i,j}(\cos \theta, \hat{s}).
\end{equation}
Of course, all the $\cos \theta$ dependence is in the parton-level differential cross-section. If only one kind of initial state parton contributes to the cross-section, then the PDF completely factors out of the differential forward-backward asymmetry, defined as a function of $\hat{s}$ by
\begin{equation}
A_{FB}(\hat{s}) = { \Sigma_{i, j} F_{i j}(\hat{s},s) \hat{\sigma}^-_{i,j}(\hat{s}) \over \Sigma_{i, j} F_{i j}(\hat{s},s) \hat{\sigma}^+_{i,j}(\hat{s})},
\end{equation}
where 
\begin{equation}\label{eq: SM contribution}
\hat{\sigma}^\pm_{i,j}(\hat{s}) \equiv \int_0^1 dz \left( \hat{\sigma}_{i,j}(z, \hat{s}) \pm \hat{\sigma}_{i,j}(-z, \hat{s}) \right).
\end{equation}

\begin{figure}
\centering
\includegraphics[width = 0.45 \textwidth]{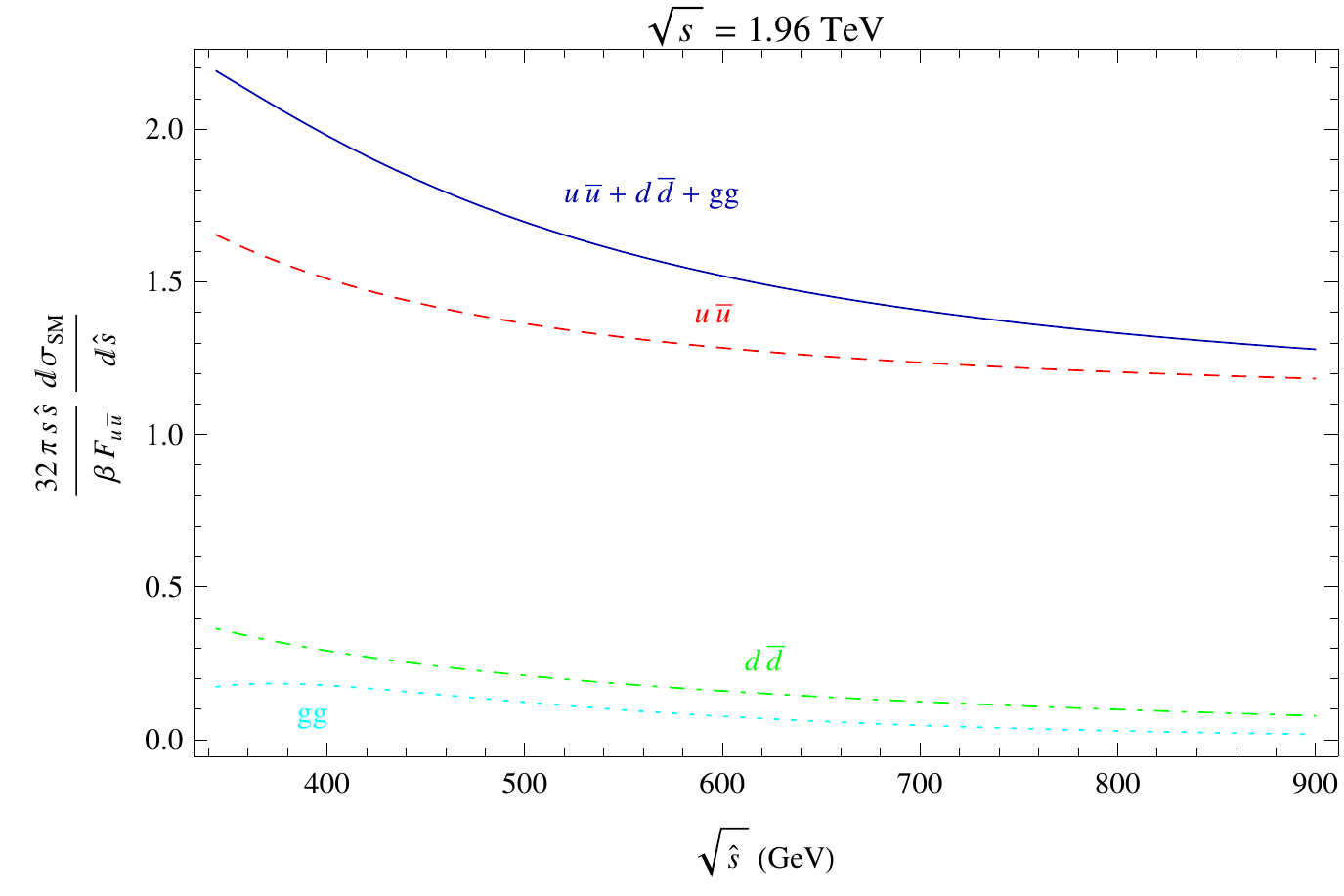}
\caption{SM contribution to the denominator in the differential forward-backward asymmetry as defined in \eqref{eq: SM contribution}. CTEQ5M parton distribution functions were used.}\label{fig: SM contribution}
\end{figure}

Suppose we are interested in a NP model with a nonzero contribution to the cross-section term generated through $u \bar{u} \rightarrow t \bar{t}$. We may then write the forward-backward asymmetry as a function of $\sqrt{\hat{s}}$ as
\begin{equation}\label{eq: AFB formula}
A_{FB}(\hat{s}) = {\hat{\sigma}_{u \bar{u}}^{NP -} \over \hat{\sigma}_{u \bar{u}}^{NP +} + \text{SM contribution}},
\end{equation} where 
\beq\label{eq: sm contribution}
\text{SM contribution} = \hat{\sigma}_{u \bar{u}}^{SM +} + {F_{d \bar{d}} \over F_{u \bar{u}}} \hat{\sigma}_{d \bar{d}}^{SM +} + {F_{g g} \over F_{u \bar{u}}} \hat{\sigma}_{g g}^{SM +},
\eeq
and is shown in Fig.~(\ref{fig: SM contribution}).
We note here that the falling SM contribution alone is not enough to give as steep a rise in the asymmetry as a function of $\hat{s}$ as is observed at CDF. The rise can steepen through a combination of the following factors: (1) ${\hat{s} \over \beta} \hat{\sigma}_{u \bar{u}}^{NP -}$ rises as a function of $\hat{s}$ and/or (2) ${\hat{s} \over \beta} \hat{\sigma}_{u \bar{u}}^{NP +}$ is comparable to the SM contribution and decreases as a function of $\hat{s}$. However, if the majority of the steepness were to come from mechanism (2), the total cross-section especially at low invariant mass would have to be comparable to the SM cross-section; this is hard to do without running into constraints on the total differential cross-section.   Thus a significant contribution must come from $\hat{\sigma}_{u \bar{u}}^{NP -}$.

There are seven kinds of terms that show up in a general cross-section involving $t$-channel mediators, including its interference with the SM:
\beq
{\hat{u}_t^2 +\hat{s} m_t^2 \over \hat{s} \hat{t}_{M}},~~{\hat{t}_t^2 + \hat{s} m_t^2 \over \hat{s} \hat{t}_{M}}, ~~{\hat{u}_t^2 \over  \hat{t}_{M}^2}, ~~{\hat{t}_t^2 \over  \hat{t}_{M}^2}, ~~{\hat{s}^2 \over  \hat{t}_{M}^2},~~{\hat{s} m_t^2 \over  \hat{t}_{M}^2}, ~~1,
\eeq
with $\hat{t} \leftrightarrow \hat{u}$ for or $u$-channel diquarks.
We examine these contributions term by term to determine which types can successfully generate a large contribution. In particular, there must be a large contribution to the asymmetry with a very modest contribution to the total cross-section.  That is to say simply that the odd contribution must be large in comparison to the even contribution.

\begin{figure}
\centering
\includegraphics[width=0.32\textwidth]{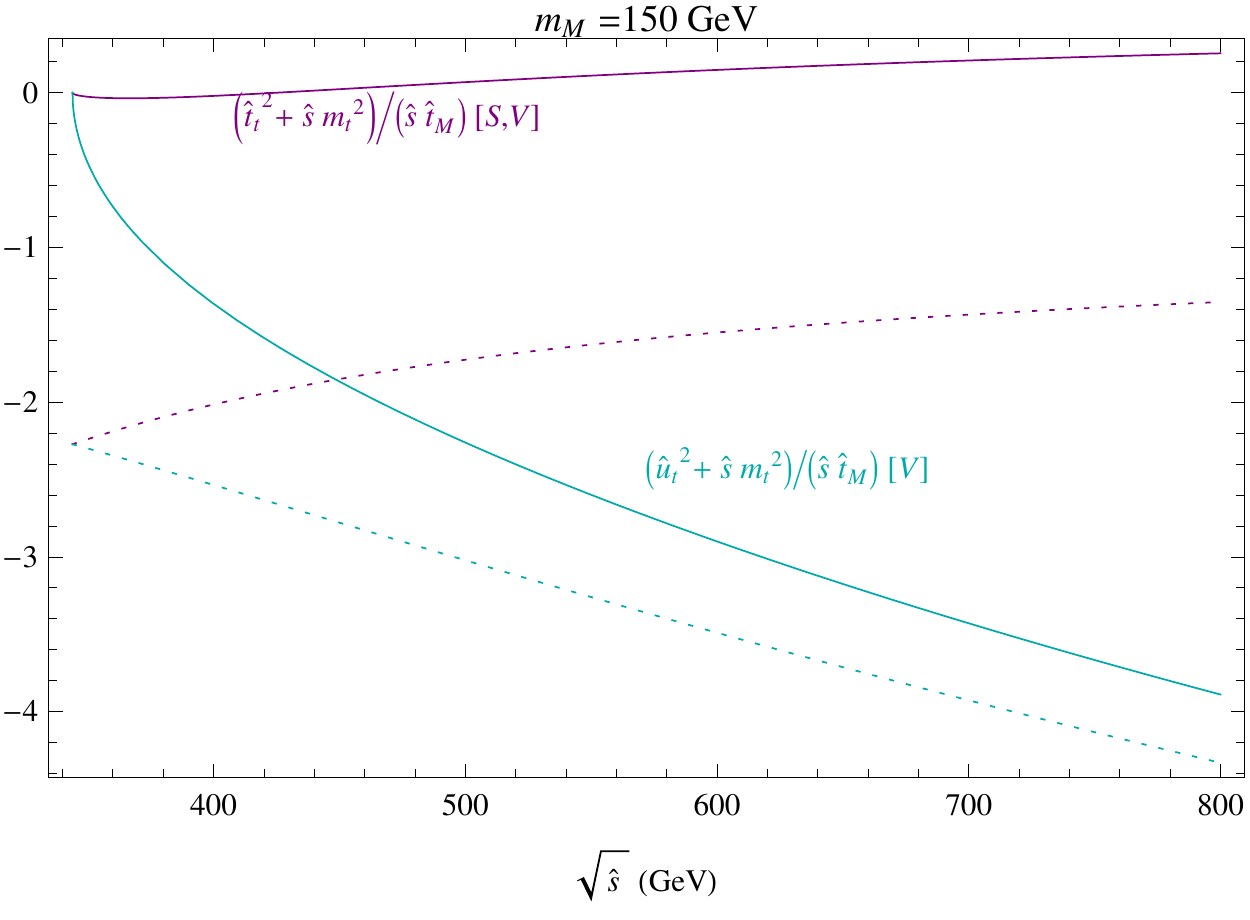}  \includegraphics[width=0.32\textwidth]{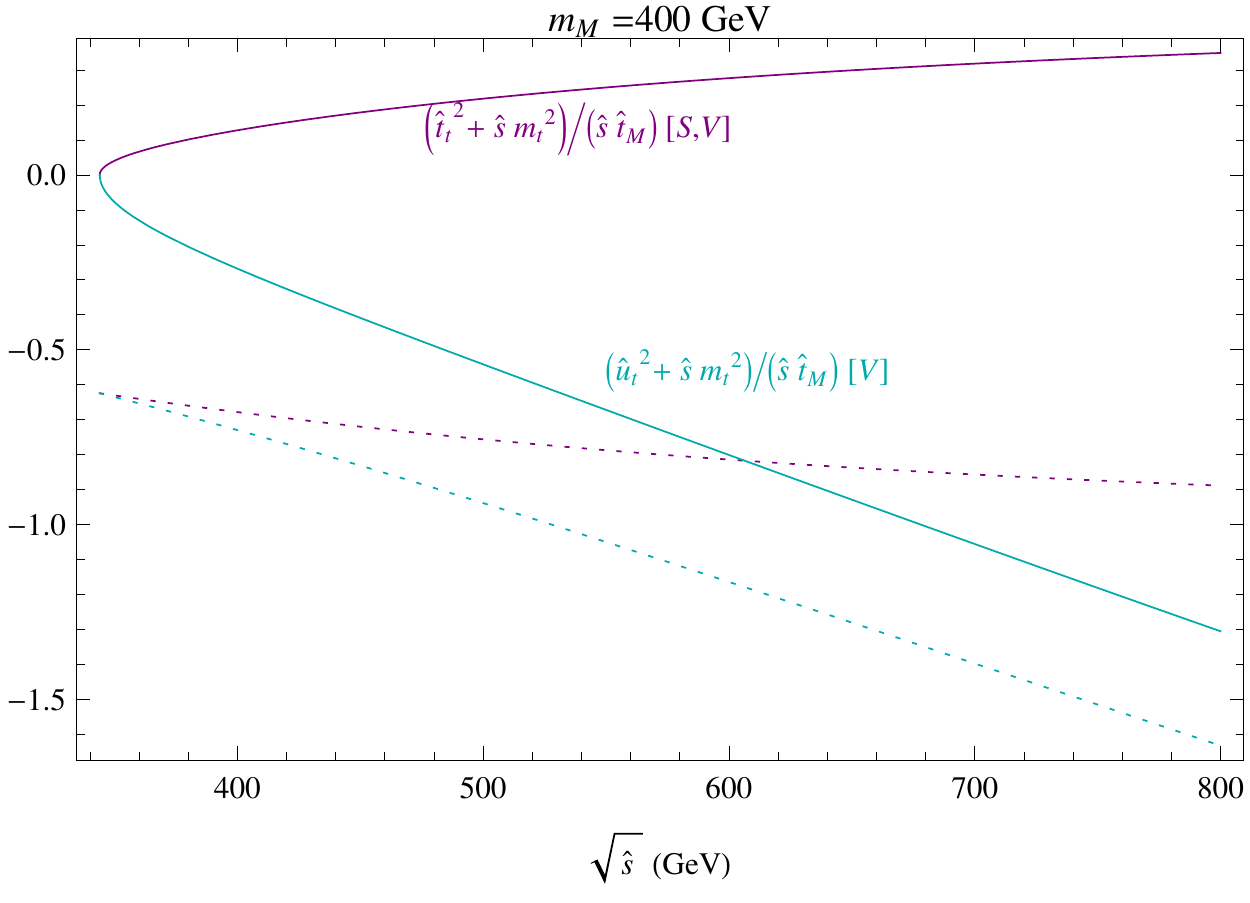} \includegraphics[width=0.32\textwidth]{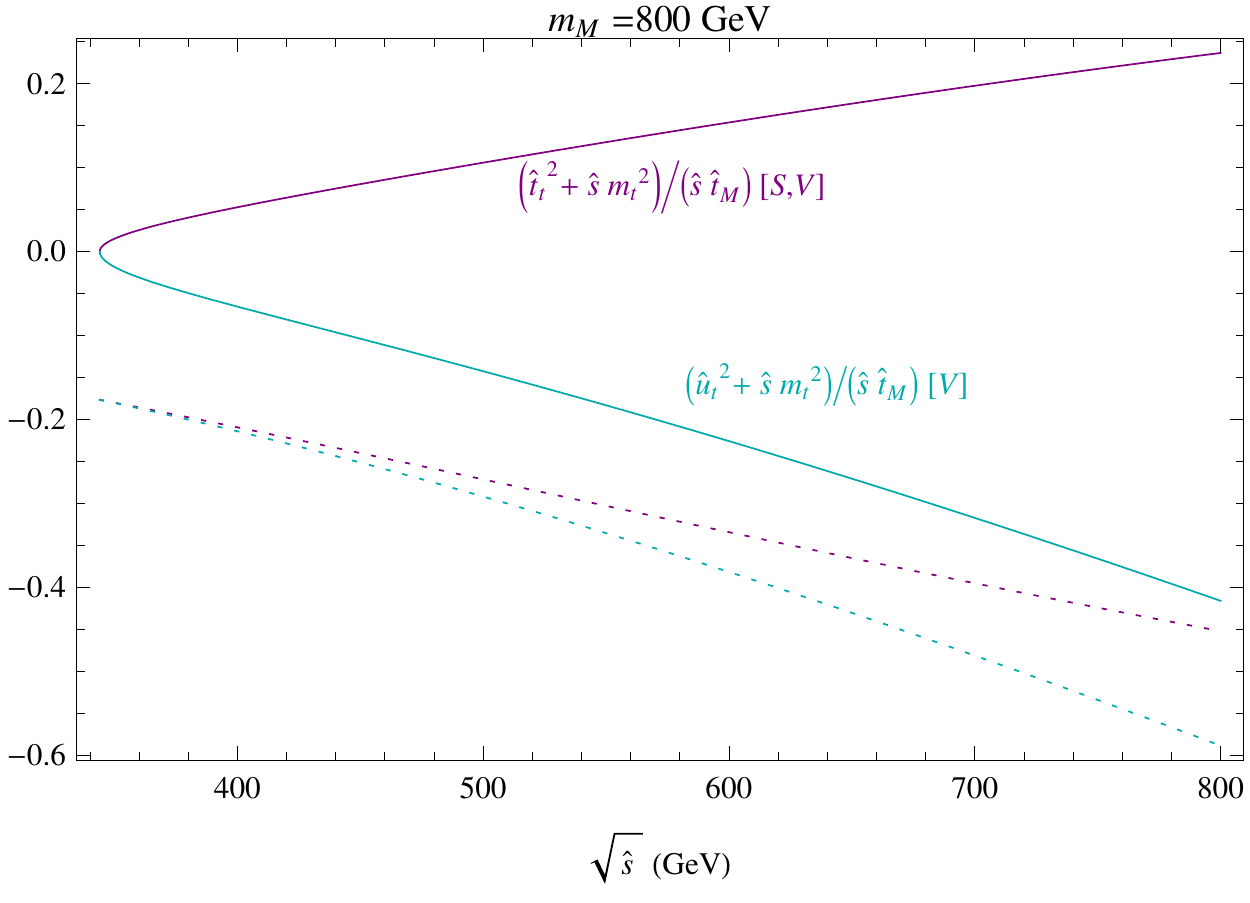}\\
\includegraphics[width=0.32\textwidth]{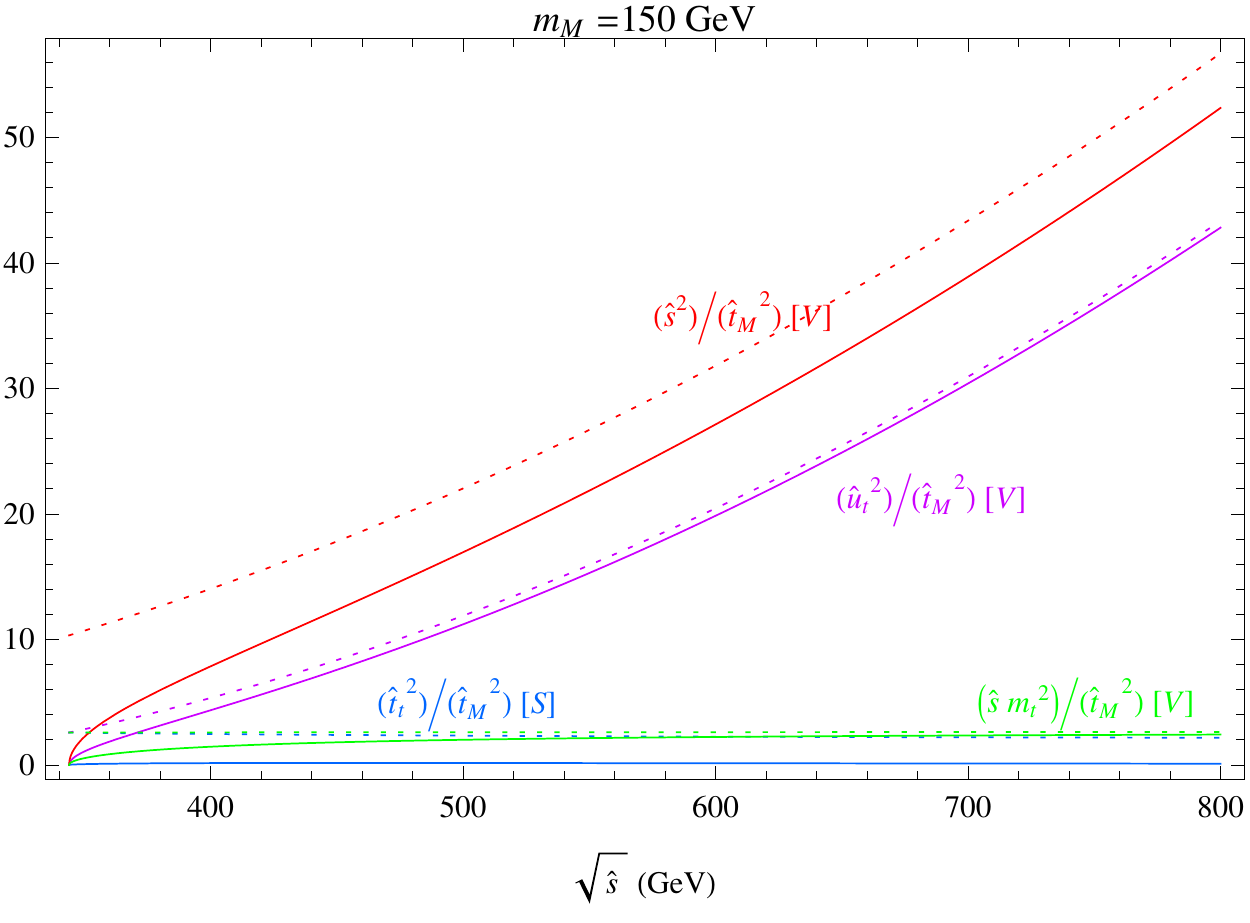}  \includegraphics[width=0.32\textwidth]{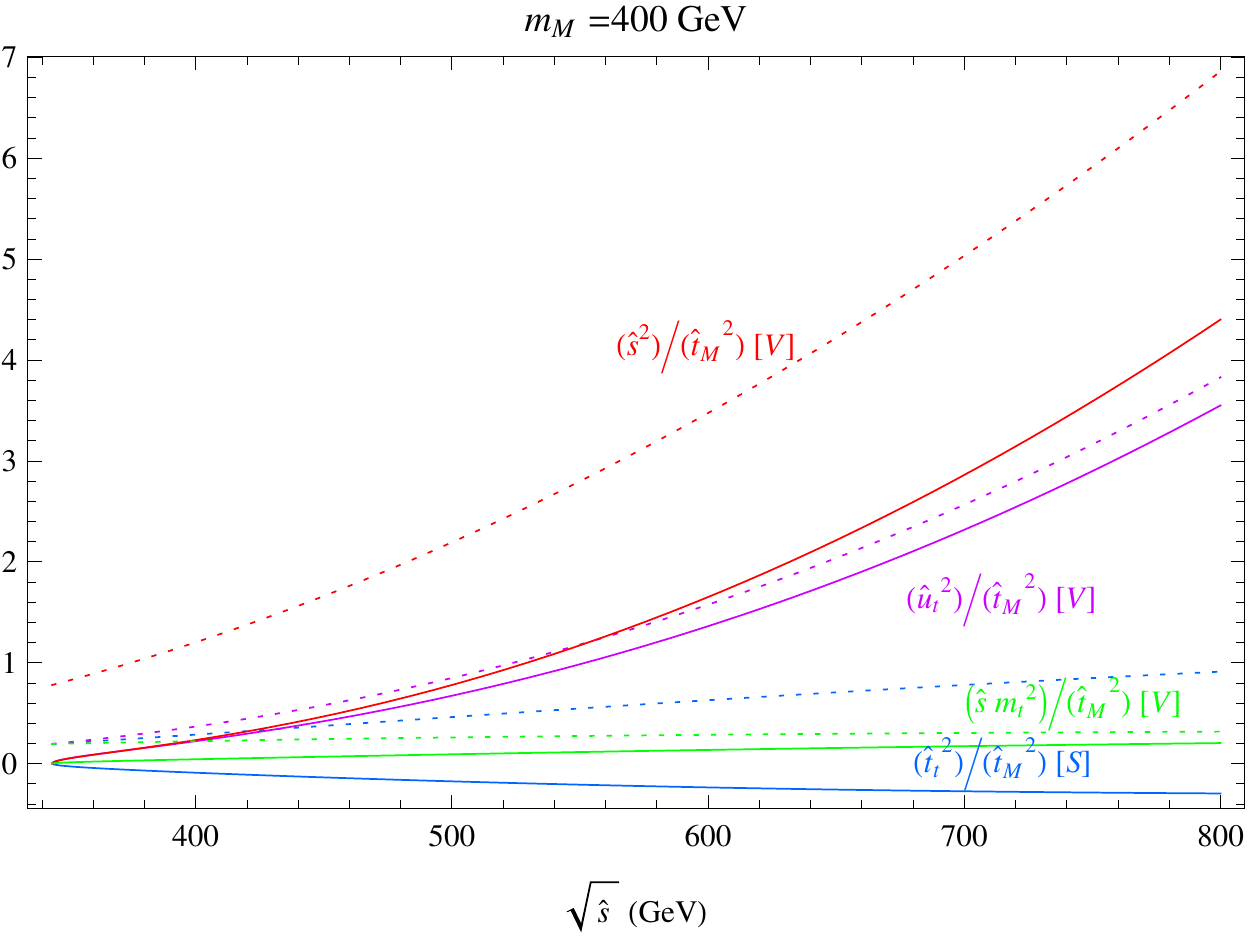} \includegraphics[width=0.32\textwidth]{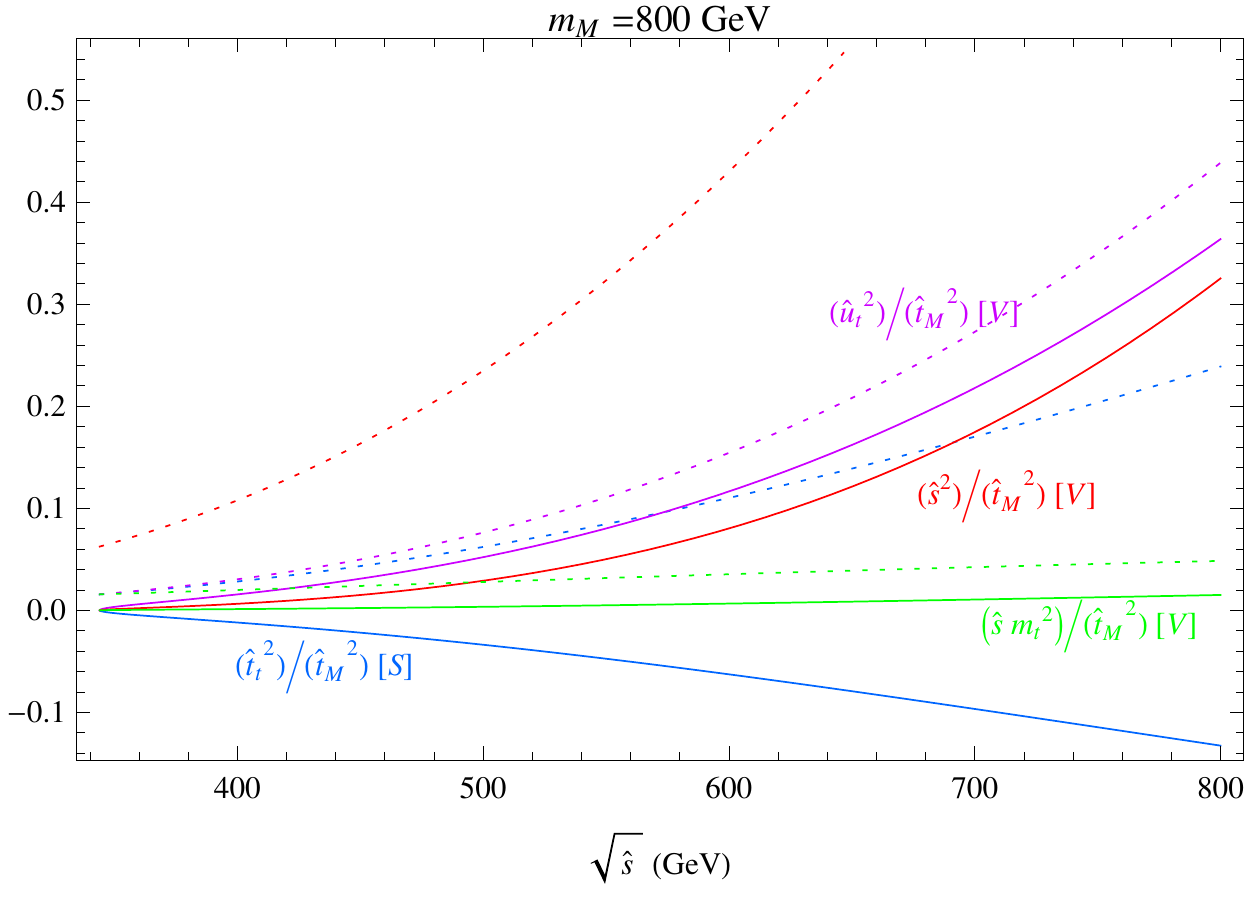}
\caption{Terms contributing to cross-sections with $t$ or $u$ channel mediators. Solid lines indicate the odd contribution and dotted the even contribution,  integrated over $\cos \theta$. The top plots include terms from the interference term, and the bottom plots from the NP squared term. For diquarks, $\hat{u} \leftrightarrow \hat{t}$, which flips the sign of the odd contribution and leaves the even contribution the same. The letters in square brackets indicate whether the term appears for scalar [S], vector [V], or both [S,V] mediators.}\label{fig: int terms}
\end{figure}

We examine this in detail in Fig.~(\ref{fig: int terms}) for different types of effective vertices.  The salient points to take away from the figures are:  (1) Scalars have odd contributions comparable to vectors only in the higher mediator mass range. (2) As a function of energy, the magnitude of the odd term for a given contribution is never greater than the magnitude of the even term, though some terms obtain much closer to equal magnitudes than others. Thus in order to best succeed in generating a sizable positive asymmetry while not destroying the invariant mass distribution, an ideal model will involve destructive interference between the even parts of the SM-NP interference and NP squared terms of the amplitude, and minimal or constructive interference between the odd part of the SM-NP interference and NP squared terms.\footnote{That some amount of destructive interference is favored by the data was noted in \cite{Grinstein:2011yv}.} By inspection, none of the scalars ($t$ or $u$-channel) can satisfy this condition. Scalar diquarks have some success in generating a substantial asymmetry in an intermediate mass range where the squared term contributes the dominant positive odd contribution. For the triplet, the interference term gives a negative odd and even contribution (so it helps to lower the cross-section but also lowers the numerator) while for the sextet the interference term enters with a minus sign and so gives a positive odd and even contribution (so it increases the numerator but also the cross-section).

\begin{figure}
\centering
\includegraphics[width=0.45\textwidth]{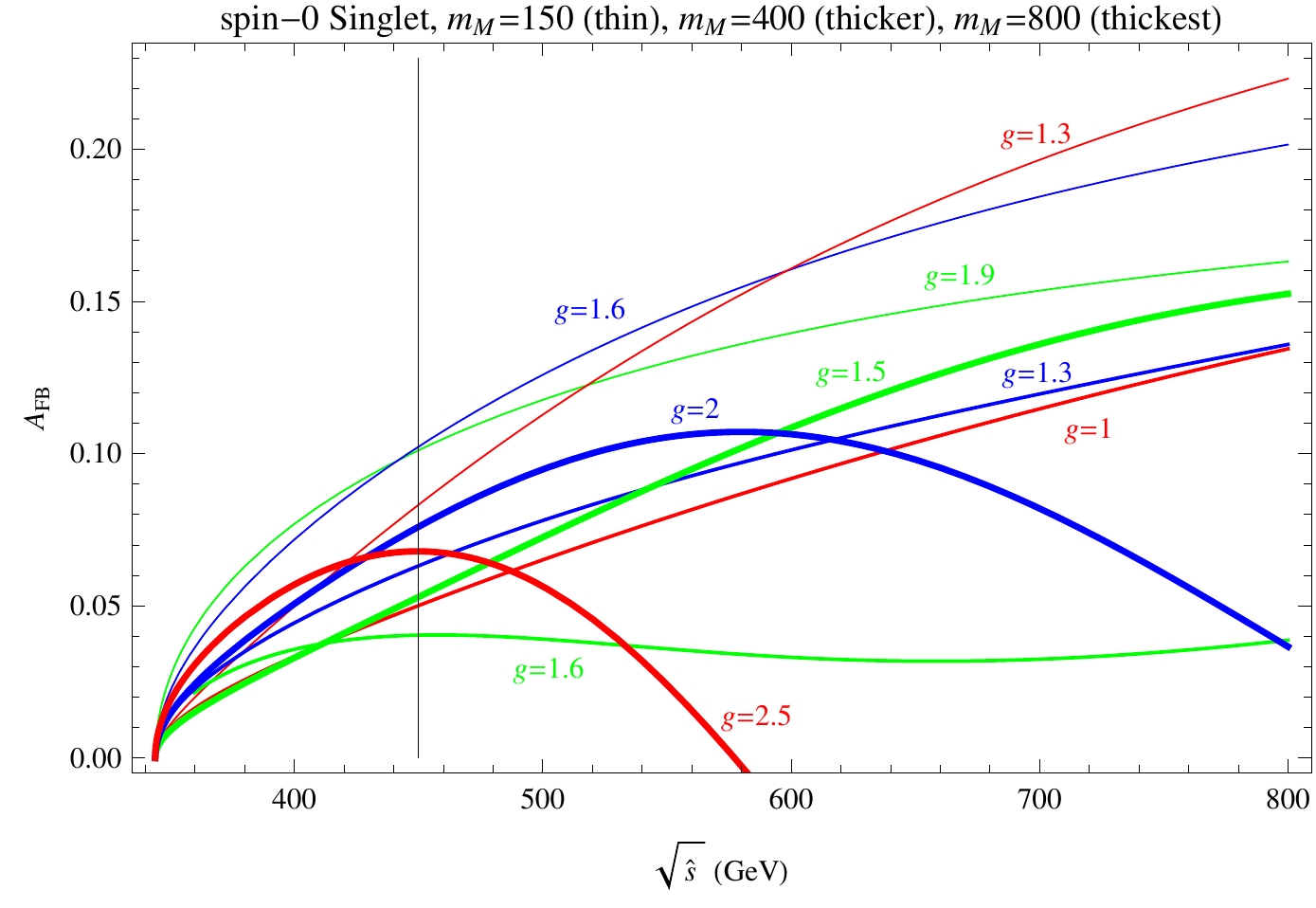} 
\includegraphics[width=0.45\textwidth]{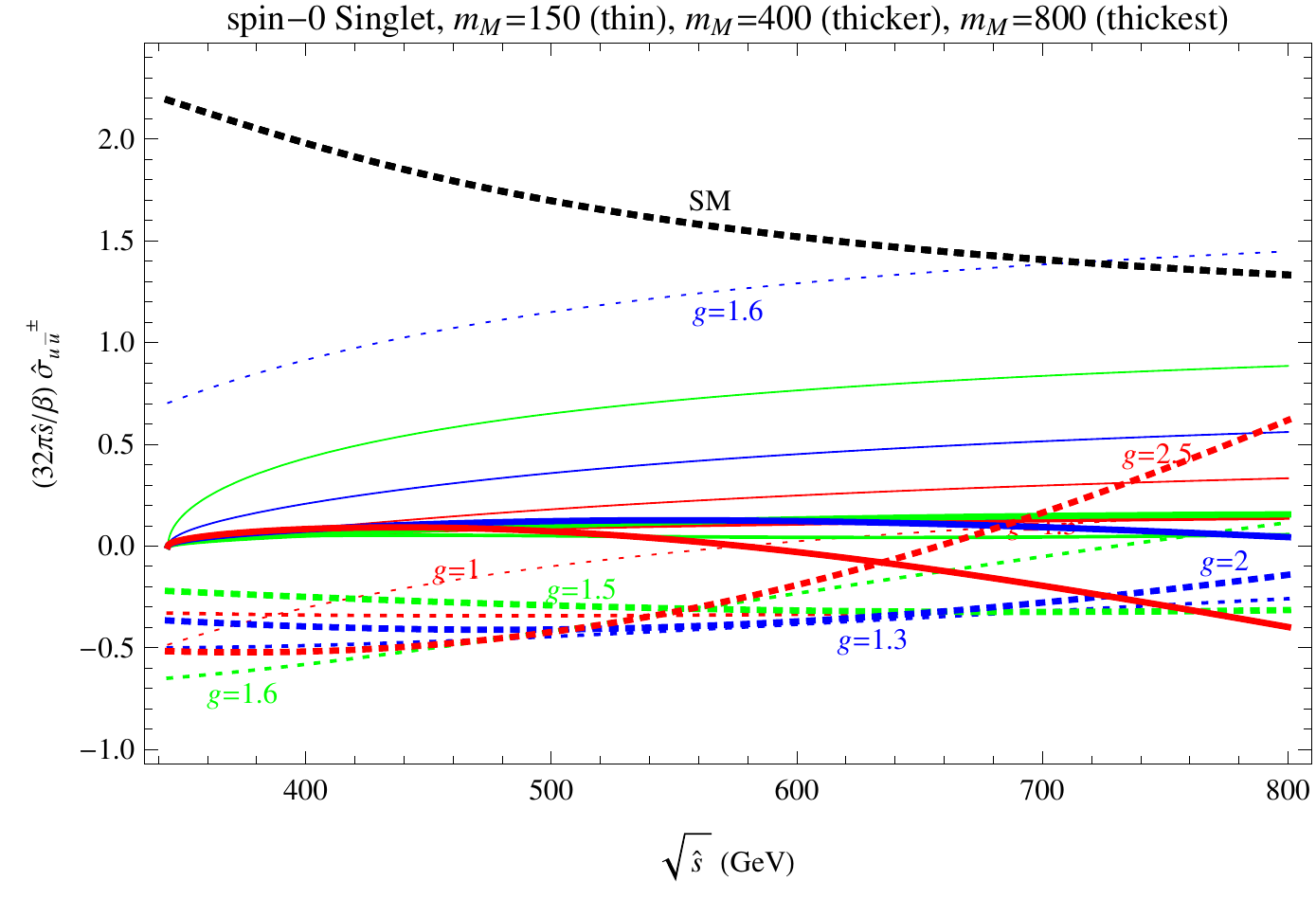}\\ 
\includegraphics[width=0.45\textwidth]{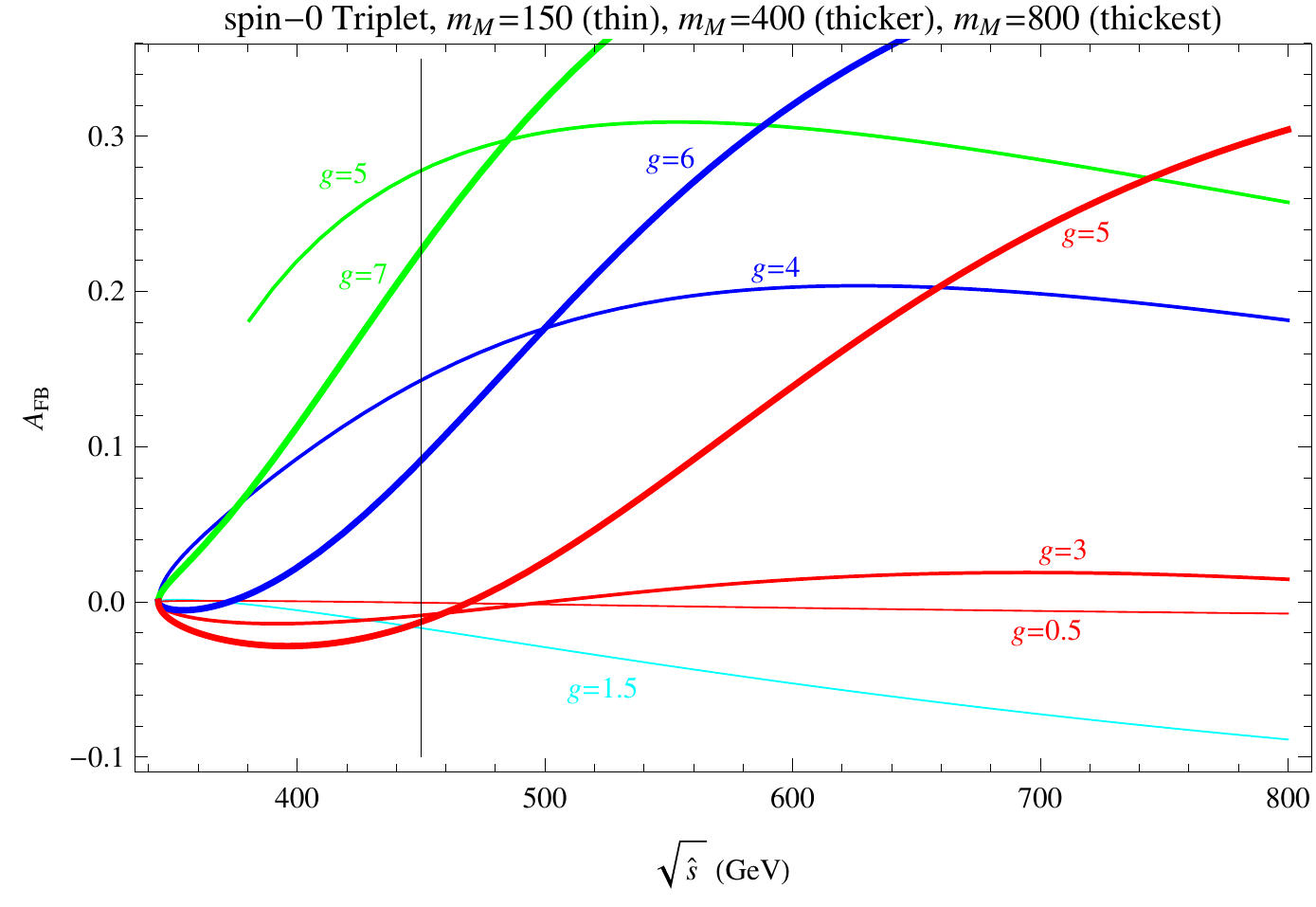} 
\includegraphics[width=0.45\textwidth]{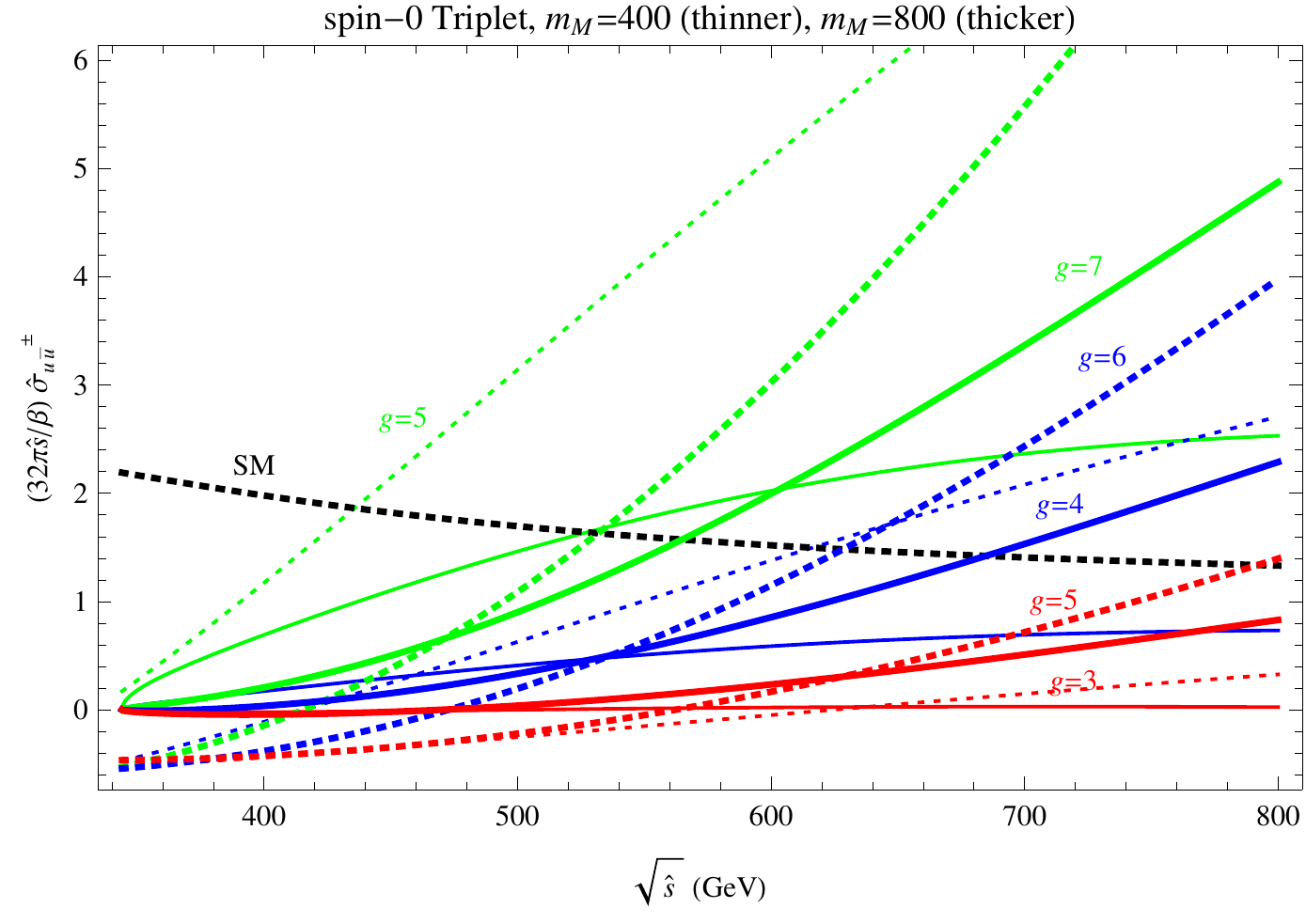}\\ 
\includegraphics[width=0.45\textwidth]{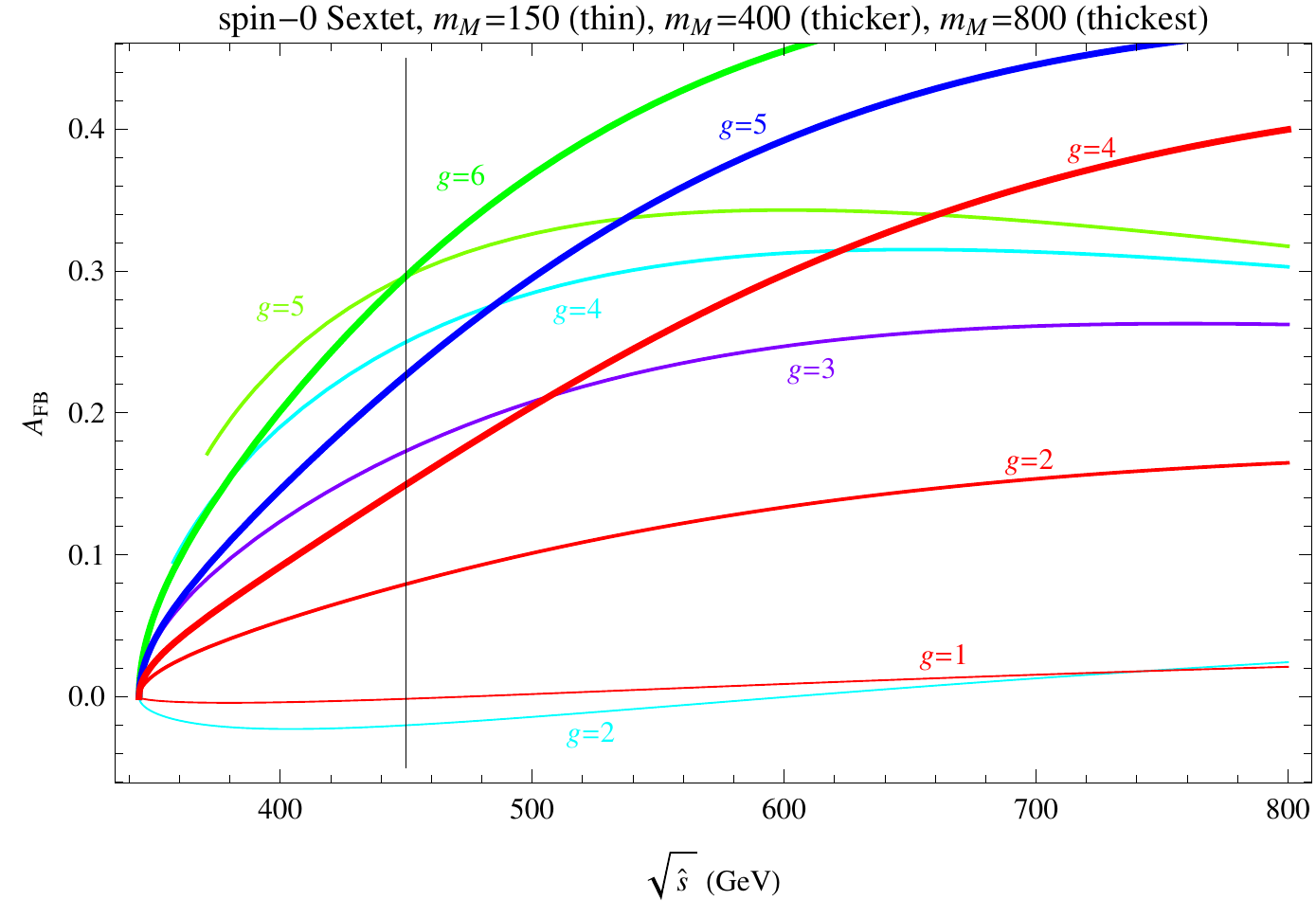} 
\includegraphics[width=0.45\textwidth]{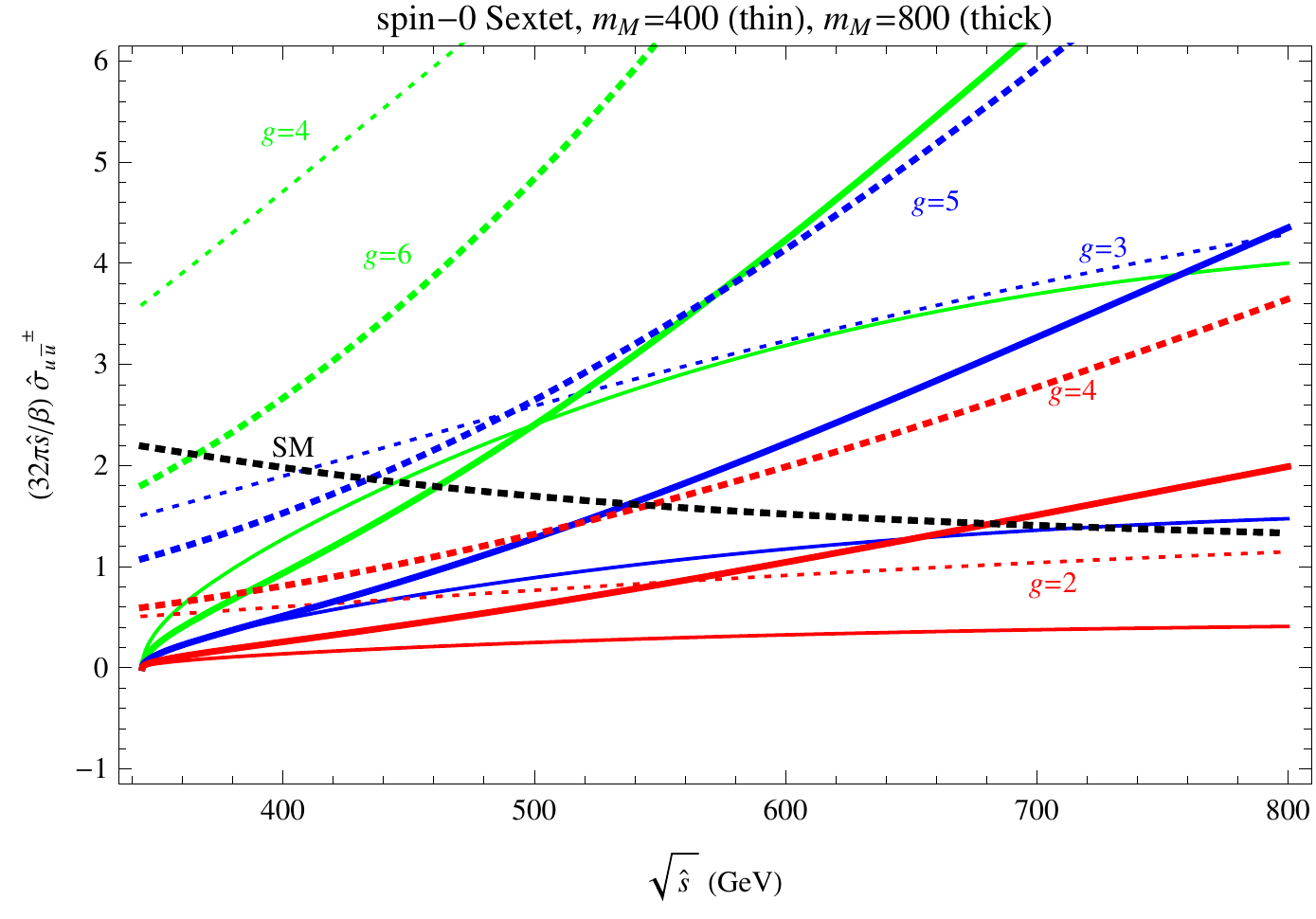} 
\caption{Spin-0 mediators. Left-hand plots show the differential asymmetry for various couplings given a 150 GeV, 400 GeV or 800 GeV mediator. A line is drawn at $\sqrt{\hat{s}} = 450~\text{GeV}$ to highlight the value of the asymmetry at the lower end of the the CDF analysis higher invariant mass bin. Due to the rapidly falling PDFs, the high invariant mass bin asymmetry will be given roughly by the value of the differential asymmetry at 450 GeV. The right-hand plots show contributions to the parton level $u \bar{u} \rightarrow t \bar{t}$ cross-section as a function of $\sqrt{\hat{s}}$ (dotted lines), and to the odd parton level cross-section (forward - backward), normalized by $32 \pi \hat{s} / \beta $ to make a dimensionless quantity. The effective Standard Model contribution as defined in \eqref{eq: sm contribution} is shown as a black dotted line.  Contributions to the total differential cross-section, $d \sigma_i (s) / d \hat{s}$, can be obtained from the dotted contributions by multiplying by the factor ${\beta F_{u \bar{u}} \over 32 \pi s \hat{s}}$. (See Eqs.~\eqref{eq: differential cross-section} and \eqref{eq: AFB formula}) }\label{fig: spin-0 mediators}
\end{figure}

\begin{figure}
\centering
\includegraphics[width=0.45\textwidth]{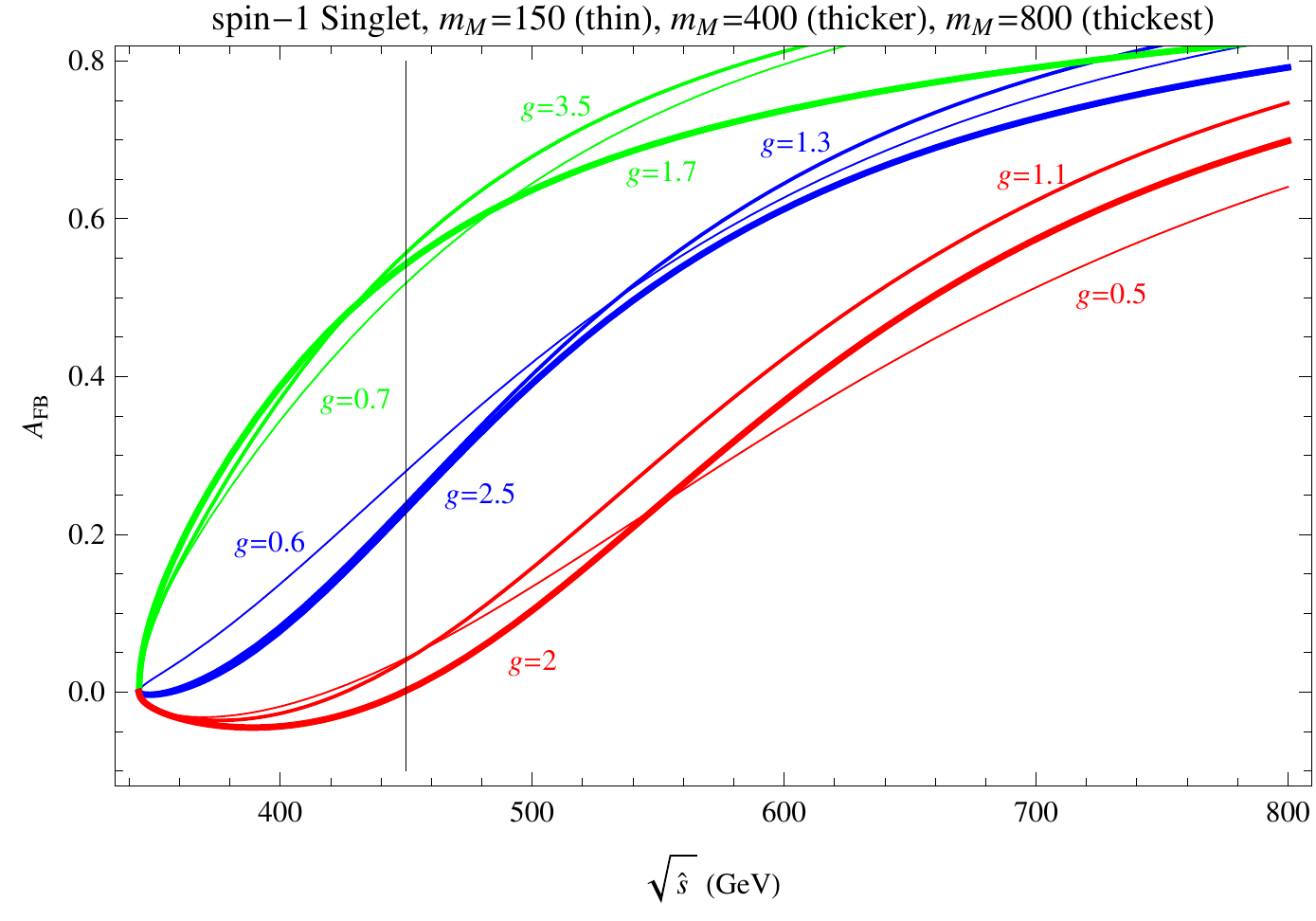} 
\includegraphics[width=0.45\textwidth]{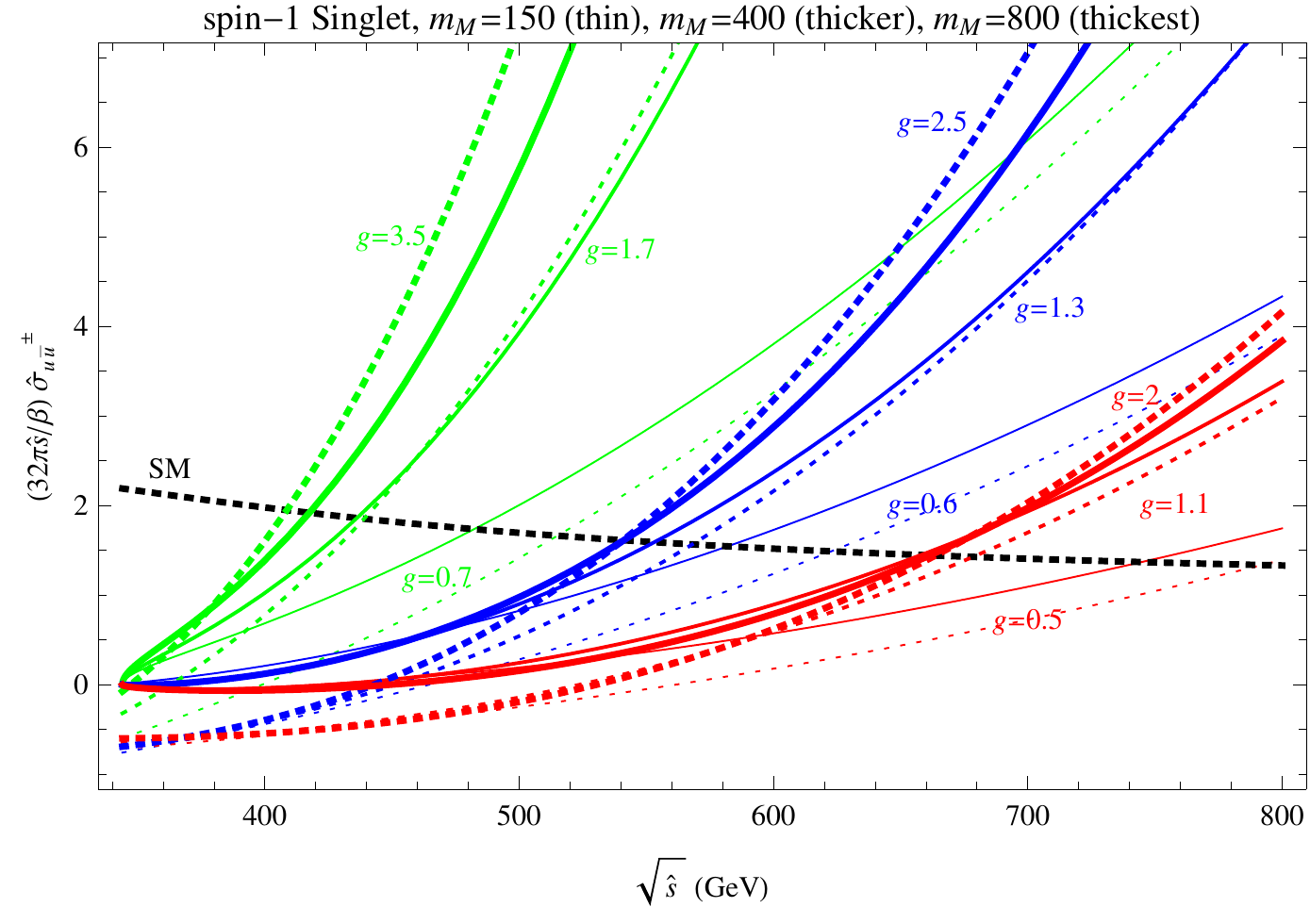}\\ 
\includegraphics[width=0.45\textwidth]{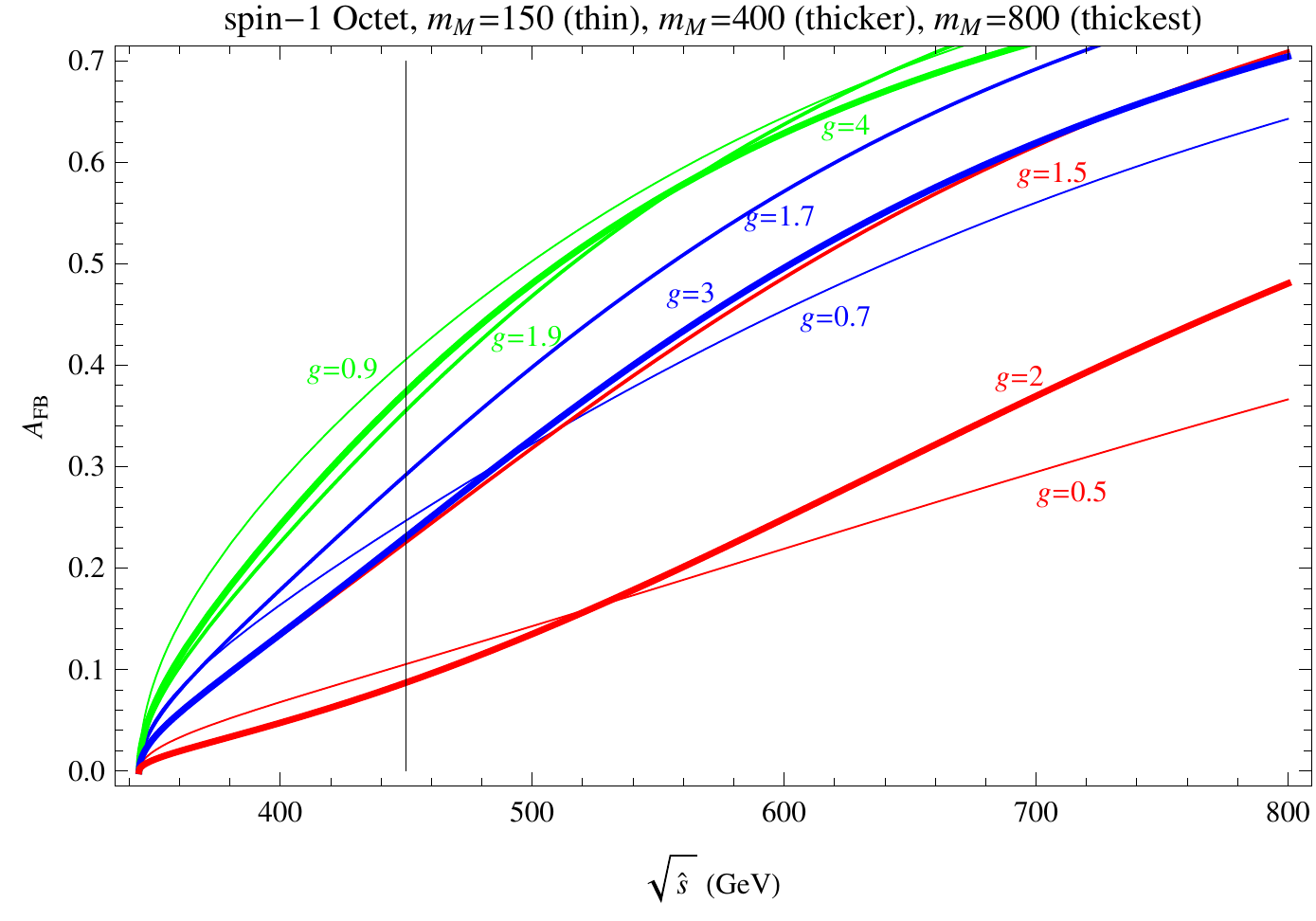} 
\includegraphics[width=0.45\textwidth]{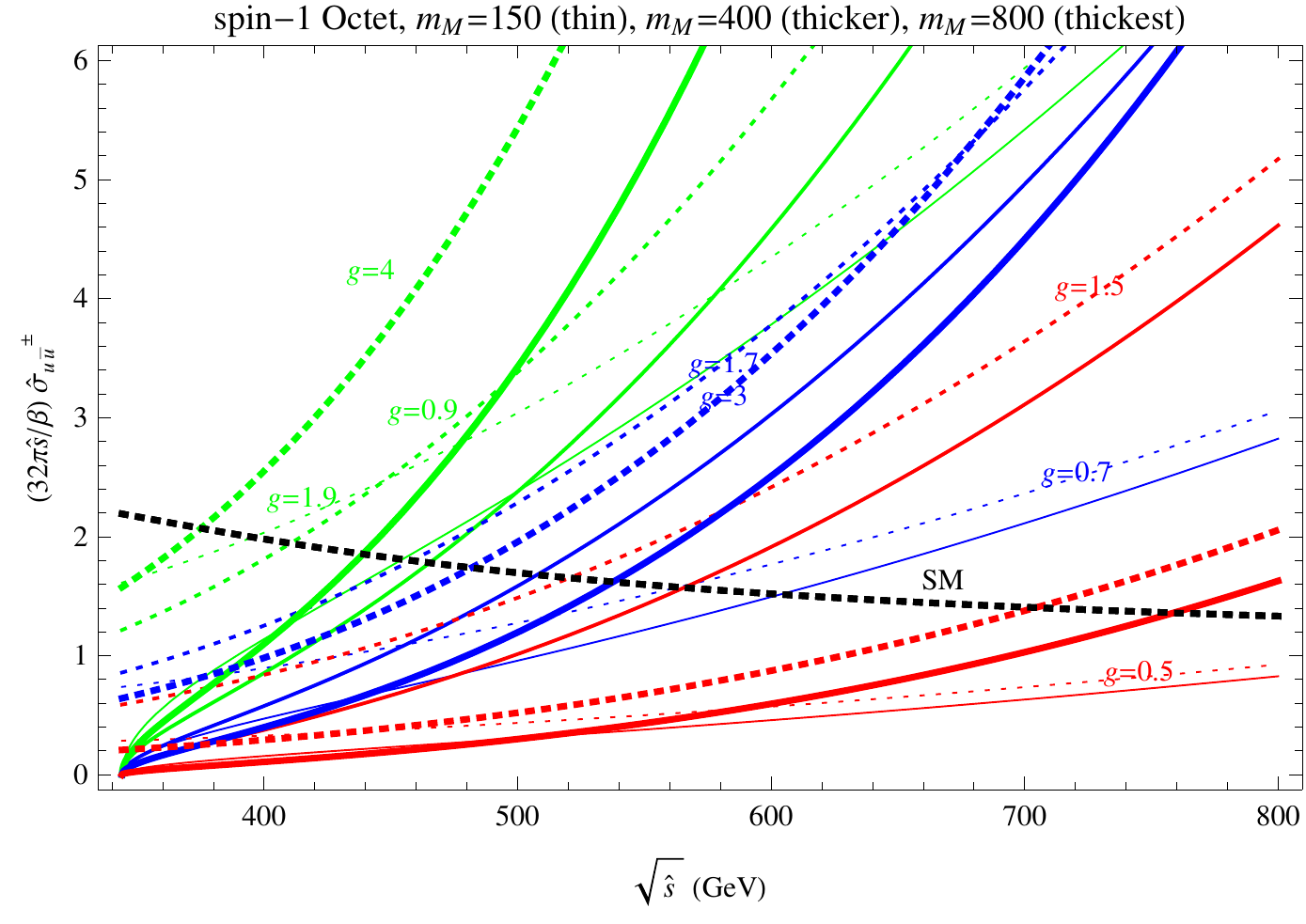} 
\caption{Spin-1 mediators. Left-hand plots show the differential asymmetry for various couplings given a 150 GeV, 400 GeV or 800 GeV mediator. A line is drawn at $\sqrt{\hat{s}} = 450~\text{GeV}$ to highlight the value of the asymmetry at the lower end of the the CDF analysis higher invariant mass bin. Due to the rapidly falling PDFs, the high invariant mass bin asymmetry will be given roughly by the value of the differential asymmetry at 450 GeV. Right-hand plots show contributions to the parton level $u \bar{u} \rightarrow t \bar{t}$ cross-section as a function of $\sqrt{\hat{s}}$, as in Fig. \ref{fig: spin-0 mediators}. The Standard Model contribution as defined in \eqref{eq: sm contribution} is shown as a black dotted line.}\label{fig: spin-1 mediators}
\end{figure}

\begin{figure}
\centering
\includegraphics[width=0.32\textwidth]{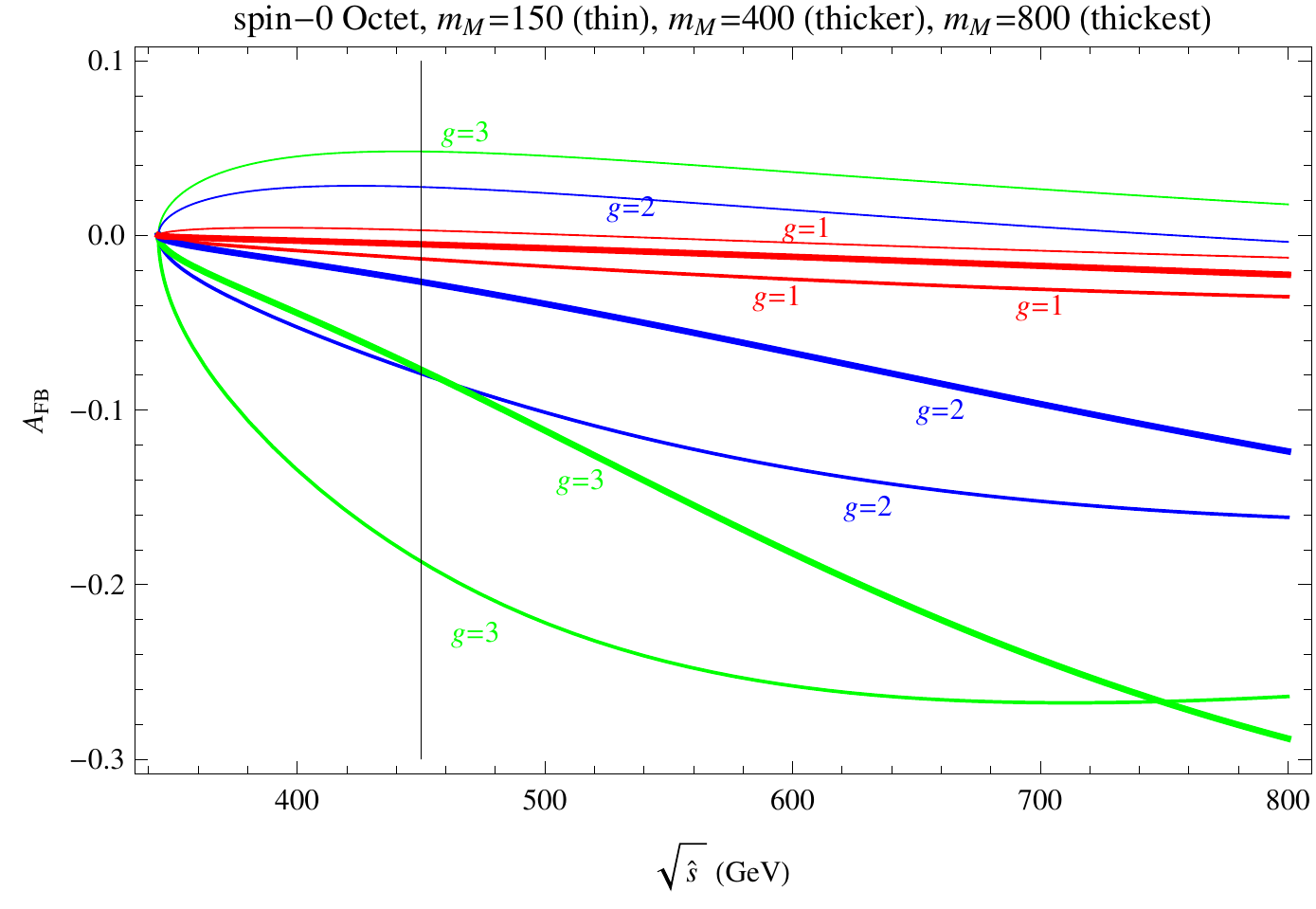}
\includegraphics[width=0.32\textwidth]{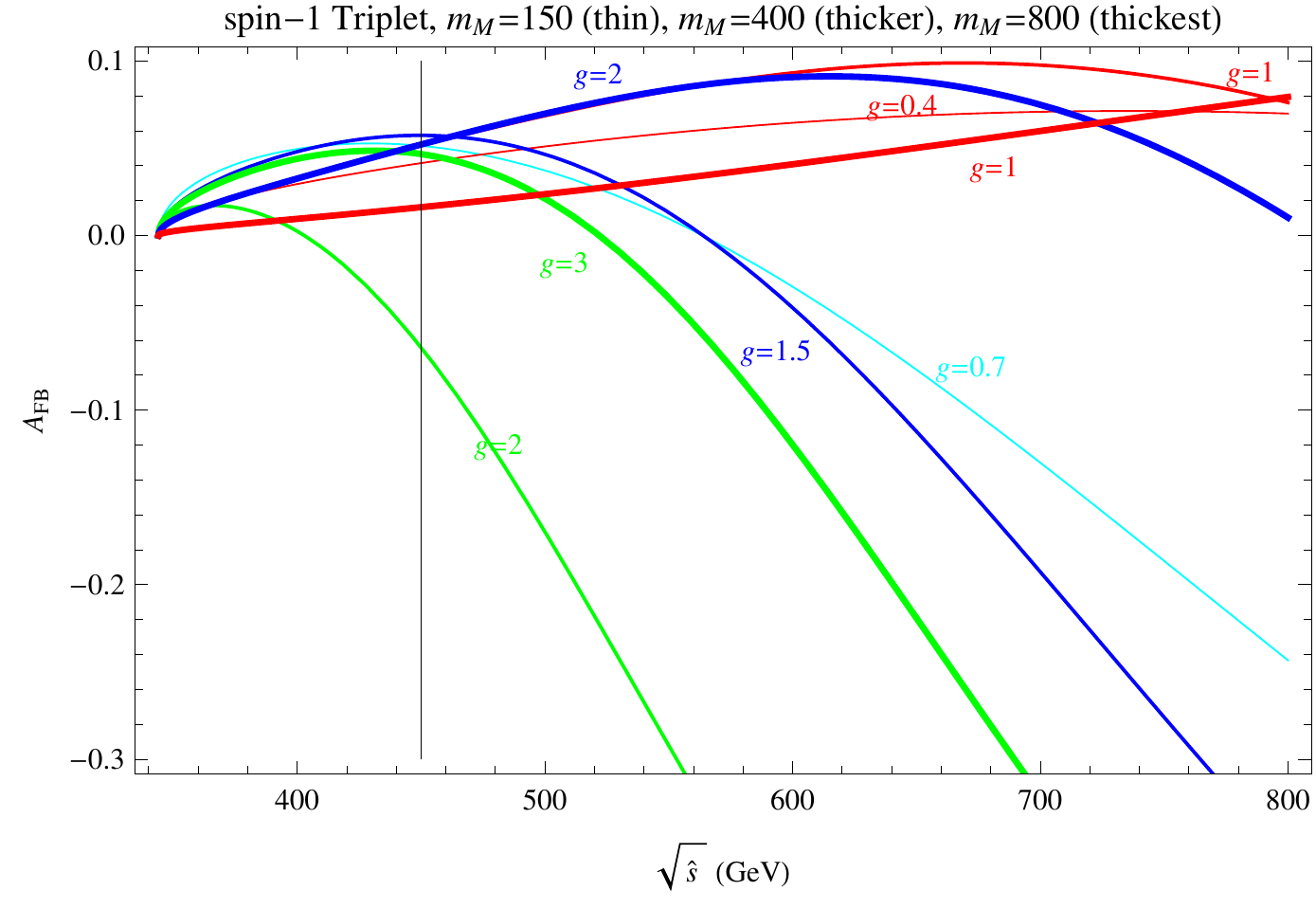}
\includegraphics[width=0.32\textwidth]{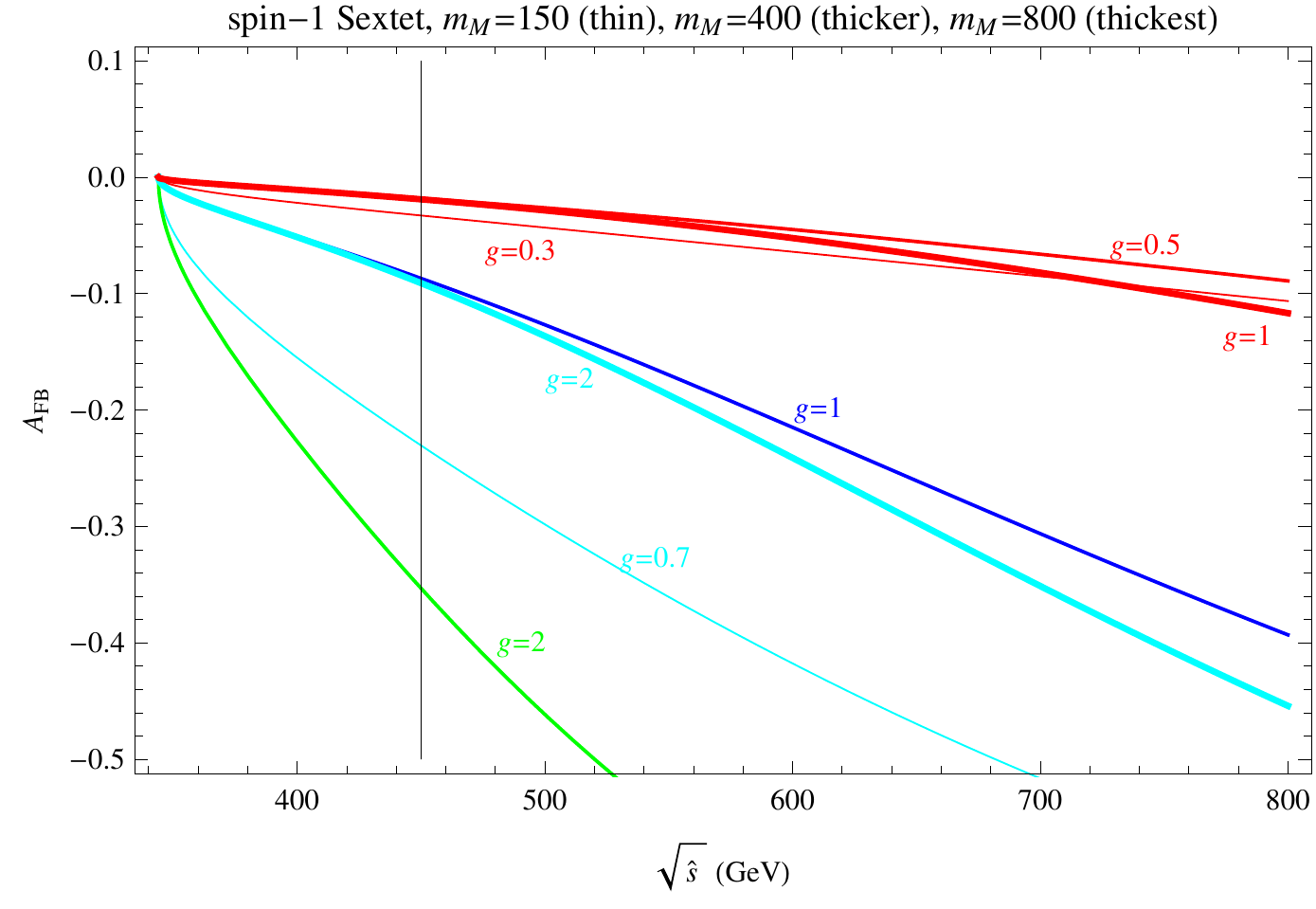}
\caption{The asymmetry for representations that cannot produce a positive asymmetry of more than a few percent.  }\label{fig: failures}
\end{figure}

For vectors there are more terms in play, so the story is a bit more complicated. To show the effects on the total asymmetry, we plot the total asymmetry (and, when relevant, cross-sections) for all $t$- and $u$-channel mediator color representations and spin combinations in Figs.~(\ref{fig: spin-0 mediators})-(\ref{fig: failures}). We show three benchmark mediator masses.   Fig.~(\ref{fig: spin-0 mediators}) shows the scalar models that succeed in generating a positive asymmetry, though in general for perturbative couplings it is not a large positive asymmetry; Fig.~(\ref{fig: spin-1 mediators}) shows the same for the vector mediator case, and it is seen that the contribution to the total asymmetry can be large for all mass ranges.  Lastly, we show in Fig.~(\ref{fig: failures}) the  mediators that fail to produce a positive asymmetry larger than 5\%.  These include the scalar color octet and vector triplet and sextet.  

\begin{figure}
\centering
\includegraphics[width=0.32\textwidth]{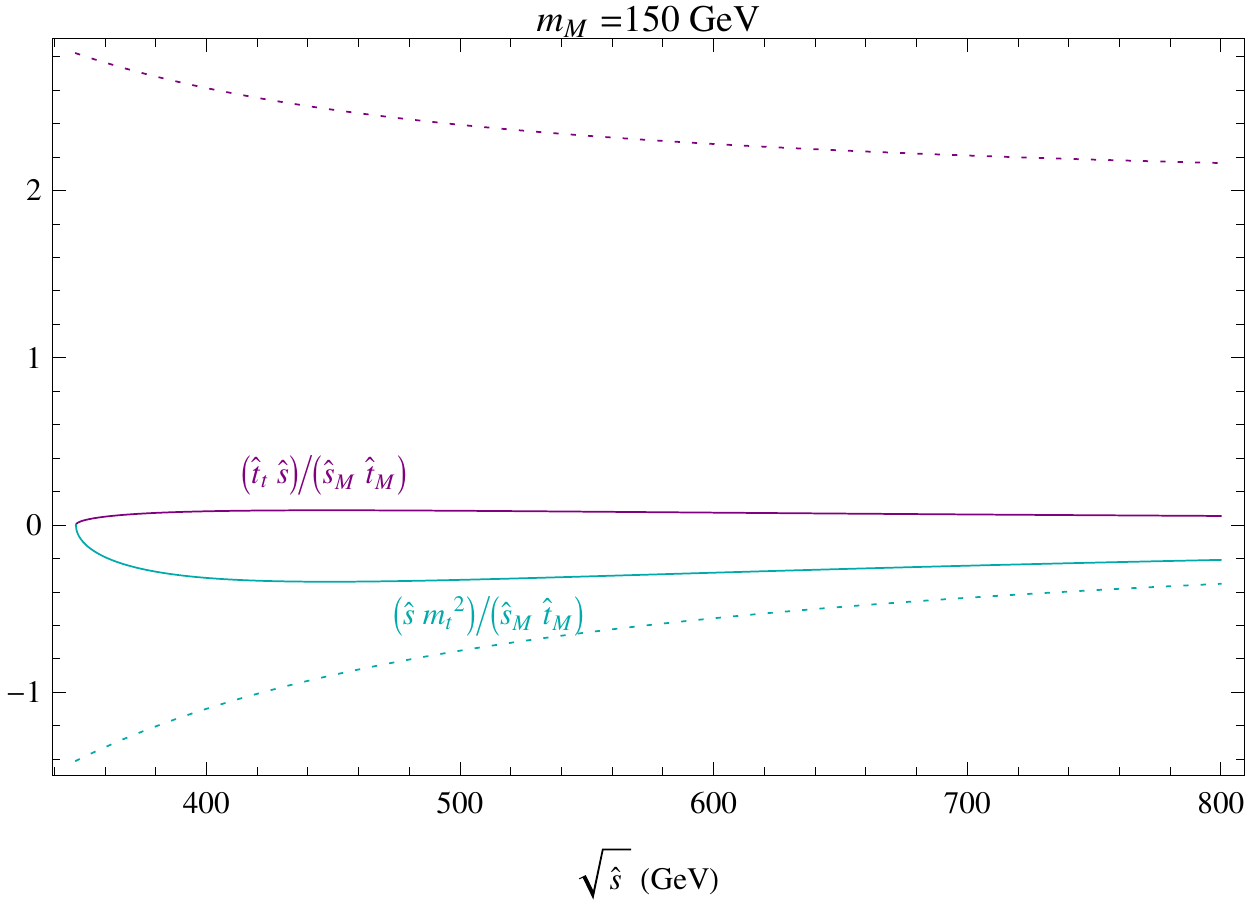}
\includegraphics[width=0.32\textwidth]{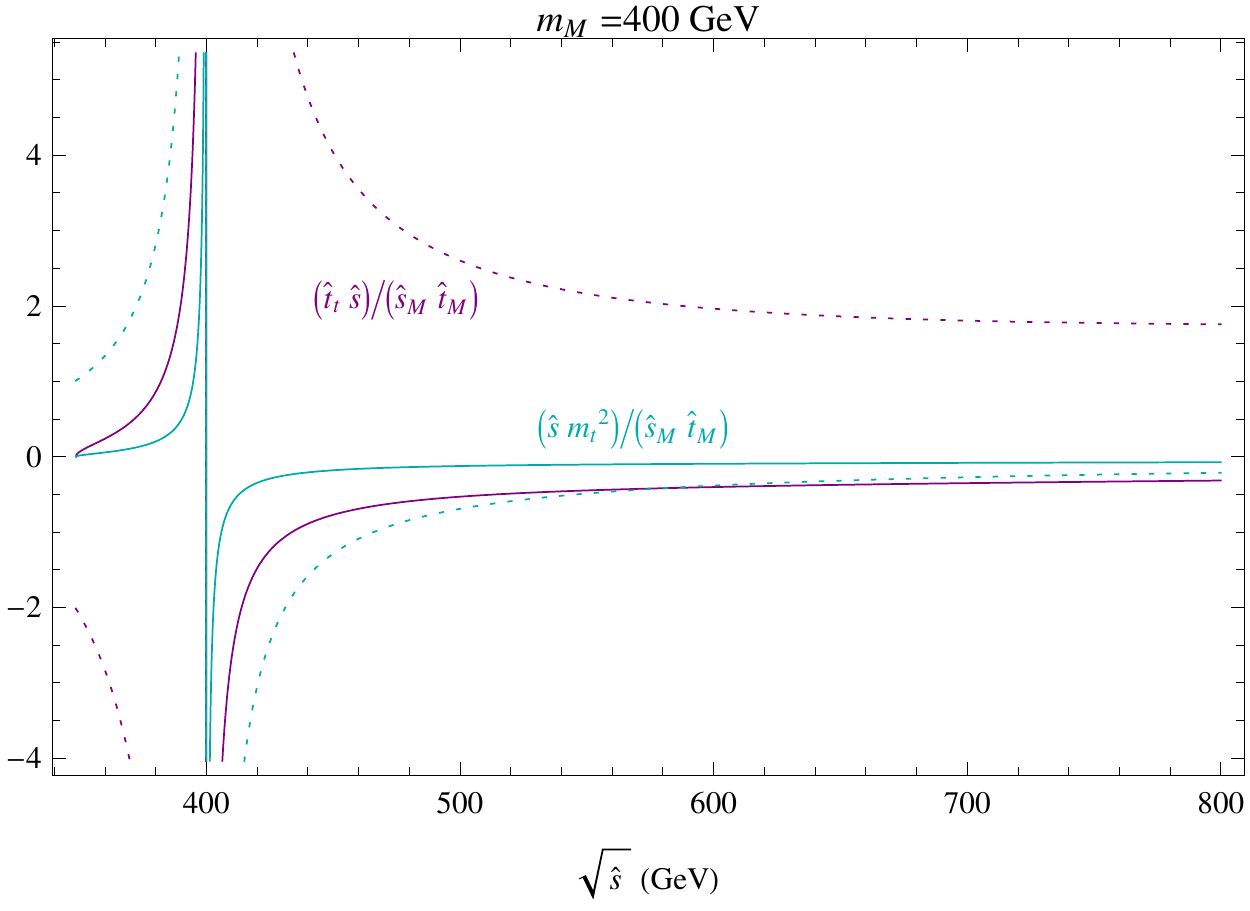}
\includegraphics[width=0.32\textwidth]{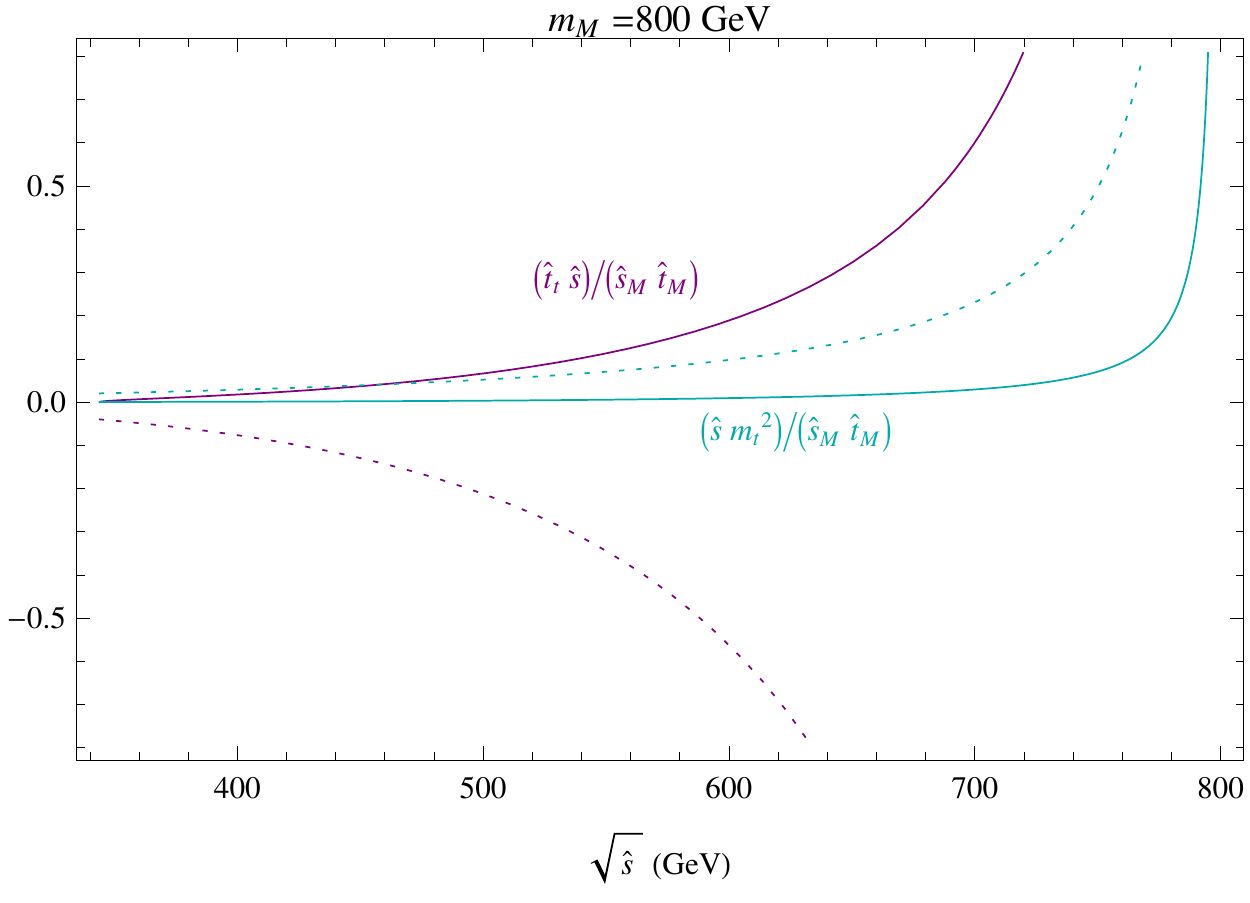}
\caption{Terms contributing to $s$-channel scalar / $t$-channel scalar interference cross-section. Solid lines are the odd contributions and dashed are the even contributions,  integrated over $\cos \theta$. Here we assume a narrow width.}\label{interference terms}
\end{figure}

One might wonder whether the asymmetry induced by scalars could be enhanced by adding another scalar with $s$-channel couplings to $u \bar{u}$ and $t \bar{t}$.  This is predicted, for example, by the flavor triplet models.  Interference between a $t$-channel scalar with mass $m_1$ and an $s$-channel scalar with mass $m_2$ would give rise to terms of the form ${ \hat{s} \hat{t}_t \over (\hat{s} - m_2^2) \hat{t}_1 }$ and ${ \hat{s} m_t^2 \over (\hat{s} - m_2^2) \hat{t}_1 }$. These contributions, assuming $m_1 = m_2$, are shown in Fig.~(\ref{interference terms}). For mediators lighter than the top quark, the odd contribution has the same sign as even for both terms, and it is hard to see how these contributions can enhance the asymmetry while not increasing the total cross-section to unacceptable levels.
For mediators heavier than the top quark, odd and even contributions for the $\hat{s} \hat{t}_t / (\hat{s}_M \hat{t}_M)$ have the opposite sign---the odd contribution is positive for energies below the mediator mass and negative above. This interference could have interesting implications for models involving both $t$-channel and $s$-channel scalars of intermediate mass.  Diquarks with $s$-channel interactions would not contribute to the $t \bar{t}$ cross-section or $A_{FB}$.

Lastly, we briefly discuss $s$-channel mediators, which can give rise to a large asymmetry for an appropriate choice of couplings and masses.  If the asymmetry is generated from $s$-channel NP interactions, then the $\cos \theta$ dependence of the  NP cross-section is a simple quadratic polynomial. The axigluon originally proposed in Ref.~\cite{Ferrario:2009bz,Frampton:2009rk} supposed a heavy axigluon to evade dijet and $t\bar{t}$ resonance searches that strongly constrain the state with masses below $\sim 2 \mbox{ TeV}$.  However, recently different regions of the axigluon parameter space have been explored.  For example, a 750 GeV state was considered in \cite{AguilarSaavedra:2011ci}, with the dijet constraints evaded by making the coupling to the top quark much larger than to the up quarks.  A 400 GeV state was considered in \cite{Tavares:2011zg}, and the $t\bar{t}$ resonance search constraints evaded by making the state sufficiently broad.  Lastly, if a vector with diagonal axial couplings to top and up has a mass slightly lighter than the top mass, then it will not show up as a resonance in the $t\bar{t}$ spectrum. We refer the reader to these references for details, though we include the axigluons in our scans of parameter space in the next section for completeness.

\section{Comprehensive Search for Models at the Tevatron}
\label{sec:tevatron}

To augment the conclusions of the previous section, we carry out a comprehensive representative scan of models using {\tt MadGraph}\cite{Alwall:2011uj}; the details of our procedure are discussed in Appendix \ref{app: madgraph}.  
We scan over $s,~t,~u$-channel models, characterized by a single new mediator of given spin and color representation \eqref{eq: NP lagrangian}; we scan over all such models that can produce a positive asymmetry of more than a few percent while remaining (somewhat) perturbative (coupling $\lesssim 6$) and contributing less than order 50\% to the total cross-section in the mass range 200 GeV - 2 TeV.\footnote{We neglect models with mass below the top ({\em e.g.} \cite{Jung:2011zv}); in general, these models will tend to rather severely overproduce the total cross-section and number of additional jets at the LHC  and/or lead to large contributions to single top production, depending on the details of the mediator decay channels.}   The models scanned are summarized in Table \ref{tab: scan summary table}. We choose representative models that generate the largest asymmetry.  For $t$-channel models we focus on mediators connecting up to top, both because they generate a large asymmetry, and also because a light neutral state runs into few constraints.   The color singlet and triplet are our representative scalar models, though neither is successful in generating a large asymmetry, as we detailed earlier.  Also note that the singlet scalar is part of an electroweak doublet, though we choose to couple this scalar to $t_L-u_R$ so that only one state is operative for the forward-backward asymmetry.  The charged component of the $SU(2)$ mediator multiplet will contribute to $b\bar{b}$ plus jet events at the LHC, but this will be easily overwhelmed by the background.  For the $t$-channel flavor-violating models, we consider both a color singlet vector (C1V) and octet vector (C8V) that couples only to right-handed states.  We also consider a flavor octet, color singlet vector (F8C1V) that couples to $\bar{U}_R \gamma^\mu U_R$, where now the up quarks are in an octet of $SU(3)_{U_R}$.  Lastly, the $s$-channel axiglue type models are considered, both in flavor universal \cite{Tavares:2011zg} and non-universal \cite{Frampton:2009rk,AguilarSaavedra:2011ci} varieties.

\begin{table}
\begin{tabular}{c c c c c c p{0.4\textwidth}}
~Model~ 	& ~Spin~ 	& ~Color~         &$SU(2)_Y$~        &Flavor 	& ~$s$-, $t$-, $u$-?~ 	& Comments and References \\
\hline \hline
C1S			& 0		& 1		& $2_{1/2}$			& 1		& $t$						& Only very moderate asymmetries achievable  $\mathcal{O}(\gtrsim 10 \%)$. Low mass ($m_M \simeq m_t$) states do slightly better  \cite{Nelson:2011us}. \\
C3S			& 0		& 3		&$1_{4/3}$  		& 1		& $u$						& a.k.a. triplet diquark. $q = 4/3$ \cite{Shu:2009xf}.  \\
C1V			& 1		& 1		&$1_0$			&1		& $t$						& a.k.a. $Z'$ or $W'$ \cite{Jung:2011zv}. \\
C8V			& 1		& 8		& $1_0$			& 1		&$t$						& \\
F8C1V		& 1		& 1		& $1_0$			& 8		&$t$, $s$					&Flavor breaking only through up Yukawa \cite{Grinstein:2011yv}.\\
schanC8V(A,R)		& 1		& 8	&$1_0$			& 1		& $s$						& a.k.a. axigluon or coloron. For $ 2 m_t < m_M \lesssim 2 {\rm TeV}$, very broad width required to avoid $t \bar{t}$ resonance searches \cite{Frampton:2009rk,Cao:2010zb,AguilarSaavedra:2011ci}.  \\
schanC8V	$\Gamma$	& 1		& 8	&$1_0$			& 1		& $s$						& $\sim 400$ GeV broad resonance via additional scalars. Universal quark couplings \cite{Tavares:2011zg}. \\
\hline
\end{tabular}
\caption{Summary of models scanned. All $t$- or $u$- channel states are taken to be non-self-conjugate.}\label{tab: scan summary table}
\end{table}

\begin{figure}
\includegraphics[width = 0.85\textwidth]{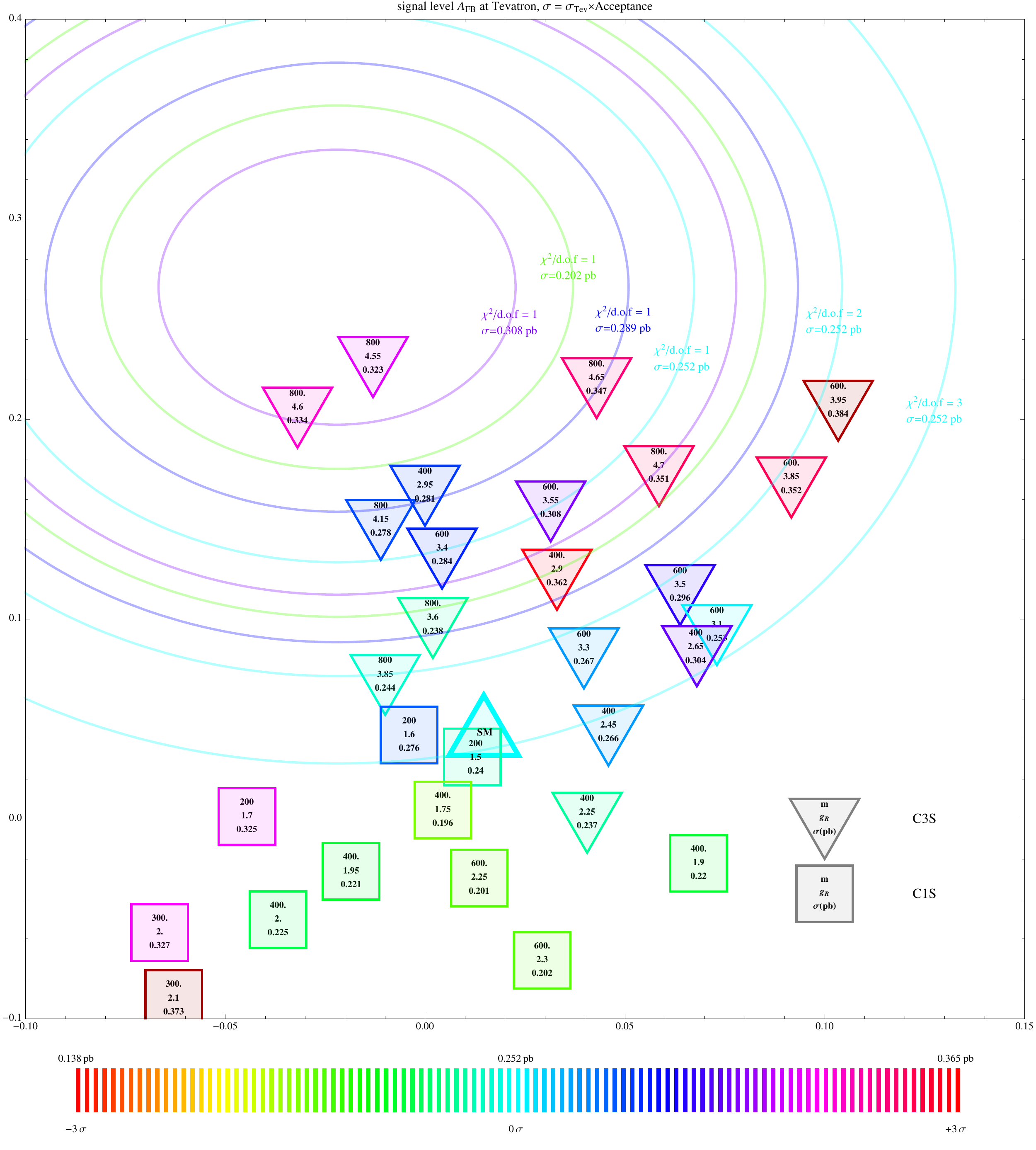}
\caption{Scatter plots depicting simulated signal level $ \{ A_{FB}(m_{t \bar{t}} < 450~\text{GeV}),  A_{FB}(m_{t \bar{t}} > 450~\text{GeV}), \sigma_{t \bar{t}} \times \text{acceptance}
\} $ at Tevatron CM energy for $t$-channel flavor-changing scalar models listed in Table \ref{tab: scan summary table}.  The models are labeled by the mass of the mediator, the coupling to right-handed quarks, and the total Tevatron production cross-section times acceptance.  The cross-sections are compared against the SM cross-section times acceptance which yields 0.252 pb at the LO; the color scales for the models indicate the deviation from the SM cross-section, as indicated by the legend at the bottom. The curves indicate constant $\chi^2$ for a given cross-section, as defined in Eq. \eqref{chisq}. Contours for four cross-section values (cyan, blue, green, purple) are shown for $\chi^2/\text{d.o.f.} =$ 1 and 2.  A single (cyan) $\chi^2/\text{d.o.f.} = 3$ contour with SM cross-section is shown. Model points of a given color should be compared to $\chi^2$ contours of the same color. }
\label{tev scatter scalar}
\end{figure}

\begin{figure}
\includegraphics[width = 0.85\textwidth]{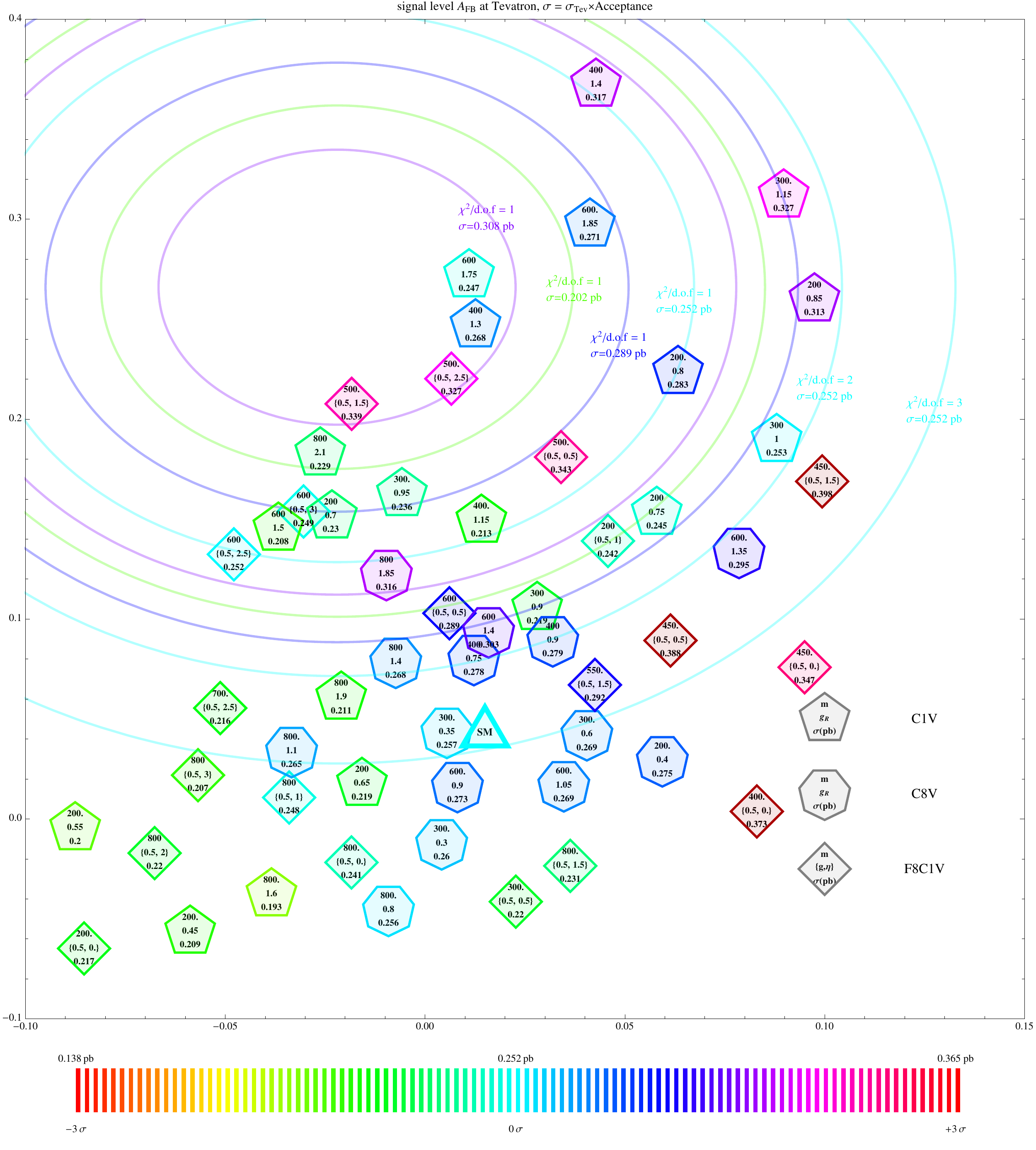}
\caption{Scatter plots depicting simulated signal level $ \{ A_{FB}(m_{t \bar{t}} < 450~\text{GeV}),  A_{FB}(m_{t \bar{t}} > 450~\text{GeV}), \sigma_{t \bar{t}} \times \text{Acceptance}
\} $ at Tevatron CM energy for $t$-channel flavor-changing vector models listed in Table \ref{tab: scan summary table}.  The models are labeled by the mass of the mediator, the coupling, and the total Tevatron production cross-section times acceptance.  The coupling conventions are discussed in detail in the text.  The cross-sections are compared against the SM cross-section times acceptance which yields 0.252 pb at the LO; the color scales for the models indicate the deviation from the SM cross-section, as indicated by the legend at the bottom.  The curves indicate constant $\chi^2$ for a given cross-section, as defined in Eq. \eqref{chisq}. Contours for four cross-section values (cyan, blue, green, purple) are shown for $\chi^2/\text{d.o.f.} =$ 1 and 2.  A single (cyan) $\chi^2/\text{d.o.f.} = 3$ contour with SM cross-section is shown. Model points of a given color should be compared to $\chi^2$ contours of the same color. }
\label{tev scatter vector}
\end{figure}

\begin{figure}
\includegraphics[width = 0.85\textwidth]{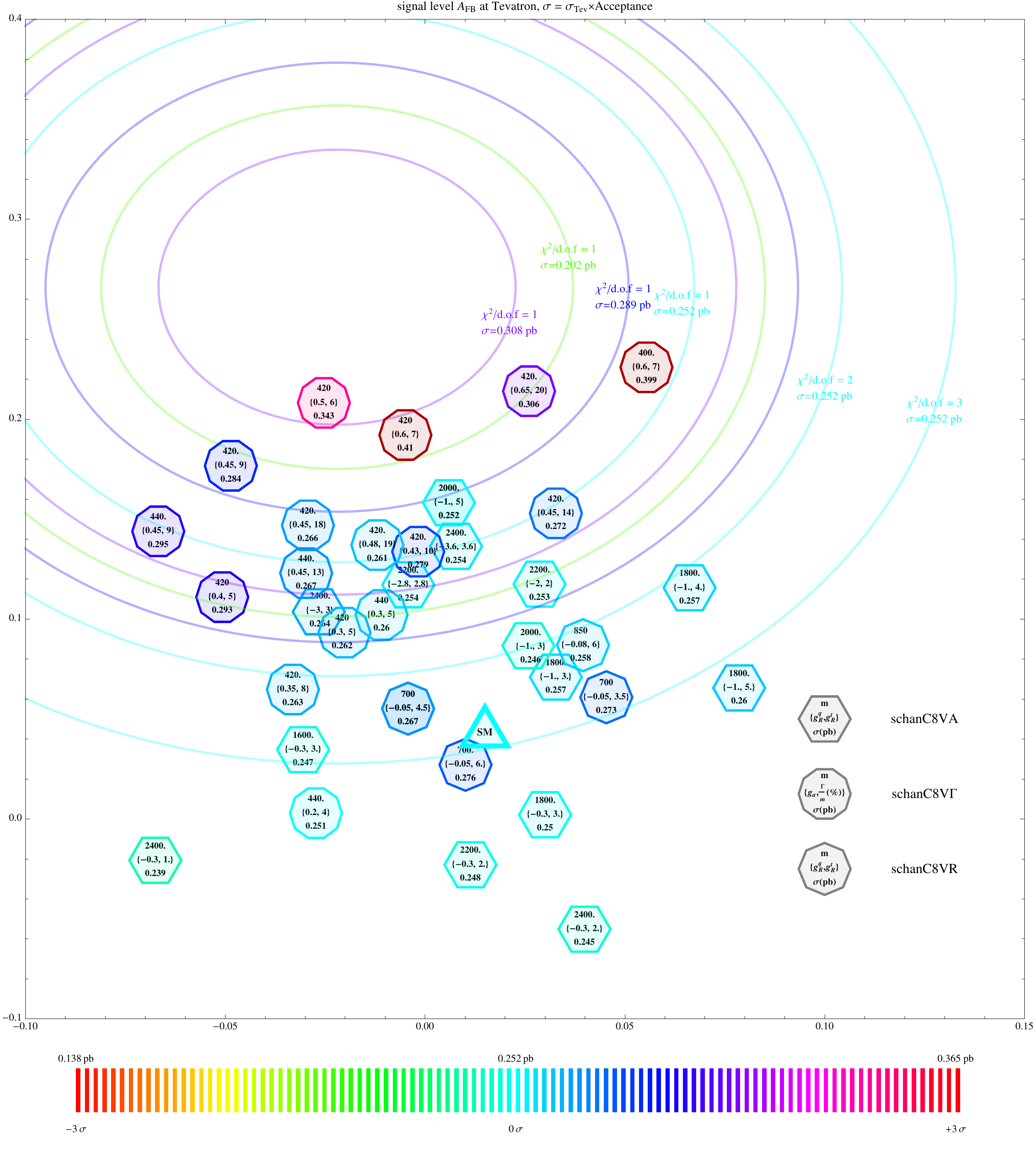}
\caption{Scatter plots depicting simulated signal level $ \{ A_{FB}(m_{t \bar{t}} < 450~\text{GeV}),  A_{FB}(m_{t \bar{t}} > 450~\text{GeV}), \sigma_{t \bar{t}} \times \text{Acceptance}
\} $ at Tevatron CM energy for axigluon models listed in Table \ref{tab: scan summary table}.  The models are labeled by the mass of the mediator, the coupling, and the total Tevatron production cross-section times acceptance.  The coupling conventions are discussed in detail in the text.  The cross-sections are compared against the SM cross-section times acceptance which yields 0.252 pb at the LO; the color scales for the models indicate the deviation from the SM cross-section, as indicated by the legend at the bottom.  The curves indicate constant $\chi^2$ for a given cross-section, as defined in Eq. \eqref{chisq}. Contours for four cross-section values (cyan, blue, green, purple) are shown for $\chi^2/\text{d.o.f.} =$ 1 and 2.  A single (cyan) $\chi^2/\text{d.o.f.} = 3$ contour with SM cross-section is shown. Model points of a given color should be compared to $\chi^2$ contours of the same color. }
\label{tev scatter schan}
\end{figure}

The results of this scan for the Tevatron are shown in Figs.~\eqref{tev scatter scalar} - \eqref{tev scatter schan}.  The coupling conventions in the figures are as follows.  The $t$-channel scalars, as well as C1V and C8V, models  are labeled by their coupling to RH quarks, with $g_L = 0$.  The flavor symmetric F8C1V model has an additional parameter $\eta$ that controls the flavor breaking coupling to the top quarks such that couplings to top-quarks have couplings  $\sim g_R + 2 \eta m_t^2/v^2$, with $v = 246 \mbox{ GeV}$.  The coupling conventions for $s$-channel models are more complicated.  The couplings in schanC8V$\Gamma$ and schanC8VA are purely axial ($g_R = -g_L$), with the former only being flavor universal.   The schanC8V$\Gamma$ model has an independent width parameter \cite{Tavares:2011zg}, which was scanned over to find models with maximally large asymmetries per unit production cross-section.  schanC8VR has non-universal couplings to right-handed quarks \cite{AguilarSaavedra:2011ci}.

We apply cuts on the simulated sample and fully reconstruct tops as described in Appendix \ref{app: madgraph} to mimmic the analysis in \cite{Aaltonen:2011kc}. 
More specifically, for Tevatron events we apply the following sets of cuts: 
\begin{itemize}
\item Exactly one electron or muon with $p_T > 20$ GeV and $|\eta| < 1.0$.
\item At least four jets with $p_T > 20$ GeV and $|\eta| < 2.0$, with at least one of the jets having a $b$-tag.  
\item $E_T^{miss} > 20$ GeV. 
\end{itemize}
We reconstruct tops as described in \cite{Gresham:2011pa}, doing a likelihood analysis on the lepton and jet kinematics to the $t \bar{t}$ hypothesis, using the algorithm described in our previous paper \cite{Gresham:2011dg}.  
 The cone jet algorithm was used for Tevatron events. Jet energy scale corrections were carried out via a procedure described in Appendix \ref{app: madgraph}.

We choose to show results after detector simulation (at the signal level) because, as discussed in \cite{Gresham:2011pa}, unfolding of data to the parton level is model dependent.   In Figs.~(\ref{tev scatter scalar}-\ref{tev scatter schan}) the axes give the signal level $\afb$ with $m_{t\bar{t}} < 450 \mbox{ GeV}$ and $m_{t\bar{t}} > 450 \mbox{ GeV}$.  The ellipses encircle the best fit points to the CDF signal level semileptonic $t\bar{t}$ $\afb$ with concentric ellipse giving $\chi^2/d.o.f. = 1,~2,~3$,  with the constraints from the total cross-section times acceptance being taken into account via
\beq
\chi^2 = \left( {\AFBL - {\AFBL}_\text{obs} \over  \sigma_{\AFBL} } \right)^2 +  \left( {\AFBH - {\AFBH}_\text{obs} \over  \sigma_{\AFBH} } \right)^2 +  \left( {\sigma_{t \bar{t}} - { \sigma_{t \bar{t},\text{SM}} }  \over  \sigma_{\sigma_{t \bar{t}}} } \right)^2.
\label{chisq}
\eeq
We use
\begin{align}
\sigma_{\AFBL} &= \sqrt{0.039^2+0.017^2+0.025^2+0.015^2}  \\
\sigma_{\AFBH} &= \sqrt{0.053^2+0.032^2+0.025^2+0.043^2} \\  
{\sigma_{\sigma_{t \bar{t}}} \over \sigma_{t \bar{t}, SM} }&= 0.10~\text{(LHC7) or}~0.15~\text{(TEV)}
\end{align}
for the error estimates and
\begin{align}
{\AFBL}_\text{obs} &= - 0.022\\
{\AFBH}_\text{obs} &= 0.266\\
\sigma_{t \bar{t},\text{SM}} &= 8.18~\text{pb (LHC7) or}~0.252~\text{pb (TEV)} \label{eq: xsec}
\end{align}
as the central values.  Note that the last value is the central LO SM cross-section times acceptance, given the cuts for Tevatron and LHC7 outlined in this section and the next.  The central values for $A_{FB}$ are the background-subtracted signal level values from \cite{Aaltonen:2011kc} and the SM values for the cross-section times acceptance are taken from our simulations of 5 million events. 
The first two contributions to the $A_{FB}$ errors are from experiment, the third for the typical statistical error from our finite-sized simulated data samples, and the last is to account for possible NLO corrections: we take this contribution to be of the same size as the NLO SM asymmetry. For the cross-section error, we take 15\% errors for the Tevatron and 10\% for LHC. 10\% is roughly the current experimental error for the Tevatron measurements, and we add a $\sim 5\%$ uncertainty due to the top mass and theory uncertainties in the NLO corrections.  For LHC, the statistics on the cross-section measurement should lead to smaller error bars and we take this into account with a smaller LHC error of 10\%.  A value $\chi^2 \sim 3$ indicates a good fit to data.   For the Standard Model with $\afb$ given by the NLO prediction and cross-section by our LO simulations, $\chi^2/3 = 2.8$. Since we take the central value for the cross-section to be the SM LO value, this value is somewhat artificially low.  These error estimates should be taken as rules of thumb to guide the eye in our figures for comparing SM against NP, rather than as hard and fast quantitative error budgets.

We discuss the scalar models first.  As can be seen from Fig.~\eqref{tev scatter scalar}, the triplet scalars generally produce larger asymmetries than singlet scalars, which generally cannot produce a larger asymmetry than the SM.  This can be qualified if the singlet scalars are lighter than the top mass, in which case signal level asymmetries as large as 10\% for $m_{t\bar{t}} > 450 \mbox{ GeV}$ can be achieved (though this is well below what is observed).  This in agreement with the parton level results of \cite{Nelson:2011us}, taking into account the factor $\sim 2$ washout translating from parton level to signal level.  The triplet scalars seem to reproduce the total asymmetry and cross-section very well.  However, it was shown in \cite{Gresham:2011pa} that these models seriously overproduce the invariant mass distribution at  large invariant mass.  We refer the reader to \cite{Gresham:2011pa} for details.

Next we discuss the $t$-channel vector mediators in Fig.~(\ref{tev scatter vector}).  As expected from the results in \cite{Gresham:2011pa}, the color singlet vector is most successful in reproducing the asymmetry at high invariant mass and satisfying the cross-section constraints.  Due to details in the form of the matrix element, the color octet is less successful.  The flavor universal octet can produce large asymmetries, but these also tend to come with fairly large contributions to the total cross-section, due to the presence of both $s$ and $t$-channel mediators.

Lastly, we discuss the $s$-channel states in Fig.~(\ref{tev scatter schan}).  The wide, low mass axiglue models, schanC8V$\Gamma$, in general are most successful at producing a large asymmetry with small contribution to the total cross-section.  The light axigluon models with couplings to right-handed quarks and masses in the 700-900 GeV range (schanC8VR) \cite{AguilarSaavedra:2011ci} do not produce a large asymmetry on the other hand; in most cases it is not larger than the SM asymmetry.  Heavy axigluon models can succeed with a large enough coupling to light quarks, but these risk being ruled out shortly by LHC dijet and $t\bar{t}$ resonance searches.

With these results in hand, we now turn to examining the implications of models that are capable of satisfying the Tevatron constraints on top analyses at LHC7.   For each class of models, and a selection of mediator masses between 200 GeV and 2 TeV, we take models with the lowest $\chi^2$ as defined by the statistic in Eq.~(\ref{chisq}).  5 million events are generated for each of these benchmark models to gain enough statistics at the high invariant mass, via the procedure in Appedix \ref{app: madgraph}.  Our benchmark models are not an exhaustive set of model choices, but they are indicative of the types of models that can generate top $A_{FB}$.  The choice of models is shown in Table~\ref{Models}.   It gives the mass and coupling of the model, the LO cross-section at Tevatron and LHC along with the acceptance A., the signal and parton level $\afb$ in the low and high invariant mass bins, along with the total asymmetry, and the $\chi^2$ at Tevatron and LHC, using the statistic discussed in the text.  These results are to be compared against the SM, shown in Table~\ref{SM}.  

Note that the models with the lowest $\chi^2$ tend to universally underproduce the total asymmetry.  The reason is that the models with the largest $\afb$ also tend to overproduce the total cross-section rather seriously, so that the $\chi^2$ prefers to take a hit on the asymmetry (which has 1$\sigma$ errors of $\sim 8\%$ at the signal level) in lieu of a large $t\bar{t}$ production cross-section.  The models that are the least successful at producing a large asymmetry with minimal impact on the total cross-section are: C8V, C1S, schanC8VR.  These models are generally able to produce little more than the SM asymmetry for $\afb$ with $m_{t\bar{t}} > 450$ GeV, and should not be considered as viable models for $\afb$.

\begin{table}

\begin{tabular}{c c c c c c}
\hline
& & & signal level & parton level & \\
parameters	& {$\sigma_\text{Tev}$ (pb), A.} & { $\sigma_\text{LHC}$ (pb), A.} & {{$A_{FB}^{<450}$, $A_{FB}^{>450}$}, $A_{FB}^\text{total}$} & {{$A_{FB}^{<450}$, $A_{FB}^{>450}$}, $A_{FB}^\text{total}$} & $\chi^2_\text{Tev}$,  $\chi^2_\text{LHC}$ \\
\hline

{$m$, $g_R$} &\multicolumn{4}{c}{C1V} \\
\hline
{200., 0.7} & {6.3, 0.037} & {146, 0.068} & {{-0.03, 0.15}, 0.06} & {{0.01, 0.39}, 0.2} & {0.8, 2.3}\\
{400., 1.3} & {7.1, 0.038} & {154, 0.073} & {{0.01, 0.25}, 0.15} & {{0.08, 0.55}, 0.35} & {0.2, 4.7}\\
{600., 1.5} & {5.3, 0.039} & {126, 0.072} & {{-0.04, 0.15}, 0.06} & {{-0.03, 0.25}, 0.1} & {1.2, 1.2}\\
{800., 2.1} & {5.8, 0.039} & {129, 0.073} & {{-0.03, 0.18}, 0.09} & {{-0.01, 0.36}, 0.18} & {0.5, 1.1}\\

\hline
{$m$, $g_R$} &\multicolumn{4}{c}{C8V} \\
\hline
{400., 0.75} & {6.8, 0.041} & {130, 0.072} & {{0.01, 0.08}, 0.04} & {{0.03, 0.1}, 0.06} & {2.1, 2.7}\\
{800., 1.4} & {6.8, 0.04} & {120, 0.072} & {{-0.01, 0.08}, 0.03} & {{0.03, 0.1}, 0.06} & {1.9, 2.}\\

\hline
{$m$, $g$, $\eta$} &\multicolumn{4}{c}{F8C1V} \\
\hline
{200., {0.5, 1.}} & {6.5, 0.037} & {148, 0.067} & {{0.05, 0.14}, 0.09} & {{0.03, 0.4}, 0.21} & {1.4, 2.8}\\
{400., {0.5, 0.}} & {9.4, 0.04} & {125, 0.069} & {{0.08, 0.}, 0.05} & {{0.23, -0.02}, 0.17} & {8.5, 5.1}\\
{600., {0.5, 3.}} & {6., 0.041} & {128, 0.071} & {{-0.03, 0.15}, 0.07} & {{-0.05, 0.31}, 0.14} & {0.7, 1.1}\\
{800., {0.5, 1.}} & {6., 0.041} & {115, 0.072} & {{-0.03, 0.01}, -0.01} & {{0., 0.03}, 0.01} & {3.5, 3.5}\\
\hline
{$m$, $g_R$} &\multicolumn{4}{c}{C1S} \\
\hline
{200., 1.5} & {5.7, 0.042} & {119, 0.072} & {{0.01, 0.04}, 0.03} & {{0., 0.06}, 0.02} & {2.9, 3.}\\
\hline
{$m$, $g_R$} &\multicolumn{4}{c}{C3S} \\
\hline
{400., 2.95} & {8.6, 0.033} & {165, 0.074} & {{0., 0.17}, 0.11} & {{0.2, 0.22}, 0.21} & {0.8, 8.4}\\
{600., 3.4} & {6.7, 0.043} & {133, 0.075} & {{0., 0.14}, 0.08} & {{0.05, 0.23}, 0.14} & {1.2, 2.6}\\
{800., 4.15} & {6.6, 0.042} & {128, 0.075} & {{-0.01, 0.15}, 0.08} & {{0.03, 0.27}, 0.15} & {0.9, 1.8}\\

\hline
{$m$, $g_R$, $\Gamma/m (\%)$} &\multicolumn{4}{c}{schanC8V$\Gamma$} \\
\hline
{420., {0.45, 18}} & {6.7, 0.04} & {116, 0.072} & {{-0.03, 0.15}, 0.05} & {{-0.03, 0.3}, 0.1} & {0.8, 0.8}\\
{440., {0.45, 13}} & {6.9, 0.039} & {118, 0.07} & {{-0.03, 0.12}, 0.04} & {{-0.11, 0.34}, 0.06} & {1.1, 1.1}\\

\hline
{$m$, $g_R^q$, $g_R^t$} &\multicolumn{4}{c}{schanC8VA} \\
\hline
{2000., {-1., 5.}} & {6.4, 0.04} & {117, 0.072} & {{0.01, 0.16}, 0.08} & {{0.06, 0.17}, 0.1} & {0.7, 0.8}\\
{2400., {-3.6, 3.6}} & {6.5, 0.039} & {119, 0.072} & {{0., 0.14}, 0.07} & {{0.07, 0.21}, 0.13} & {1., 1.}\\

\hline
{$m$, $g_R^q$, $g_R^t$} &\multicolumn{4}{c}{schanC8VR} \\
\hline
{700., {-0.05, 4.5}} & {6.7, 0.04} & {116, 0.07} & {{0., 0.06}, 0.02} & {{0.02, 0.07}, 0.04} & {2.4, 2.4}\\
{850., {-0.08, 6.}} & {6.7, 0.039} & {117, 0.072} & {{0.04, 0.08}, 0.06} & {{0.02, 0.08},0.04} & {2.2, 2.2}\\

\end{tabular}

\caption{A representative set of models chosen for LHC analysis. Acceptance is labeled ``A.''.  }
\label{Models}

\end{table}

\begin{table}

\begin{tabular}{c c c c c }
\hline
 \multicolumn{4}{c}{LO SM cross-section, NLO $A_{FB}$} \\
\hline
  & & signal level & parton level &  \\
	 {$\sigma_\text{Tev}$ (pb), A.} & { $\sigma_\text{LHC}$ (pb), A.} & {{$A_{FB}^{<450}$, $A_{FB}^{>450}$}, $A_{FB}^\text{total}$} & {{$A_{FB}^{<450}$, $A_{FB}^{>450}$}, $A_{FB}^\text{total}$} & $\chi^2_\text{Tev}$,  $\chi^2_\text{LHC}$ \\
\hline

 {6.3, 0.04} & {115, 0.071} & {{0.015, 0.043}, 0.024} & {{0.040, 0.088}, 0.058} & {2.8, 2.8}\\

\hline

\end{tabular}

\caption{The SM LO cross-section at Tevatron and LHC along with the acceptance, A., the signal and parton level $\afb$ in the low and high invariant mass bins, along with the total asymmetry, and the $\chi^2$ at Tevatron and LHC, using the statistic discussed in the text.}
\label{SM}

\end{table}

\begin{figure}
\includegraphics[width = 1.0\textwidth]{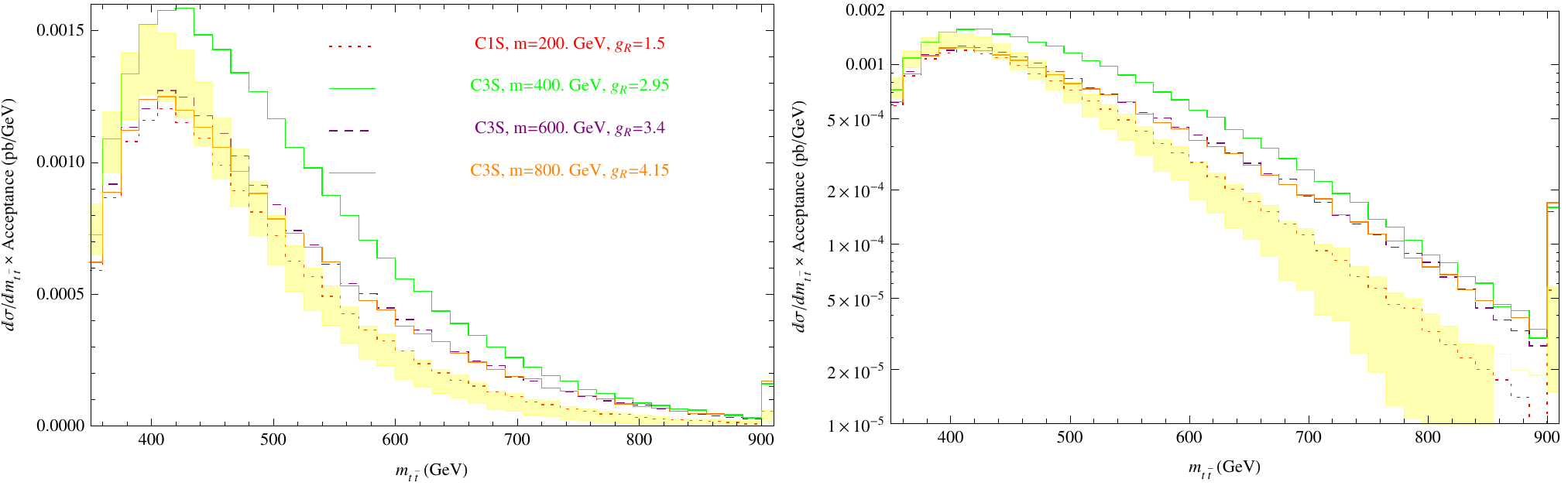} \\
\includegraphics[width = 1.0\textwidth]{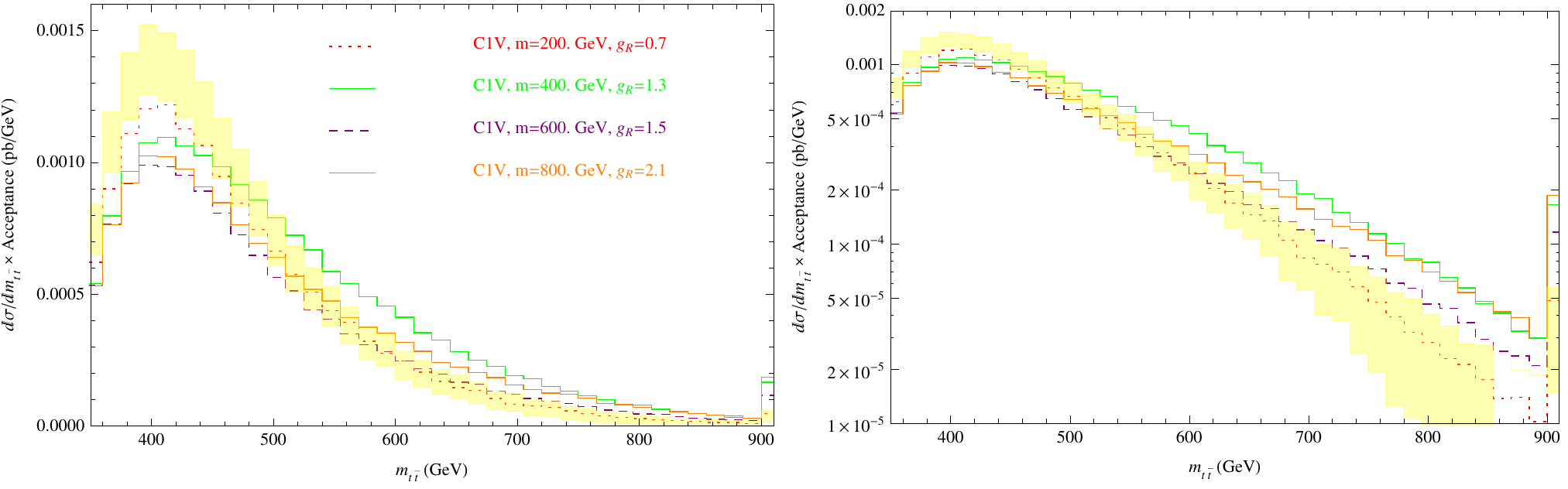} \\
\includegraphics[width = 1.0\textwidth]{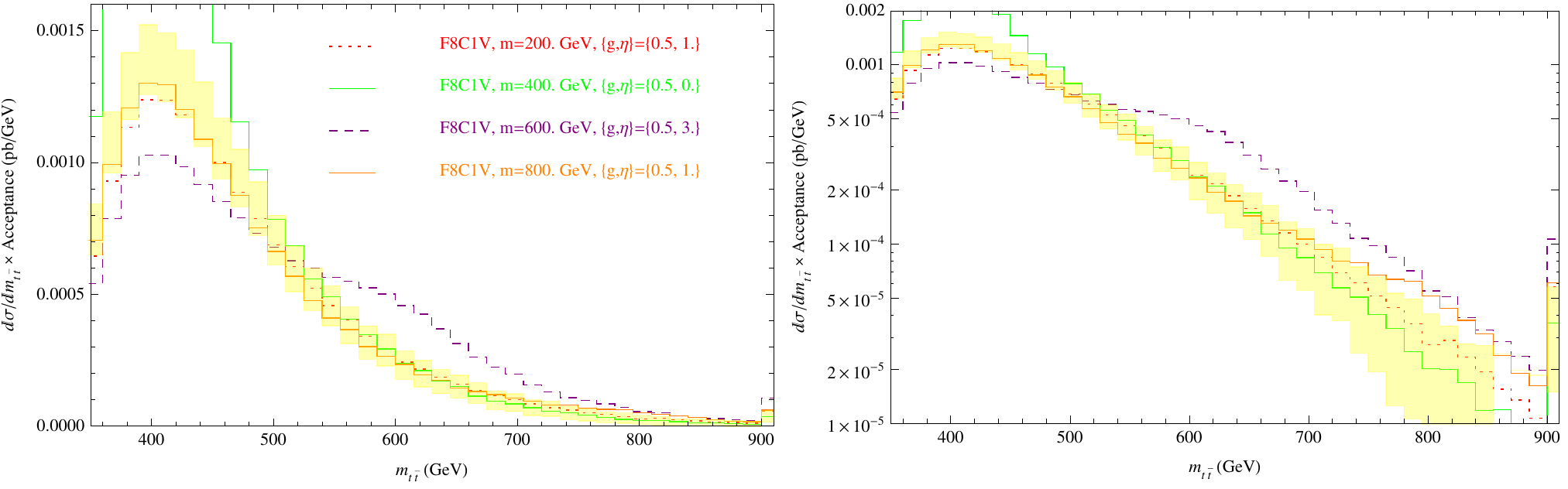} \\
\caption{Tevatron invariant mass distributions, on both linear and log scales, for our benchmark models choices.  The SM is shown in the yellow band, with statistical errors for 5.3 fb$^{-1}$ of data. }
\label{TevInvMassA}
\end{figure}

\begin{figure}
\includegraphics[width = 1.0\textwidth]{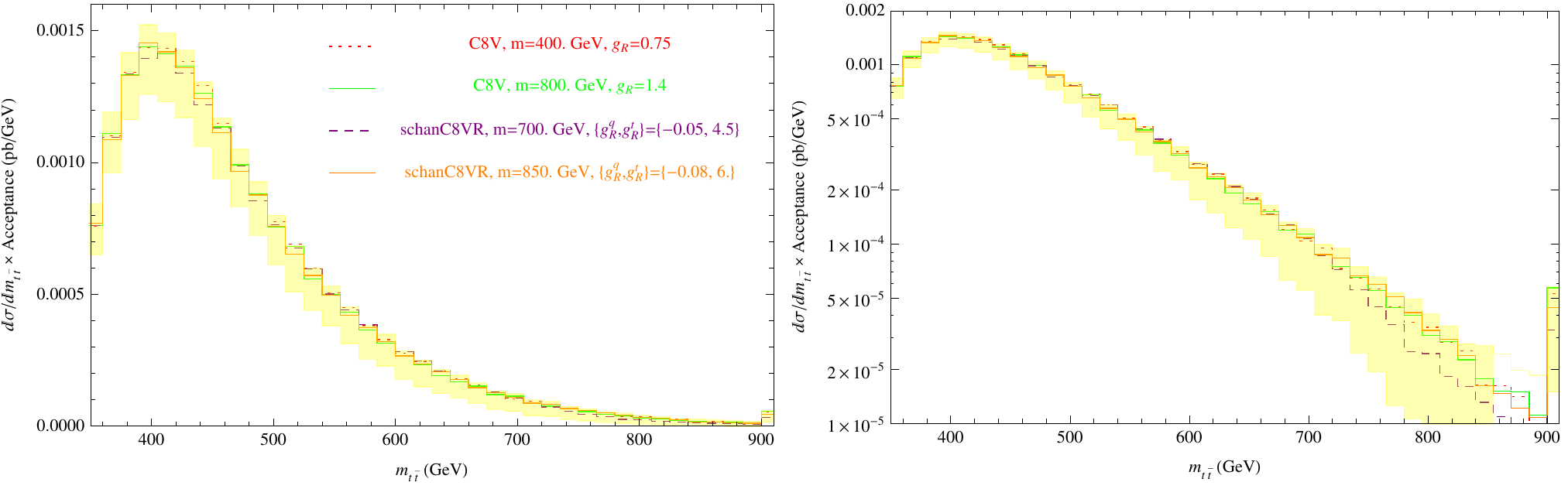} \\
\includegraphics[width = 1.0\textwidth]{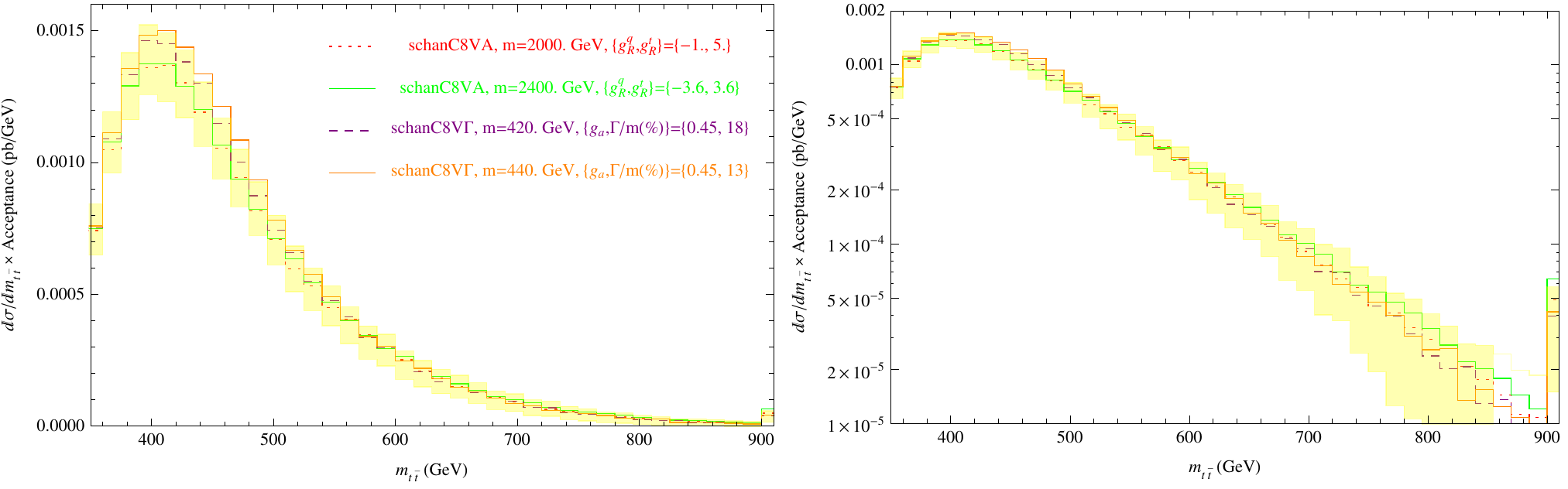} \\
\caption{Tevatron invariant mass distributions, on both linear and log scales, for our benchmark models choices.  The SM is shown in the yellow band, with statistical errors for 5.3 fb$^{-1}$ of data.}
\label{TevInvMassB}
\end{figure}

Before moving on to the LHC analysis, we check the Tevatron invariant mass distributions for the classes of models that we examine more carefully.  As we learned in \cite{Gresham:2011pa}, acceptance effects can be important in bringing NP models into agreement with the Tevatron invariant mass distributions.  We show the Tevatron invariant mass distributions in Figs.~\eqref{TevInvMassA}-\eqref{TevInvMassB}, for comparison to the LHC results we discuss next.

\section{Implications for Top Physics at the LHC}
\label{Sec: LHC}

For the LHC benchmark points analysis, we generate 5 million events for each model, as we did for the Tevatron analysis.  We also modified the {\tt PGS} code to implement the anti-$k_T$ algorithm \cite{Cacciari:2008gp} to mimmic ATLAS as detailed in Appendix \ref{app: madgraph}.
 In the following, closest attention should be paid to the C1V, F8C1V, C3S, schanC8V$\Gamma$ and schanC8VA models, as these, among the models in the literature we have considered, are able to generate the top $\afb$ to a reasonable degree. 

The variables that we focus on at the LHC are:
\begin{itemize}
\item Total cross-section. The chief uncertainties here come from NLO corrections from both the SM and NP, and the uncertainty in the top mass;
\item Invariant mass distribution.  Here again NLO corrections will play an important role;
\item Number of additional jets.  In $t$-channel models, single production of the mediator in conjunction with the top is an important process at the LHC.  A gluon and light quark in the initial state will exchange a top in the $t$-channel and produce a top along with a mediator as in Fig.~(\ref{phiproduction}).  The mediator will prefer to decay to a top and another jet, leading to a potential enrichment of events with an extra jet.  The direct search for the top-jet resonance as a signature for these models was studied in \cite{Gresham:2011dg}, but its presence may be known through counting the number of additional jets in $t\bar{t}$ events.
\item Rapidity distribution of the lepton.  Especially for models with a $t$-channel resonance, the leptons may be produced in a more forward direction at the high invariant mass. On the other hand, single mediator production leading to $t \bar{t} + \text{jets}$ events can lead to more central leptons.
\end{itemize}

We follow the cuts discussed in the $200 \mbox{ pb}^{-1}$ ATLAS semileptonic top analysis \cite{InvariantMass}.  We require:
\begin{itemize}
\item exactly one electron with $p_T > 25 \mbox{ GeV}$ and $|\eta| < 2.5$, or exactly one muon with $p_T > 20 \mbox{ GeV}$ and $|\eta| < 2.5$;
\item at least four jets with $p_T > 25 \mbox{ GeV}$ and $|\eta| < 2.5$, one of which must be $b$-tagged;
\item if the lepton is an electron, we require $E_T^{miss} > 35 \mbox{ GeV}$ and the transverse mass of the lepton and $E_T^{miss}$ be greater than 25 GeV; if the lepton is a muon, we require $E_T^{miss} > 20 \mbox{ GeV}$ and the transverse mass of the lepton with $E_T^{miss}$, plus the $E_T^{miss}$, be greater than 60 GeV;
\item jets within $\Delta R < 0.2$ are removed so as to avoid double-counting of electrons as jets.
\end{itemize}
In addition, ATLAS demands isolation cuts; since we do clustering in {\tt PGS} before placing the cuts, we do not apply them. 
$m_{t\bar{t}}$ is re-constructed in the same way as ATLAS, carried out without a full top reconstruction.  The neutrino momentum is found assuming the $W$ mass and massless neutrino conditions.  For some events there is no positive energy solution, in which case the event is discarded. According to the ATLAS analysis, we take the longitudinal neutrino momentum to be the real part of the mass constraint solution in the case of imaginary solutions and we take the solution with smallest absolute value if there are two solutions. 

\begin{figure}
\includegraphics[width = 0.85\textwidth]{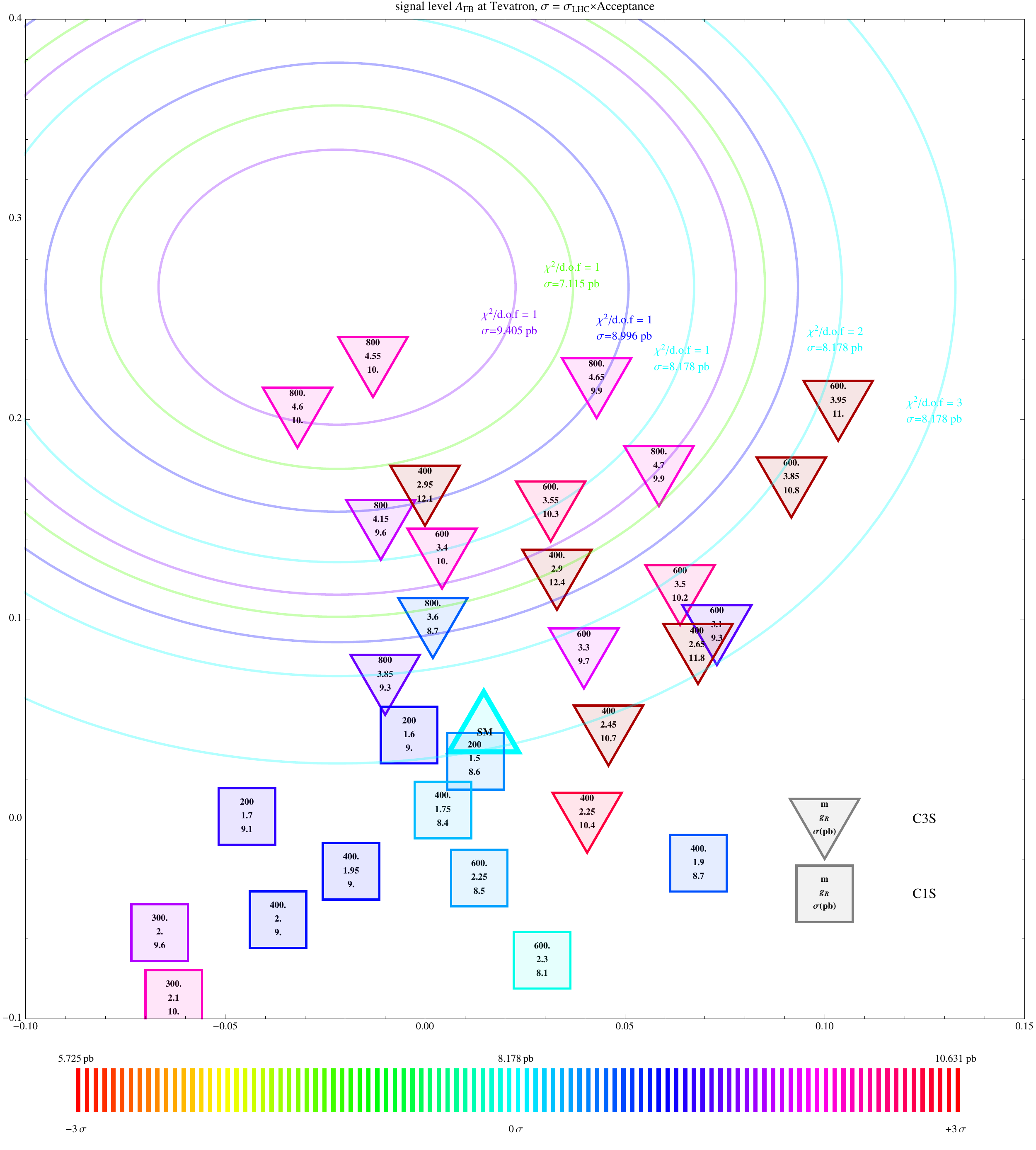}
\caption{Scatter plot depicting simulated signal level $ \{ A_{FB}(m_{t \bar{t}} < 450~\text{GeV}),  A_{FB}(m_{t \bar{t}} > 450~\text{GeV}), \sigma_{t \bar{t}} \times \text{acceptance}
\}$ for $t$-channel flavor-changing scalar models listed in Table \ref{tab: scan summary table}.  The models are labeled by the mass of the mediator, the coupling to right-handed quarks, and the total LHC production cross-section times acceptance.  The cross-sections are compared against the SM cross-section times acceptance which yields 8.178 pb at the LO; the color scales for the models indicate the deviation from the SM cross-section, as indicated by the legend at the bottom.  The curves indicate constant $\chi^2$ for a given cross-section, as defined in Eq. \eqref{chisq}. Contours for four cross-section values (cyan, blue, green, purple) are shown for $\chi^2/\text{d.o.f.} =$ 1 and 2.  A single (cyan) $\chi^2/\text{d.o.f.} = 3$ contour with SM cross-section is shown. Model points of a given color should be compared to $\chi^2$ contours of the same color. }
\label{LHC scatter scalar}
\end{figure}

\begin{figure}
\includegraphics[width = 0.85\textwidth]{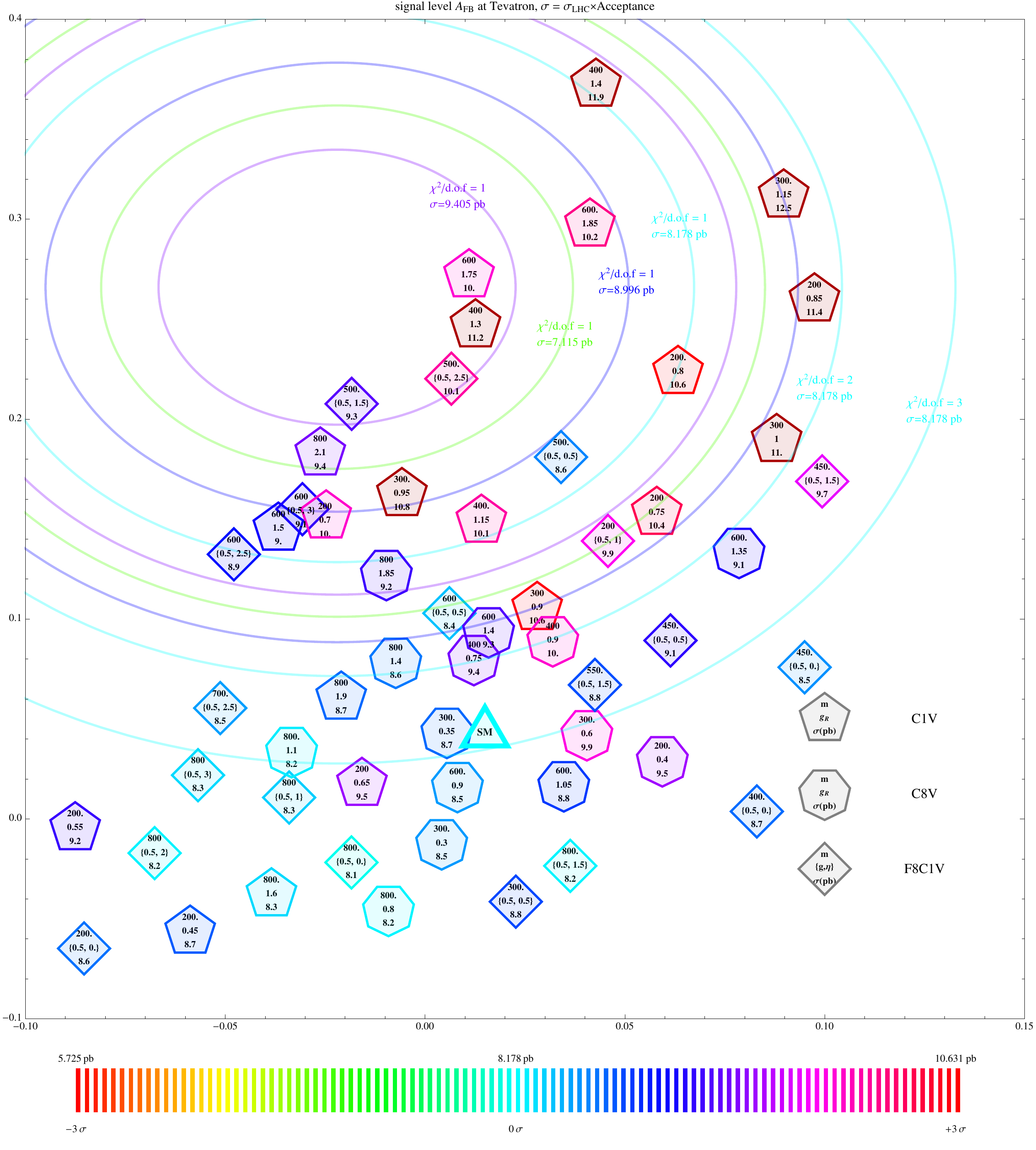}
\caption{Scatter plot depicting simulated signal level $ \{ A_{FB}(m_{t \bar{t}} < 450~\text{GeV}),  A_{FB}(m_{t \bar{t}} > 450~\text{GeV}), \sigma_{t \bar{t}} \times \text{Acceptance}
\}$ for $t$-channel flavor-changing vector models models listed in Table \ref{tab: scan summary table}.  The models are labeled by the mass of the mediator, the coupling, and the total LHC production cross-section times acceptance.  The coupling conventions are discussed in detail in the text.  The cross-sections are compared against the SM cross-section times acceptance which yields 8.178 pb at the LO; the color scales for the models indicate the deviation from the SM cross-section, as indicated by the legend at the bottom.  A single (cyan) $\chi^2/\text{d.o.f.} = 3$ contour with SM cross-section is shown. The curves indicate constant $\chi^2$ for a given cross-section, as defined in Eq. \eqref{chisq}. Contours for four cross-section values (cyan, blue, green, purple) are shown for $\chi^2/\text{d.o.f.} =$ 1 and 2.  Model points of a given color should be compared to $\chi^2$ contours of the same color. }
\label{LHC scatter vector}
\end{figure}

\begin{figure}
\includegraphics[width = 0.85\textwidth]{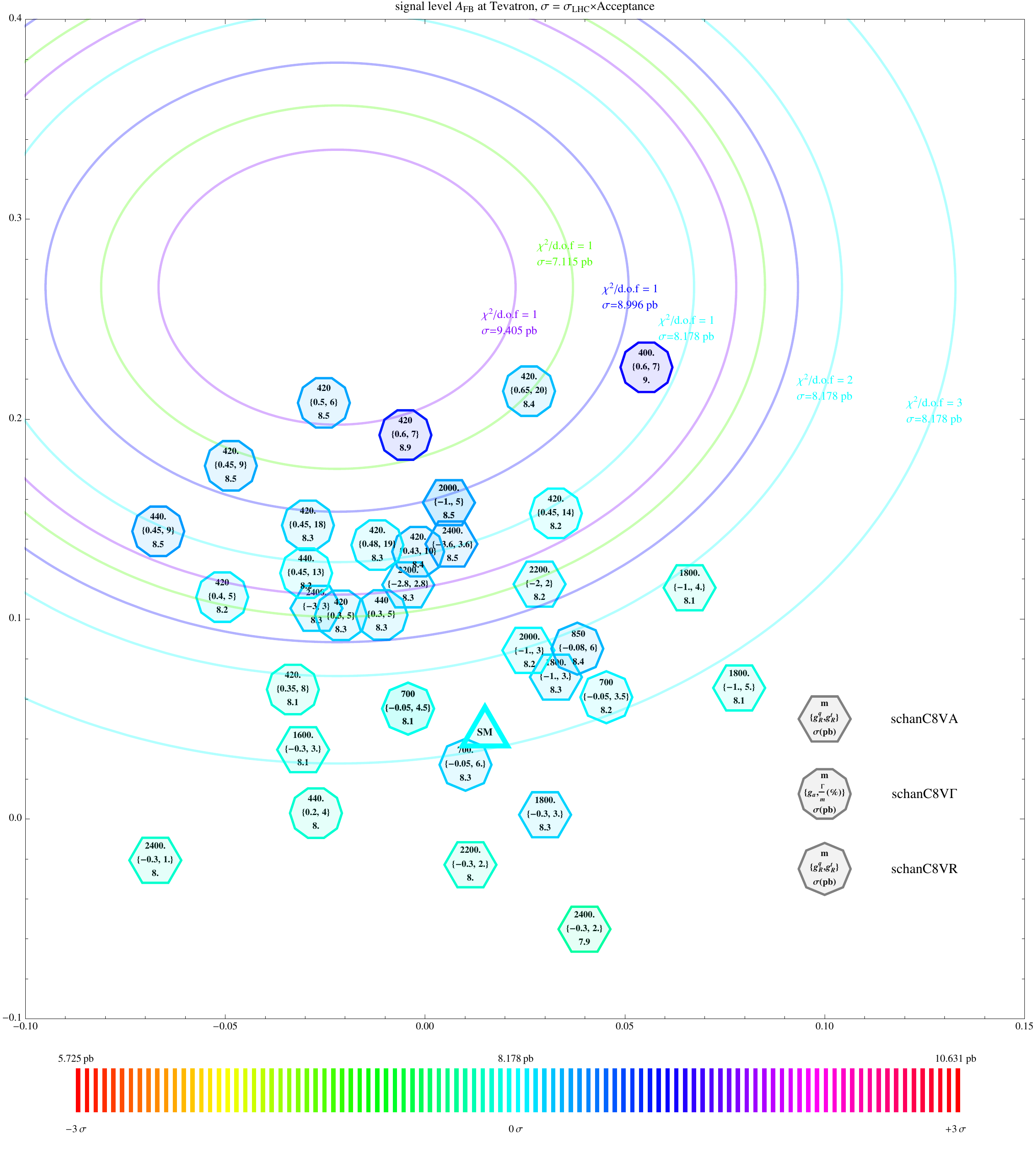}
\caption{Scatter plot depicting simulated signal level $ \{ A_{FB}(m_{t \bar{t}} < 450~\text{GeV}),  A_{FB}(m_{t \bar{t}} > 450~\text{GeV}), \sigma_{t \bar{t}} \times \text{Acceptance}
\} $ for axigluon models listed in Table \ref{tab: scan summary table}.  The models are labeled by the mass of the mediator, the coupling, and the total LHC production cross-section times acceptance.  The coupling conventions are discussed in detail in the text.  The cross-sections are compared against the SM cross-section times acceptance which yields 8.178 pb at the LO; the color scales for the models indicate the deviation from the SM cross-section, as indicated by the legend at the bottom.  A single (cyan) $\chi^2/\text{d.o.f.} = 3$ contour with SM cross-section is shown. The curves indicate constant $\chi^2$ for a given cross-section, as defined in Eq. \eqref{chisq}. Contours for four cross-section values (cyan, blue, green, purple) are shown for $\chi^2/\text{d.o.f.} =$ 1 and 2.  Model points of a given color should be compared to $\chi^2$ contours of the same color. }
\label{LHC scatter schan}
\end{figure}

The first and simplest measure is the top forward-backward asymmetry versus the total production cross-section at the LHC.  There is a trade-off between models with a large enough coupling to produce the observed forward-backward asymmetry, while simultaneously having a small enough coupling that single mediator production at the LHC does not lead to a large contribution to the $t\bar{t}$ cross-section.   However, given that the higher mass models in particular have large couplings, one expects the next-to-leading-order (NLO) corrections to play a significant role in both the total cross-section and invariant mass distributions.  Given the gross-overproduction of some models of the total cross-section, some may, however, be reasonably eliminated.  This can be seen in Figs.~(\ref{LHC scatter scalar}-\ref{LHC scatter schan}), where we plot the Tevatron $A_{FB}$ in low and high invariant mass bins, with total production cross-section at LHC times acceptance indicated by color.  We again compare LO MadGraph results against the LO SM cross-section times acceptance \eqref{eq: xsec}. We find a total LO matched SM $t \bar{t} + 0~\text{or}~1$ jets cross-section of 115 pb for $m_t = 172$ GeV. Note that there is a large K-factor of $\sim 1.6$ expected at LHC7 which enters to match the total cross-section observed (of about 180 pb) against the LO contribution.   The LO cross-section times acceptance is 8.178 pb.

A couple of features in Figs.~\ref{LHC scatter scalar}-\ref{LHC scatter schan}) in particular are worth highlighting.  $t$-channel models at the LHC overproduce the total cross-section much more than at the Tevatron.  This is because single production of the $t$-channel mediators gives rise to a significant contribution to the total cross-section.  This effect is more important for lighter mediators, so that light mediators become much more disfavored at the LHC.

Note that this brings in a significant tension for $t$-channel models of $\afb$ between the constraints from Tevatron and the LHC.  The Tevatron $t\bar{t}$ invariant mass distribution tended to favor lighter mass mediators because they lead to less distortion of the invariant mass distribution at high invariant mass \cite{Gresham:2011pa}, while LHC favors heavier mediators because they lead to less distortion in the total cross-section.

\begin{figure}
\includegraphics[width = 1.0\textwidth]{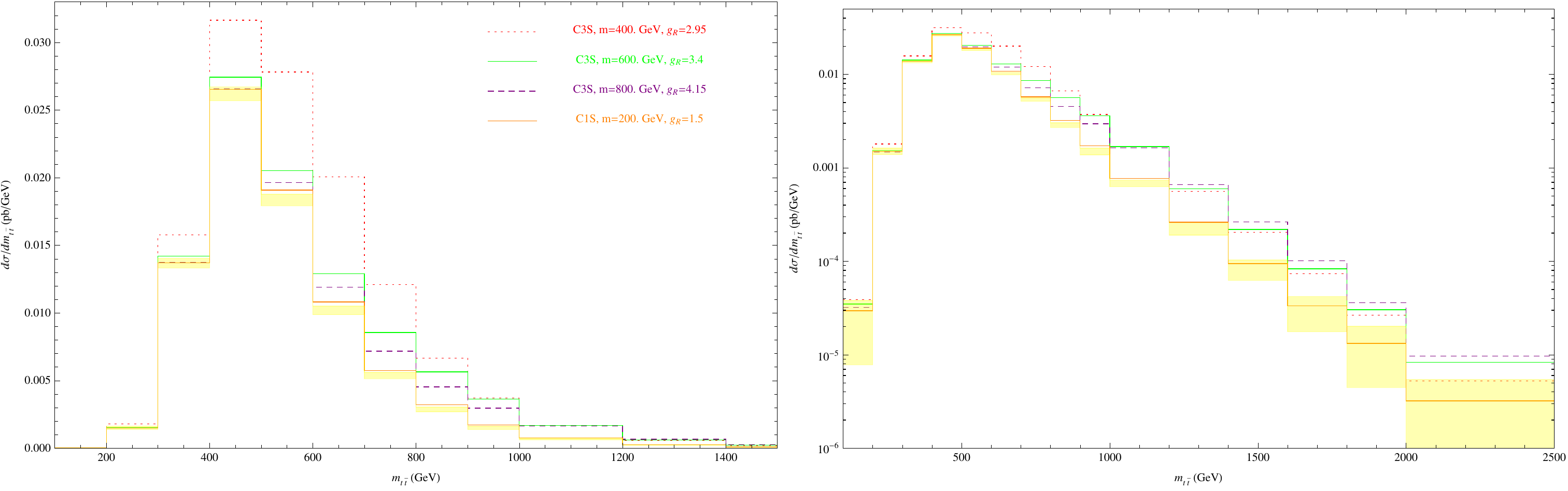} \\
\includegraphics[width = 1.0\textwidth]{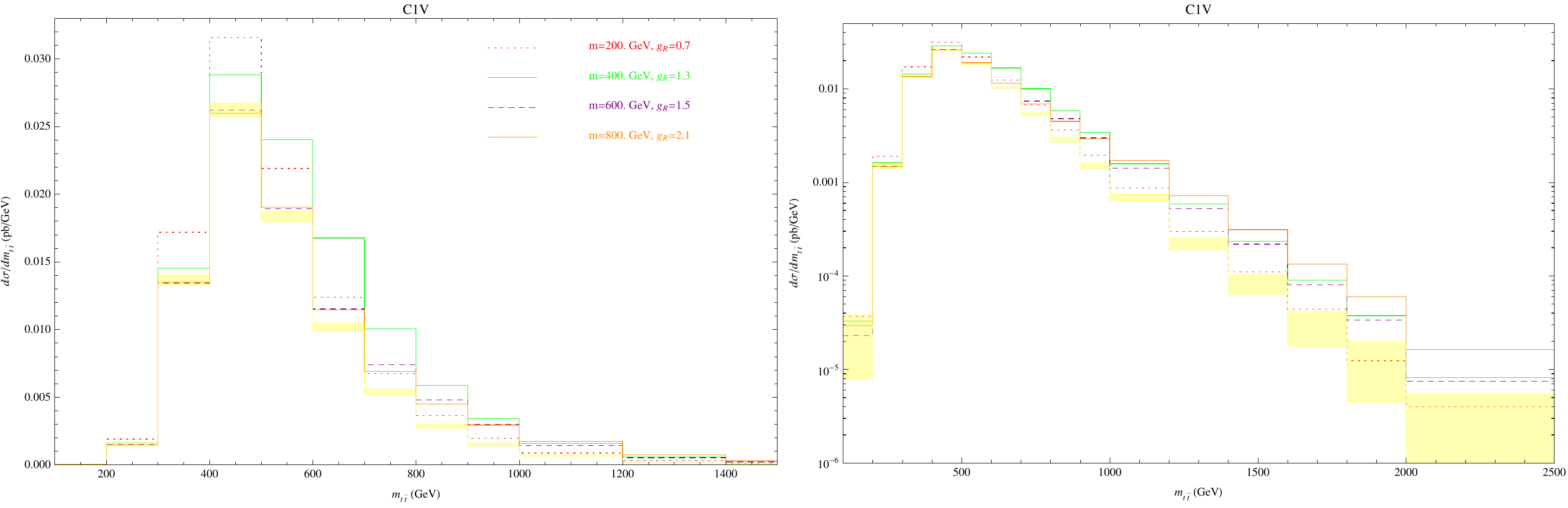} \\
\includegraphics[width = 1.0\textwidth]{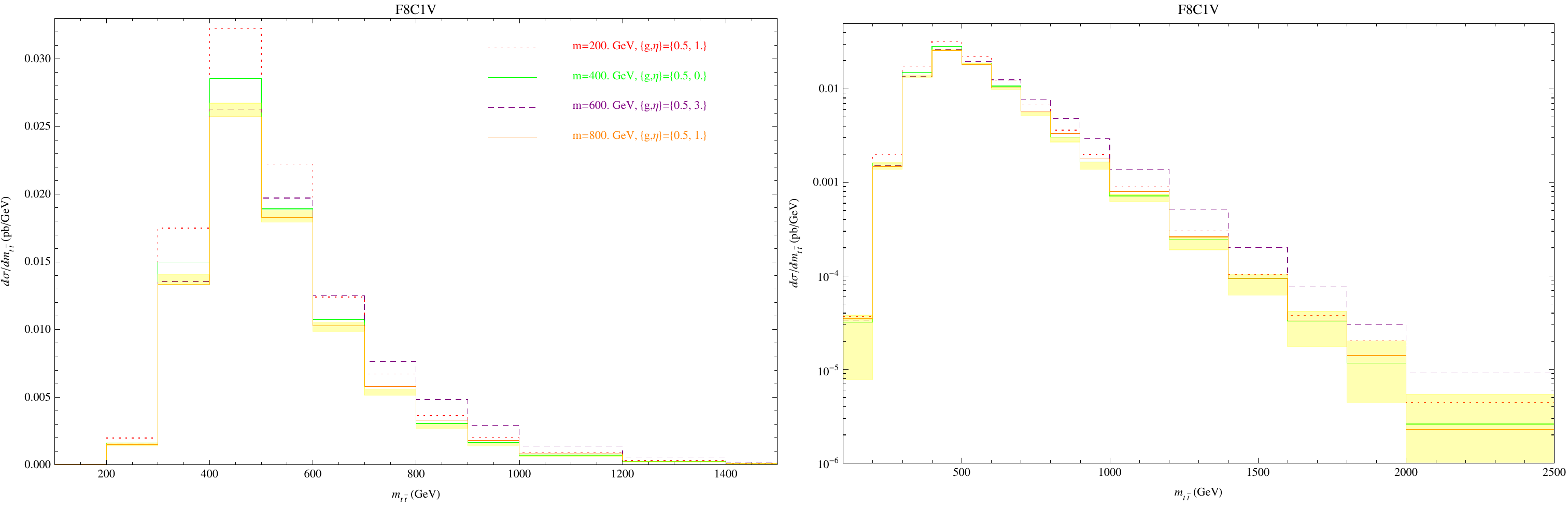} \\
\caption{C1S, C3S, C1V, F8C1V models differential cross-section times acceptance (on both linear and log scales) at LHC7 versus reconstructed $m_{t \bar{t}}$, as compared to SM LO expectation, with $\pm 1 \sigma$ yellow bands corresponding to statistical error given $1 \text{fb}^{-1}$. Models shown are those with the lowest $\chi^2$ for a given mass as defined in Eq.~\eqref{chisq}, except for the 600 and 800 GeV  C1V models, which were chosen to have the lowest $\chi^2$ and be within 10\% of the SM cross-section times acceptance at Tevatron. }
\label{MttA}
\end{figure}

\begin{figure}
\includegraphics[width = 1.0\textwidth]{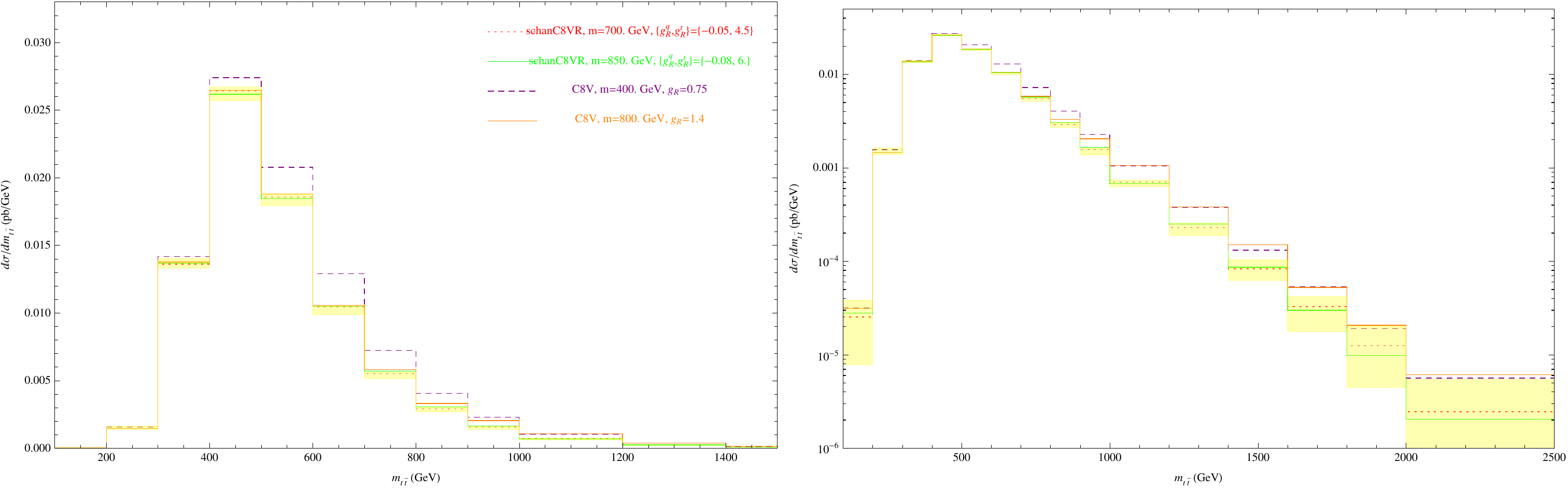} \\
\includegraphics[width = 1.0\textwidth]{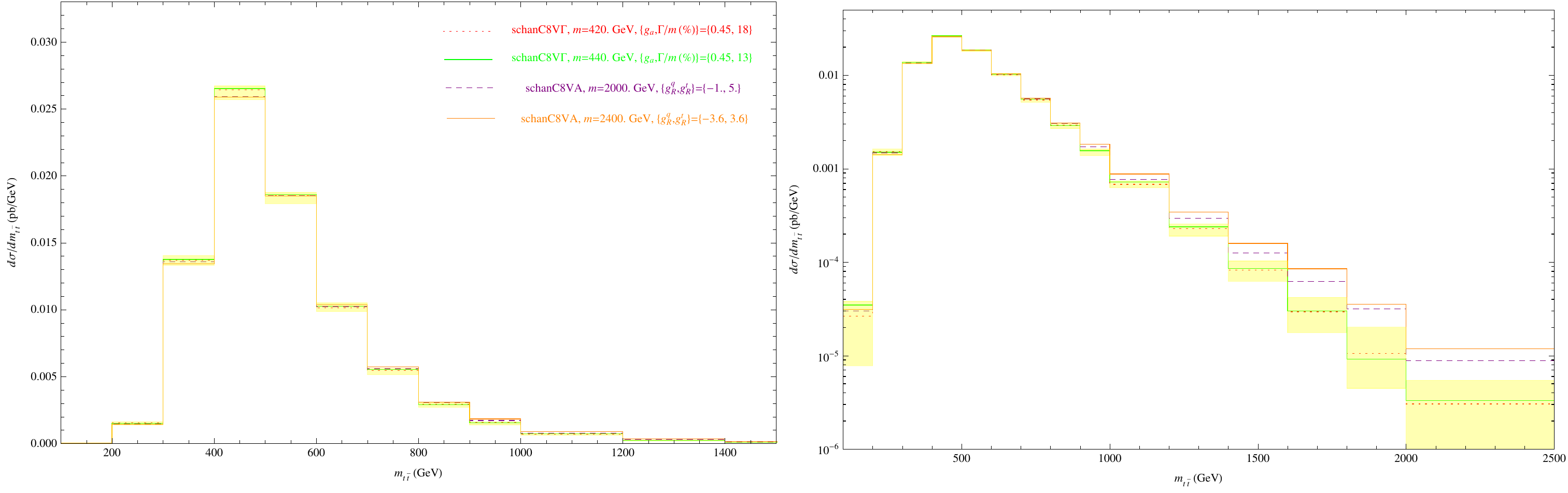} \\
\caption{C8V and schanC8 models differential cross-section times acceptance (on both linear and log scales) at LHC7 versus reconstructed $m_{t \bar{t}}$, as compared to SM LO expectation, with $\pm 1 \sigma$ yellow bands corresponding to statistical error given $1 \text{fb}^{-1}$. Models shown are those with the lowest $\chi^2$ as defined in Eq.~\eqref{chisq}.}
\label{MttB}
\end{figure}

We next consider the effect on the invariant mass distribution for our benchmark models. The results are shown in Figs.~\eqref{MttA},~\eqref{MttB}.  The effect of the NP on the shape of the invariant mass distribution is very different at the LHC than at the Tevatron.  At the Tevatron, the effects of the new mediator become most pronounced at the high invariant mass---for the $t$-channel models in particular.  At the LHC, this effect is not present, because most of the impact of the new mediators on the cross-section is simply single mediator production.  The $s$-channel models with a sufficiently broad width have little impact on the invariant mass distribution.  Thus it appears that most of the constraint on the new models comes simply from the total cross-section measurement.  We also note that acceptance effects explored in \cite{Gresham:2011pa} are not as important at LHC as at Tevatron, both because NP $t\bar{t}$ events at LHC are more central, and because the rapidity coverage for leptons at ATLAS and CMS is better than at Tevatron.

\begin{figure}
\includegraphics[width = 1.0\textwidth]{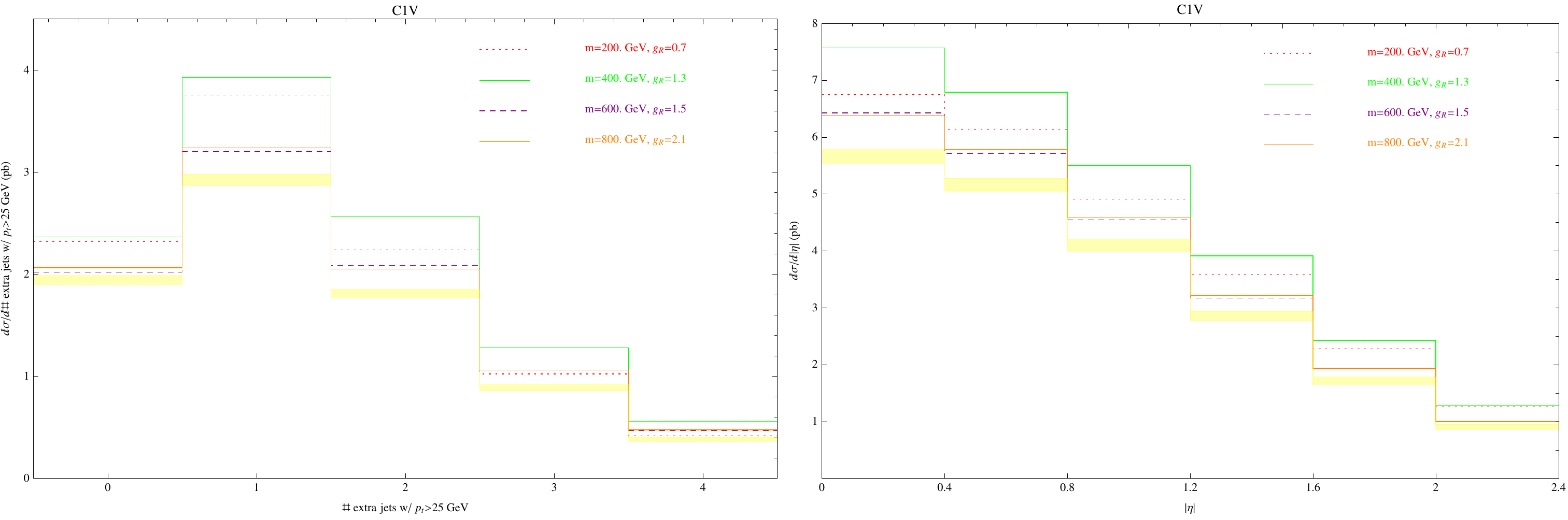}\\
\includegraphics[width = 1.0\textwidth]{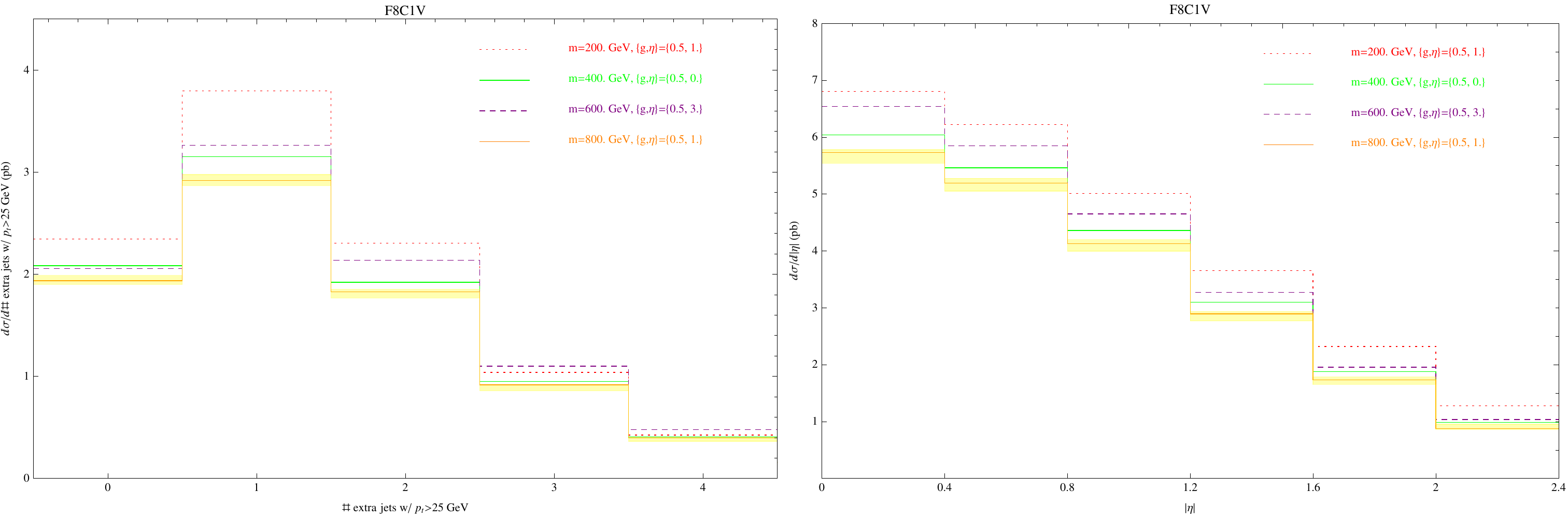}\\
\includegraphics[width = 1.0\textwidth]{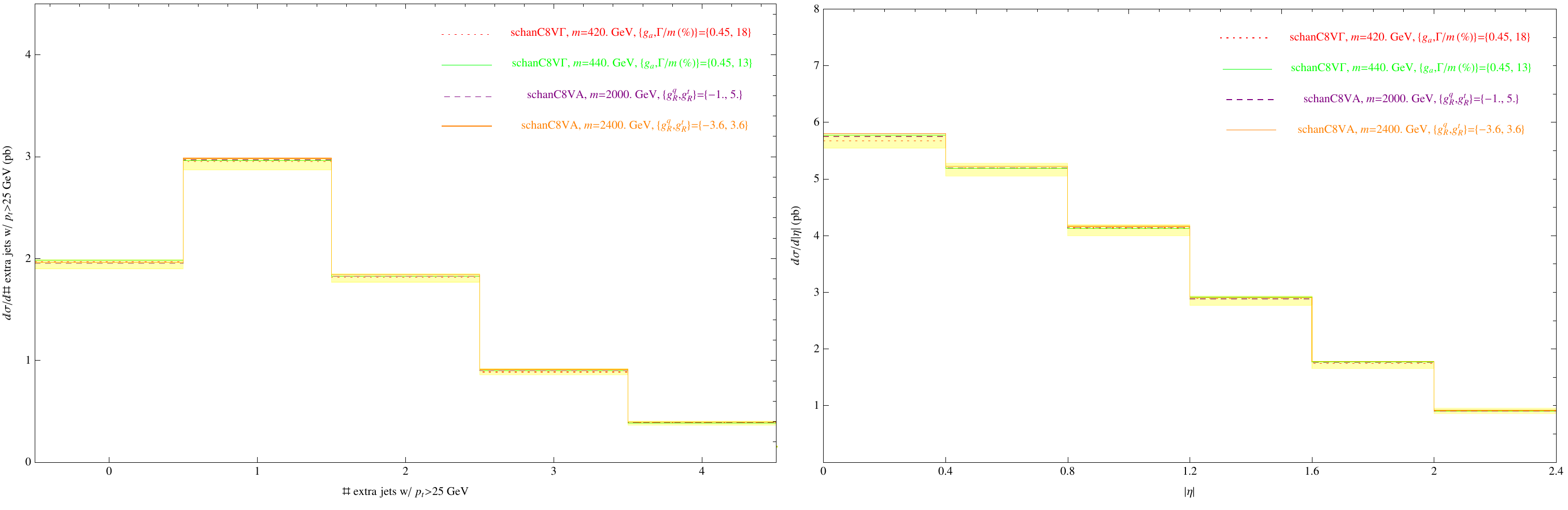}\\
\caption{Number of additional jets and lepton differential rapidity distribution of C1V, F8C1V and schanC8 models at LHC7, as compared to SM LO expectation, with $\pm 1 \sigma$ yellow bands corresponding to statistical error given $1 \text{fb}^{-1}$. Models shown are those with the lowest $\chi^2$ as defined in Eq.~\eqref{chisq}.}
\label{jetsA}
\end{figure}

\begin{figure}
\includegraphics[width = 1.0\textwidth]{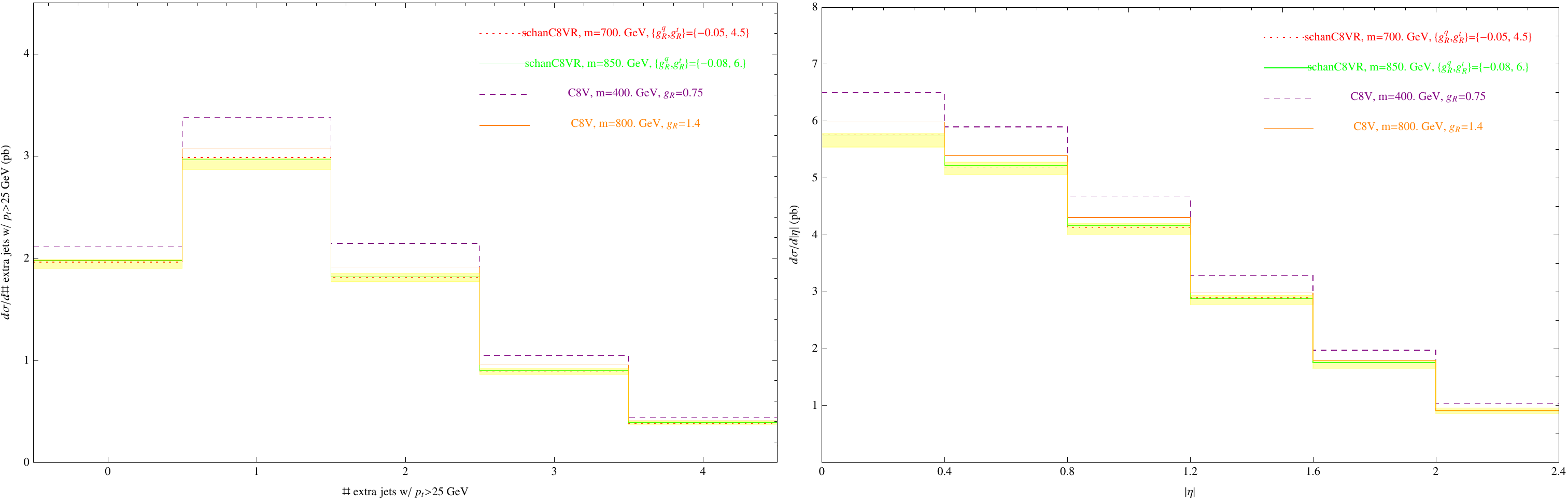}\\
\includegraphics[width = 1.0\textwidth]{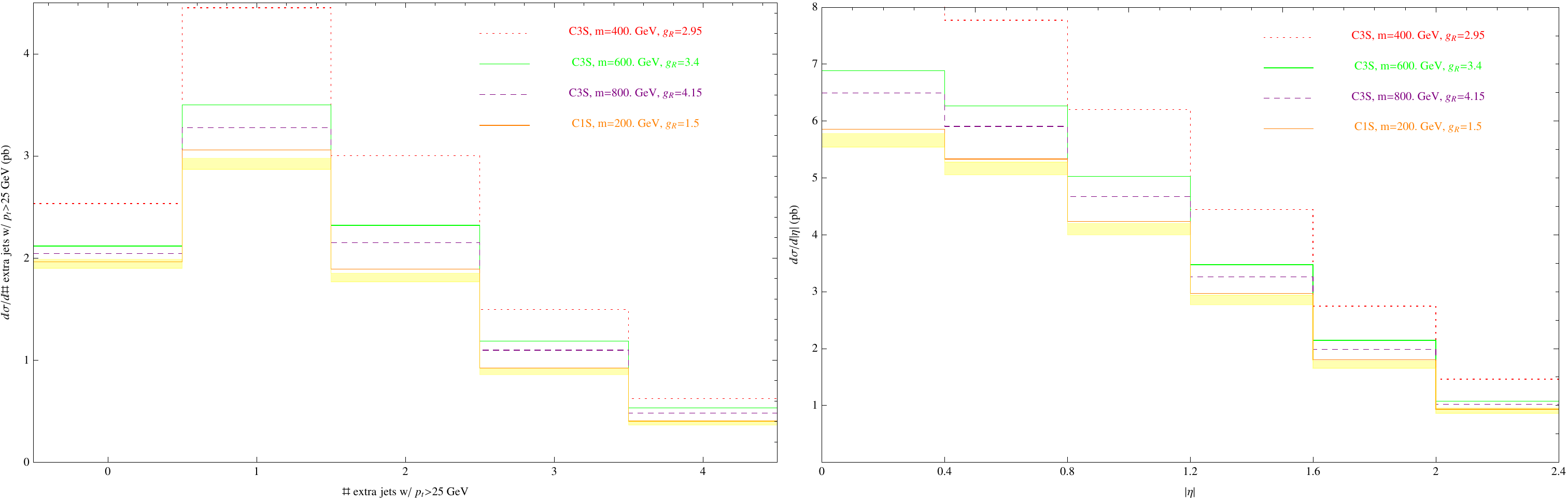}\\
\caption{Number of additional jets and lepton differential rapidity distribution of schanC8, C8V, C1S and C3S models at LHC7, as compared to SM LO expectation, with $\pm 1 \sigma$ yellow bands corresponding to statistical error given $1 \text{fb}^{-1}$. Models shown are those with the lowest $\chi^2$ as defined in Eq.~\eqref{chisq}.}
\label{jetsB}
\end{figure}

Perhaps the leading discriminant  is simply the number of additional jets in the event, shown in Figs.\eqref{jetsA},~\eqref{jetsB}.  While the overall production cross-section may be somewhat uncertain due to NLO corrections, leading to an uncertainty in the overall normalization of the NP curves, there is a significant difference in the ratio of the number of events with one extra jet to the number of events with no extra jets.  In fact, all of the $t$-channel models that generate a large asymmetry significantly overproduce the number of events with one additional jet.  One might wonder whether this effect could be reduced by allowing a significant branching fraction to light quarks; however in this case these events will contribute significantly to the single top analyses, which already with only $200 \mbox{ pb}^{-1}$ of data have an uncertainty of only 40 pb \cite{singletop}.  In the case of a significant branching fraction of the mediator to light quarks, single mediator production will easily contribute a significant fraction of this cross-section, with even more severe constraints arising in the high $H_T$ tail of the distribution, as pointed out in \cite{Craig:2011an}.  For reference in these figures we have also shown the rapidity distribution of the leptons; single mediator production results in a more central lepton rapidity distribution.

\section{Conclusions}

We have carried out a comprehensive analysis of NP models for top $\afb$ utilizing both Tevatron and the prospective LHC7 constraints with $1 \mbox{ fb}^{-1}$.   We considered effective vertices for all possible spin, color and flavor representations connecting top quarks with up quarks.  We were able to show on general grounds why scalar mediated models have difficulty reproducing the observed asymmetry.  We revisited the Tevatron signal level invariant mass distributions, as investigated in our earlier paper \cite{Gresham:2011pa}.  We found that the prospective LHC constraints on the total cross-section offer complimentary constraints to the Tevatron invariant mass distribution.  In the case of $t$-channel mediators, the LHC total cross-section places a strong constraint on light mediators, while the Tevatron invariant mass distributions place strong constraints on heavy mediators that are able to produce the asymmetry.  The vanilla $t$-channel models thus seem disfavored at present.  Heavy, narrow axigluons (with masses $\sim 2 \mbox{ TeV}$) are currently becoming more tightly constrained with the recent LHC7 top results.  A 400 GeV axigluon with large width and universal couplings to quarks appears at present to evade all existing constraints.

The LHC is rapidly closing the window on viable models for the top forward-backward asymmetry.  More non-generic features, such as large widths as in the light axigluon discussed here, will be necessary to make viable models consistent with both the Tevatron top $\afb$ and LHC top observables.

{\em Acknowledgments}: We thank Sunghoon Jung, Daniel Whiteson, and Dirk Zerwas for discussions.  KMZ thanks the Aspen Center for Physics for hospitality while part of this work was being completed.  The work of KMZ was supported in part by NSF CAREER award PHY 1049896. MG is supported by the Michigan Society of Fellows.

{\em Note added 1}: After the first version of this manuscript was completed,
a new result on top $A_{FB}$ from the D0 collaboration appeared \cite{Abazov:2011rq}, in which
$A_{FB}$ in the high $t\bar{t}$ invariant mass bin is significantly lower than
that of the CDF result.  As a result, the concentric $\chi^2$
contour ellipses in Figs.~(\ref{tev scatter scalar}-\ref{tev scatter schan}),~(\ref{LHC scatter scalar}-\ref{LHC scatter schan}), and (\ref{tev scatter scalar parton}-\ref{tev scatter schan parton}) will move down and to the right when the CDF and D0 results are combined, so that
many model points in danger with CDF alone will have a significantly lower $\chi^2$.  As a result, the best model point may change. We leave the analysis to a future
publication.

Also after the first version of this manuscript was completed, new results on $t \bar{t}$ cross-sections at LHC7 were released at the 2011 \href{http://eps-hep2011.eu/}{International Europhysics Conference on High Energy Physics}. No significant deviation from the SM expectation was measured \cite{atlasEPS}.

{\em Note added 2}: Because of the rapidly changing experimental results, we
will periodically update our results on a web page. This
website will also provide some figures not included in this paper.  See
\href{http://susy.physics.lsa.umich.edu/TopPhysics}{http://susy.physics.lsa.umich.edu/TopPhysics}.

\appendix
\section{Parton Level Asymmetries}\label{app: parton level}

\begin{figure}
\includegraphics[width = 0.85\textwidth]{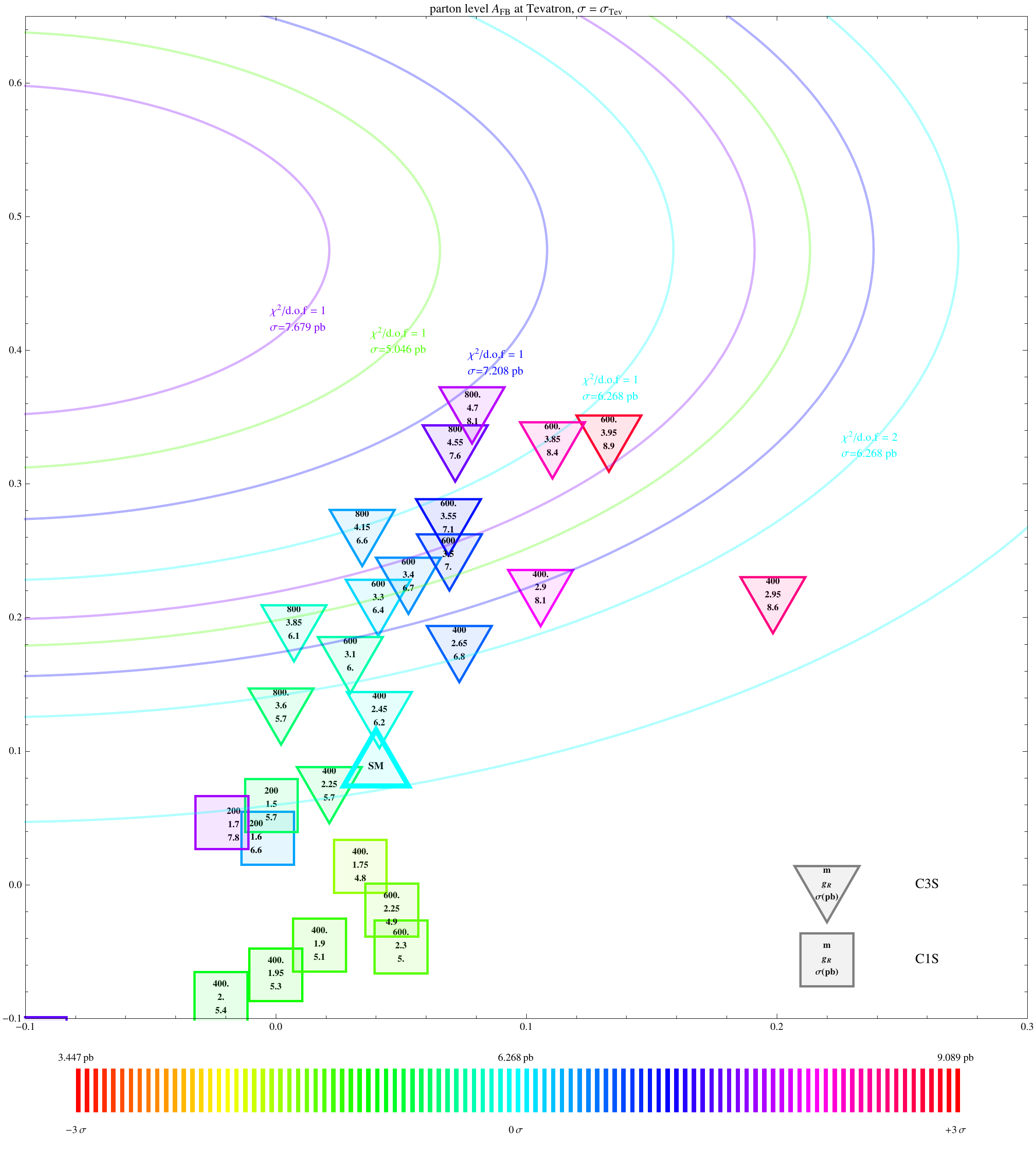}
\caption{Scatter plots depicting simulated {\em parton} level $ \{ A_{FB}(m_{t \bar{t}} < 450~\text{GeV}),  A_{FB}(m_{t \bar{t}} > 450~\text{GeV}), \sigma_{t \bar{t}} 
\} $ at Tevatron CM energy for $t$-channel flavor-changing scalar models listed in Table \ref{tab: scan summary table}.  The models are labeled by the mass of the mediator, the coupling to right-handed quarks, and the total Tevatron production cross-section.  The cross-sections are compared against the SM cross-section which yields 6.3 pb at the LO; the color scales for the models indicate the deviation from the SM cross-section, as indicated by the legend at the bottom. The curves indicate constant $\chi^2$ for a given cross-section, as defined in Eq. \eqref{chisq}. Contours for four cross-section values (cyan, blue, green, purple) are shown for $\chi^2/\text{d.o.f.} =$ 1 and 2.  A single (cyan) $\chi^2/\text{d.o.f.} = 3$ contour with SM cross-section is shown. Model points of a given color should be compared to $\chi^2$ contours of the same color. }
\label{tev scatter scalar parton}
\end{figure}

\begin{figure}
\includegraphics[width = 0.85\textwidth]{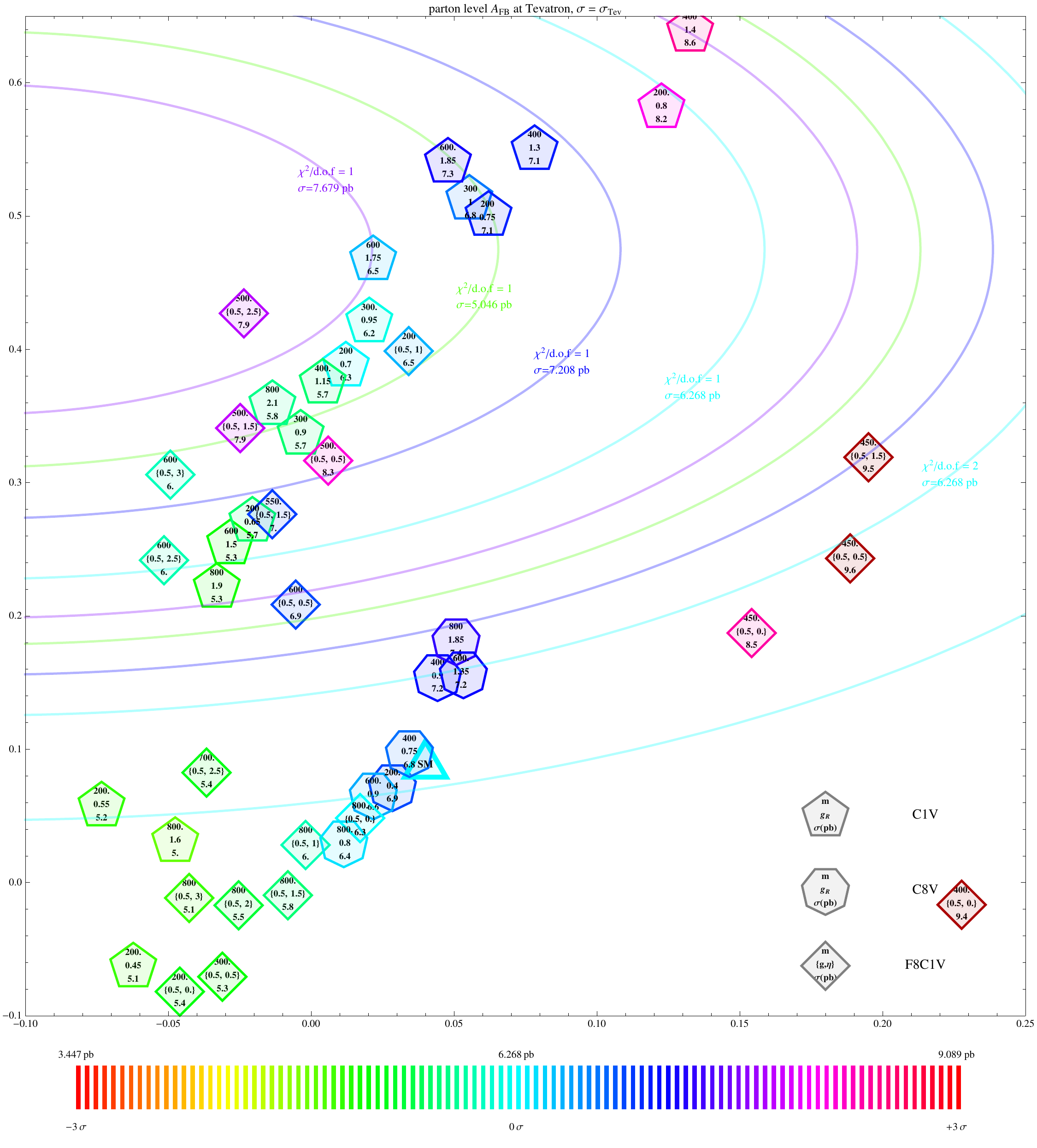}
\caption{Scatter plots depicting simulated {\em parton} level $ \{ A_{FB}(m_{t \bar{t}} < 450~\text{GeV}),  A_{FB}(m_{t \bar{t}} > 450~\text{GeV}), \sigma_{t \bar{t}}
\} $ at Tevatron CM energy for $t$-channel flavor-changing vector models listed in Table \ref{tab: scan summary table}.  The models are labeled by the mass of the mediator, the coupling, and the total Tevatron production cross-section.  The coupling conventions are discussed in detail in the text.  The cross-sections are compared against the SM cross-section which yields 6.3 pb at the LO; the color scales for the models indicate the deviation from the SM cross-section, as indicated by the legend at the bottom.  The curves indicate constant $\chi^2$ for a given cross-section, as defined in Eq. \eqref{chisq}. Contours for four cross-section values (cyan, blue, green, purple) are shown for $\chi^2/\text{d.o.f.} =$ 1 and 2.  A single (cyan) $\chi^2/\text{d.o.f.} = 3$ contour with SM cross-section is shown. Model points of a given color should be compared to $\chi^2$ contours of the same color. }
\label{tev scatter vector parton}
\end{figure}

\begin{figure}
\includegraphics[width = 0.85\textwidth]{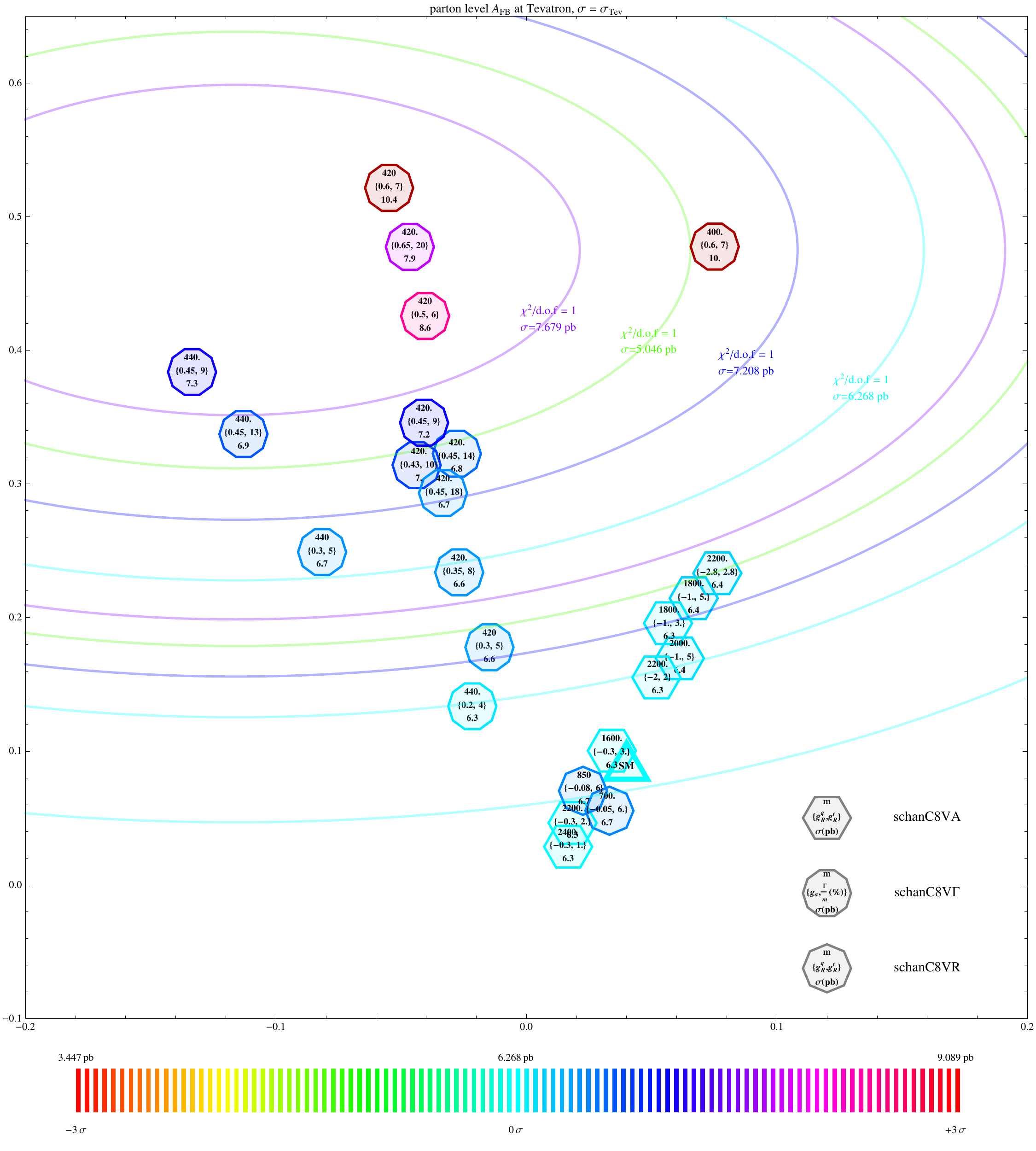}
\caption{Scatter plots depicting simulated {\em parton} level $ \{ A_{FB}(m_{t \bar{t}} < 450~\text{GeV}),  A_{FB}(m_{t \bar{t}} > 450~\text{GeV}), \sigma_{t \bar{t}} 
\} $ at Tevatron CM energy for axigluon models listed in Table \ref{tab: scan summary table}.  The models are labeled by the mass of the mediator, the coupling, and the total Tevatron production cross-section.  The coupling conventions are discussed in detail in the text.  The cross-sections are compared against the SM cross-section which yields 6.3 pb at the LO; the color scales for the models indicate the deviation from the SM cross-section, as indicated by the legend at the bottom.  The curves indicate constant $\chi^2$ for a given cross-section, as defined in Eq. \eqref{chisq}. Contours for four cross-section values (cyan, blue, green, purple) are shown for $\chi^2/\text{d.o.f.} =$ 1 and 2.  A single (cyan) $\chi^2/\text{d.o.f.} = 3$ contour with SM cross-section is shown. Model points of a given color should be compared to $\chi^2$ contours of the same color. }
\label{tev scatter schan parton}
\end{figure}

As a complement to the Tevatron signal level asymmetries shown in Figs.~(\ref{tev scatter scalar})-(\ref{tev scatter schan}), we show the parton level asymmetries, so that theorists can easily map signal level onto parton level for a broad range of models.  These are shown in Figs.~(\ref{tev scatter scalar parton})-(\ref{tev scatter schan parton}) for the same model points as in Figs.~(\ref{tev scatter scalar})-(\ref{tev scatter schan}).  Note in comparing the signal and parton level plots that a number of points are deleted in the parton level plot in cases where they cluster strongly around the SM point and become indistinguishable.  For these parton level plots the $\chi^2$ statistic used to draw contours is defined in Eq.~(\ref{chisq}), but with
\begin{align}
\sigma_{\AFBL} &= \sqrt{0.146^2+0.047^2+0.005^2+0.040^2}  \\
\sigma_{\AFBH} &= \sqrt{0.101^2+0.049^2+0.005^2+0.088^2} \\  
{\sigma_{\sigma_{t \bar{t}}} \over \sigma_{t \bar{t}, SM} }&= 0.15
\end{align}
for the error estimates, and with
\begin{align}
{\AFBL}_\text{obs} &= - 0.116\\
{\AFBH}_\text{obs} &= 0.475\\
\sigma_{t \bar{t},\text{SM}} &= 6.27 \label{eq: sm parton xsec}
\end{align}
for the central values.  The central values for $A_{FB}$ are the parton level values from \cite{Aaltonen:2011kc} and the SM values for the cross-section are taken from our simulations of 5 million events. 
The first two contributions to the $A_{FB}$ errors are from experiment, the third for the typical statistical error from our finite-sized simulated data samples, and the last is to account for possible NLO corrections: we take this contribution to be of the same size as the NLO SM asymmetry. These error estimates should be taken as rules of thumb to guide the eye in our figures for comparing SM against NP, rather than as hard and fast quantitative error budgets.

\section{Event Generation}\label{app: madgraph}

In this appendix, we describe our event generation setup and 
strategies for data analysis presented in the main text of this 
paper. 
We employ \texttt{FeynRules v1.4.10} for model file generation\cite{feynrules}, 
 \texttt{MadGraph5 1.3.3} for event generation\cite{Alwall:2011uj}, 
\texttt{PYTHIA v6} 
for parton showering and hadronization\cite{Sjostrand:2000wi}, and a modified \texttt{PGS4}
for fast detector simulation\cite{PGS4}. 

This work involves a large survey of different 
models and model parameters, and model-dependent acceptance in 
detection is an important issue in interpreting experimental observations. 
Thus fast detector simulation on a large number of events is necessary.
Although there are criticisms on the credibility of fast detector 
simulation tools, fast detector simulation tools like \texttt{PGS4} are 
indispensible for this paper.

To obtain more realistic and reliable results, we tune 
the detector simulation and our analysis to the 
current experiments in such a way that performance is not harmed.  
For comparison of our results to data, we show NP models compared to 
the SM with the same analysis setup.  Then we can draw
conclusions on the status of NP models, since all experimental 
analyses are accompanied with their own SM simulation. 
In the following section, we summarize our considerations. 
 
\subsection{Fast Detector Simulation and Object Reconstruction}

We simulate our model points given the specifications of the CDF detector at the Tevatron and from the ATLAS detector at the
LHC. We use the default detector parameters for CDF and ATLAS given 
in the official distribution of \texttt{PGS4}. Some important detector 
parameters used in \texttt{PGS4} are summarized in Table \ref{PGSdetector}. 

\begin{table}
\begin{tabular}{|c|c|c|}
\hline
 Detector & CDF & ATLAS \\
\hline
 $(\eta,\phi)$ cells in cal & (80,24) & (81,63)  \\
 $\eta$ width of cal cells for $|\eta| < 5$ & 0.1 &  0.1 \\
 $\phi$ width of cal cells & 0.262 & 0.1 \\
 EM cal resolution (GeV) & $0.01\oplus 0.2\sqrt{E/{\rm GeV}}$ & $0.01 \oplus 0.1 \sqrt{E/{\rm GeV}}$ \\
 had cal resolution (GeV) & $0.8 \sqrt{E/{\rm GeV}}$ & $0.8 \sqrt{E/{\rm GeV}}$ \\
 MET resolution & 0.2 & 0.2 \\
 cal trigger cluster threshold  & 3 GeV & 3 GeV \\
 outer radius of tracker & 1.0 m & 1.0 m \\
 magnetic field & 1.4 T & 2 T \\
 sagitta resolution  & $4\times 10^{-5}$ m & $5\times 10^{-6}$ m \\
 track finding efficiency & 0.98 & 0.98 \\
 minimum track $P_T$  & 0.30 GeV/c & 0.3 GeV/c \\
 tracking eta coverage & 2.0 & 2.5 \\
 $e/\gamma$ eta coverage & 2.0 & 3.0 \\
 $\mu$ eta coverage & 2.0 & 2.4 \\
 $\tau$ eta coverage & 2.0 & 2.0 \\ 
\hline
\end{tabular}
\caption{\label{PGSdetector} Detector Parameters of \texttt{PGS4} simulation 
for the Tevatron CDF and LHC ATLAS detectors  }
\end{table}

While we have not modified the detector parameters, the {\tt PGS} algorithm used 
for object reconstruction is rather outdated and therefore can give rise to
significantly different results. We summarize our changes in the following. 

\paragraph{Jet Reconstruction and Jet Energy Scale Correction:}

\begin{figure}
\centering
\includegraphics[width=0.45\textwidth]{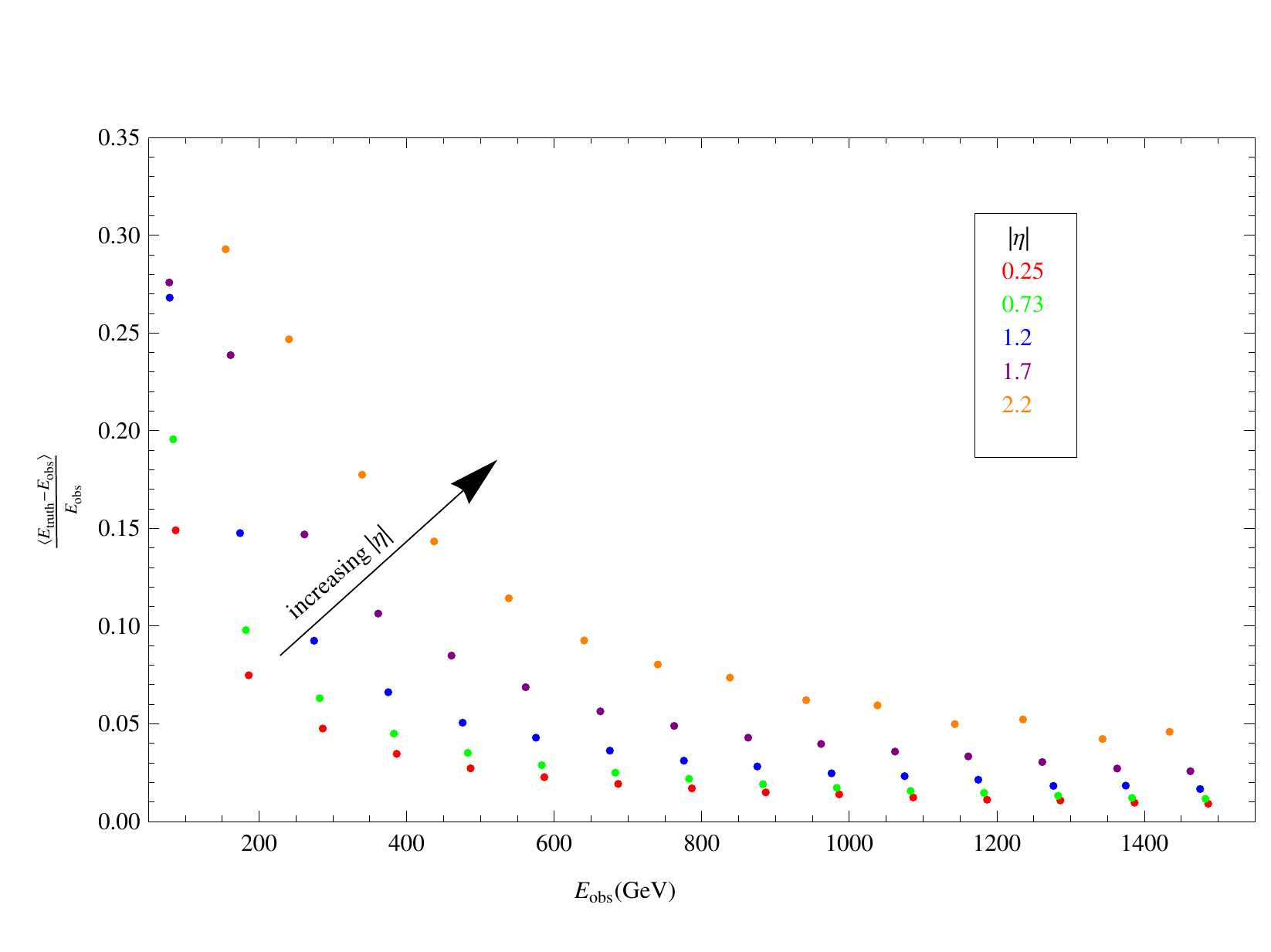}
\caption{\label{CDFATLASjetenergyscale} Jet energy scale pseudo-experiment
using $e^+ e^- \to q\bar{q}$ process.  } 
\end{figure}

The official version of \texttt{PGS} supports two jet algorithms: 
the cone algorithm and $k_T$-jet algorithm.  While the CDF analysis on $A_{FB}$ 
employed the cone algorithm, the LHC analyses use the anti-$k_T$-jet 
algorithm for jet reconstruction \cite{Cacciari:2008gp}. We modified the $k_T$-jet algorithm 
implementation
in \texttt{PGS} by changing particle-particle and particle-beam distance 
measures according to the anti-$k_T$-jet algorithm definition. 
For comparison with Tevatron results, 
we use the cone algorithm with $\Delta R$ = 0.4, and for comparison with LHC results, we use the anti-$k_T$-jet algorithm with 
$\Delta R$ = 0.4. 

After parton showering/hadronization and detector simulation which includes
calorimeter errors in the measurement, the jet energy
obtained from the jet reconstruction algorithm will be significantly
different from the true value of the original parton. 
Note that this can be a significant source of distortion in event 
distributions with respect to energy scale variables such as invariant 
masses or transverse momenta. Therefore, such ``measured'' values
of the jet energy must be corrected by performing standard candle
experiments. The experiments have published their 
jet energy scale correction procedure in the literature, and we carry out our own 
jet energy scale correction for the {\tt PGS} detector simulation. 
For this purpose, we generated the SM dijet event samples  from 
electron-position collisons ($e^+ e^- \to q \bar{q}$) with 
$\sqrt{s}$ = 200 GeV  to 3000 GeV in 200 GeV increments  for the PGS implementation of the ATLAS detector, with 100,000 events 
for each energy sample.  We generated similar samples for the PGS implementation of the CDF detector from $\sqrt{s}$ = 200 GeV  to 1000 GeV in 100 GeV increments.

From the samples, we compared the mean value of measured jet energy scales 
to the designated parton energy, and extracted the variance of the
probabilistic jet energy measurement. In Fig.~(\ref{CDFATLASjetenergyscale}), 
we present the jet energy shift in this pseudo-experiment. 
The resultant jet energy scale distortion is reasonably matched with
that of real detectors. The jet energy scale correction 
for CDF is given by 
\begin{eqnarray}
\frac{\Delta p_T}{p_{T\rm obs}} &=& \frac{1.63}{\sqrt{p_{T\rm obs}}}, 
\quad
\Delta \eta = 0,  \label{jesCDF}
\end{eqnarray} 
where $p_{T \rm obs}$ is the nominal $p_T$ value of a constructed jet and
$\Delta p_T  = p_{T \rm true} - p_{T \rm obs}$ is in GeV. 
For ATLAS,  
\begin{eqnarray}
\frac{\Delta p_T}{p_{T\rm obs}} &=& {\Delta E \over E_{\rm obs}} = \frac{14.23 + 7.53 \eta_{\rm obs}^2 }
      {\sqrt{ p_{T \rm obs}^2 \cosh^2\eta_{\rm obs}+ m_{\rm obs}^2 }} = {\Delta m \over m_{\rm obs}}, \quad
\Delta \eta = 0,
\label{jesATLAS}
\end{eqnarray} 
where $m_{\rm obs}$ is a jet mass. 
Here, for the CDF analysis, we ignore the jet mass, and jet momentum is
parameterized only by $p_T,~\eta,~\phi$, while  we retain a non-zero jet mass 
for the LHC analysis, since jet mass is a variable used in $m_{t \bar{t}}$ reconstruction.
The variance of jet energy and angular parameters from 
SM dijet simulation for CDF is: 
\begin{eqnarray}
\frac{\sigma(p_T)}{p_{T\rm obs}} &=&  0.0593 
+ \frac{1.21}{\sqrt{p_{T\rm obs}}}, \quad
\sigma(\eta) = 0.0112 + \frac{0.65}{\sqrt{p_{T \rm obs}}}. \label{sigmaCDF} 
\end{eqnarray}

\paragraph{Top Reconstruction:} 

\begin{figure}
\centering
\includegraphics[width=0.5\textwidth]{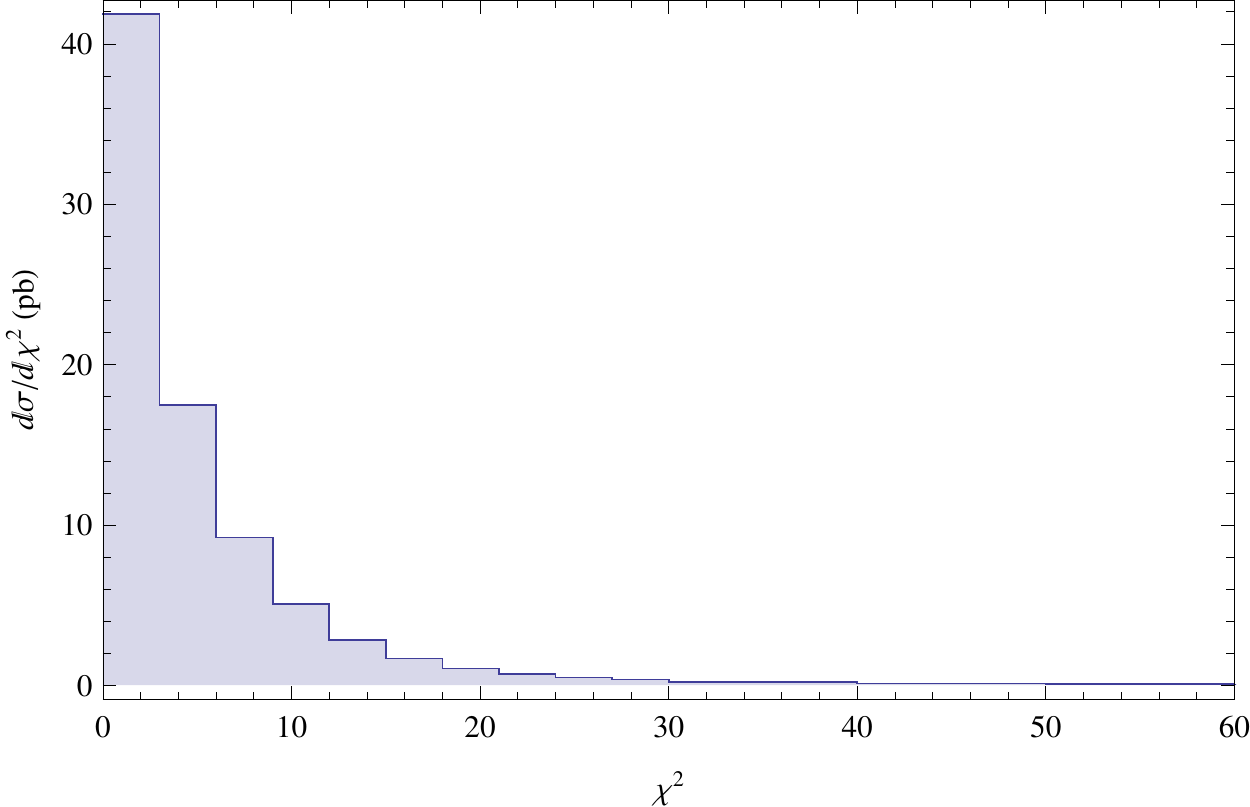}
\caption{Cross-section versus $\chi^2$ of top pair reconstruction for simulated LO SM events at $\sqrt{s} = 1.96$ TeV.  The bin size below $\chi^2=30$ is in increments of 3, corresponding to the number of degrees of freedom in the $\chi^2$ fit. \label{ttbarchisqr}}
\end{figure}

To extract $A_{FB}$, we reconstruct particle momenta of a semileptonic top pair
using the $\chi^2$ method, as in the CDF analysis.   We find the missing 
neutrino momentum and fix combinatorics by minimizing  
$\chi^2$ of over-constrained on-shell mass relations: 
\begin{eqnarray}
y_1 = p^2_\nu =0, \quad y_2 = (p_\ell + p_\nu )^2 - m_W^2 = 0, \quad
y_3 = (p_{b_l}+p_\ell + p_\nu)^2 - m_t^2 = 0, \\
y_4 = (p_{j_1}+p_{j_2})^2 - m_W^2 = 0,  \quad
y_5 = (p_{j_1}+p_{j_2}+p_{b_h})^2 - m_t^2 = 0. 
\end{eqnarray}
$\chi^2$ is defined by $\chi^2 = \vec{y}^T \cdot V^{-1} \cdot \vec{y}$ 
where $y = (y_1,y_2,y_3,y_4,y_5)$ and $V$ is the covariance matrix of 
$y_i$s. The detailed method is presented in Appendix A of \cite{Gresham:2011pa}.

Since we use a modified jet energy scale for the analysis, jet energies
and the covariance matrix $V$ are corrected correspondingly. 
$V$ depends on individual jet-jet covariance matrix, which is related to
$\sigma (p_T)$ and $\sigma (\eta)$ obtained in 
Eq.~(\ref{sigmaCDF}). 

 As a consistency check of our jet energy scale correction and top 
reconstruction algorithm, we 
show the resultant $\chi^2$ distribution of SM $t\bar{t}$ events in 
Fig.~(\ref{ttbarchisqr}). Taking into account degradation due to the
combinatoric background, 
the result is reasonably well-matched with a theoretical curve and with the 
CDF analysis (shown as Fig. 15  in \cite{Aaltonen:2011kc}).

\subsection{Event Generation and Parameter Scan Strategy}

\begin{table}
\begin{tabular}{|c|c|c|}
\hline 
Model & Parameter Scan Range (mass in GeV unit) \\
\hline 
C1V & Rough:  $ \{ (m,g_R) | m \in \{ 200,400,600,800,1000 \}, g_R \in \{0.5,1.0\dots5.0\}$  \\
    & Fine:  $ (m,g_R) = (200,\{0.4,0.45\dots0.95\}), 
               (300,\{0.4,0.45\dots1.30\}),$ \\
    & $ (400,\{0.6,0.65\dots1.40\}), 
        (600,\{1.0,1.05\dots1.90\}), (800,\{1.30,1.35\dots2.2\} $ \\
\hline
C8V & Rough: $\{ (m,g_R) | m \in \{ 200,400,600,800,1000 \}, g_R \in \{0.5,1.0\dots5.0\} $   \\
    & Fine:  $ (m,g_R) = (200,\{0.2,0.25\dots0.40\}), 
                         (300,\{0.3,0.35\dots0.80\}),$ \\
    &        $ (400,\{0.4,0.45\dots0.90\}),
               (600,\{0.5,0.55\dots1.50\}), 
               (800,\{0.7,0.75\dots2.0\})$ \\
\hline
F8C1V & Rough: $\{ (m,g,\eta) | m \in \{200,400\dots800\}, g=0.5,\eta\in 
\{ 0,0.5\dots3.0 \} \} $   \\
& Fine: $\{ (m,g,\eta) | m \in \{300,350\dots700\}, g=0.5,\eta\in\{0,0.5\dots3.0\} \}$ \\
\hline 
C1S & Rough: $\{(m,g_R) | m \in \{200,400,600,800,1000\}, g\in\{0.5,1.0\dots5.0\} \}$  \\
    & Fine: $(m,g_R) = (200,\{1.5,1.55\dots2.0\}), (300,\{1.5,1.55\dots2.20),$ \\ 
    & $(400,\{1.5,1.55\dots2.30\}), (600,\{2.0,2.05\dots3.0\}), (800,\{2.5,2.55\dots4.0\})$ \\
\hline
C3S & Fine: $(m,g_R) = (400,\{1.5,1.55\dots3.50\}), (600,\{2.5,2.55\dots4.50\}), (800,\{3.5,3.55\dots,5.5\})$   \\
\hline
schanC8V$\Gamma$ & Fine: $\{ (m,g_R,n_\phi,m_\phi) | m\in\{420,440\},
g_R\in\{0.35,0.45\dots0.65\},n_\phi \in \{4,5,6,7\},m_\phi=100  \}$   \\
\hline
schanC8VA & Rough: $\{(m,g^q_R,g^t_R)| m\in\{1600,1800\dots2400\},g^q_R=-0.3,g^t_R\in \{1,2\dots5\} \}$ \\
\hline
schanC8VR & 
 Fine: $(m,g^q_R,g^t_R) = (700,-0.05,\{2.0,2.5\dots6.0\}), (850,-0.08,\{2.0,2.5\dots8.0\}), $ \\
& $(1000,-0.15,3),(1000,-0.125,5),(1000,-0.1,8),(1500,-0.4,5.5),(1500,-0.3,8)$
\\
\hline
\end{tabular}
\caption{\label{modelpointscan} Summary of model points scanned. For the $s$-channel model with large decay width,  schanC8V$\Gamma$,
we take an {\em additional} contribution to the width of the mediator into scalars $\phi$ which is  $\Gamma_\phi/m \approx (g_s^2 n_\phi^2 / 16\pi) (1-4 m_\phi^2/m^2)^{3/2}$ \cite{Tavares:2011zg}.  }
\end{table}

We analyzed eight classes of models : C1V, C8V, F8C1V, C1S, 
C3S, schanC8V$\Gamma$, schanC8VA, schanC8VR, as discussed in the main body of text.  We consider the $t\bar{t}$ pair production cross section at the LHC and 
Tevatron, and $A_{FB}$ at the Tevatron as the test of different models.   We generate events for the process $t\bar{t}$ + 0 or 1 jets with MLM matching. 
The renormalization group and factorization scales are fixed to be 200 GeV, 
and the top quark mass is 172 GeV. We employ CTEQ6L parton distribution 
functions. 

Our analysis has been done in three steps: rough scan, fine scan and benchmark 
point analysis. For the rough and fine scans, we generate 100,000 
\texttt{MadGraph} events for each point, passing them through the \texttt{PYTHIA}
and \texttt{PGS} pipelines. At the stage of the fine scan, we are able to see 
which models look most promising as shown in Figs.~(\ref{tev scatter scalar})-(\ref{tev scatter schan}).  
For each benchmark model point, we generate five million \texttt{MadGraph}
events followed by \texttt{PYTHIA} and \texttt{PGS}. 
Note that the number of generated events is reduced by 20\% - 40\% 
due to the MLM matching procedure.   We summarize the model points in Table \ref{modelpointscan}. 

\subsection{Cluster Pipeline Setup}

Although not a physics problem, generating event sets for a large 
number of model points is an intensive computing task with many engineering issues. The difficulty arises particularly
with a cluster computing setup because it poses a new 
paradigm for software design.   
We share our experience in 
addressing such issues and suggest a common infrastructure.

We utilize a cluster server named \texttt{Flux} in the 
Center for Advanced Computing at the University of Michigan. 
Ideally, the event generation
software \texttt{MadGraph} could handle the cluster server configuration
seamlessly, but in practice it is not easily implementable. 
The major challenge is that the disk I/O speed of a shared file system is not 
fast enough for \texttt{MadGraph} event generation. \texttt{MadGraph} creates 
a large number of small files, and the delay in writing to the shared 
file system causes the program to crash. A solution is found by making use of local storage; in most modern 
cluster computers, each node is provided with its own (fast) local storage 
space. To utilize this, the desired workflow must be to send
a job which installs \texttt{MadGraph} on the temporary local disk space 
in the cluster node, followed by event generation tasks, uploading 
generated files, and finally erasing the temporary files. 

Since parallel computing is, in some sense, highly nondeterministic (on account of network latency or cluster usage traffic), 
jobs routinely fail. Therefore, it is important to make a highly resilient system for 
reducing the burden of bookkeeping of failed jobs. We design each job 
as a smart agent program which autonomously tests and monitors its own 
progression status. This requires us to make a central server for controlling
assignment and checking the status of each job by having each job client 
report its status and wait for a new assignment for the next job if failed 
or finished. 

Many high energy physics programs contain legacy codes, and \texttt{MadGraph}
is no exception. Due to incompatibilities or missing features, these codes often must be modified by users. However, quick-and-dirty code repairs usually 
increase the complexity of a system. To control this, we make
 wrapping modules for external programs which is under our version control. 
By making those modules easily installable, the overall development becomes 
much simpler and easier in error control. We call this system 
\texttt{pipeline} which is essentially a set of installable high energy physics
program modules.

For inter-process communication, we choose a standard web service interface,
since HTTP protocols are not blocked in the usual firewall setup of 
a cluster. By standardizing the job specification interface and each 
computer configuration, one can achieve flexibility and extensibility in 
routine high energy  physics jobs. The job queue server retains information for 
each task for future documentation, and also effectively dispatches jobs. 
The web service choice has been superior in making a good user interface
and utilizing common available tools.

\begin{figure}
\subfigure[~Server configuration]{ 
\includegraphics[width=0.35\textwidth]{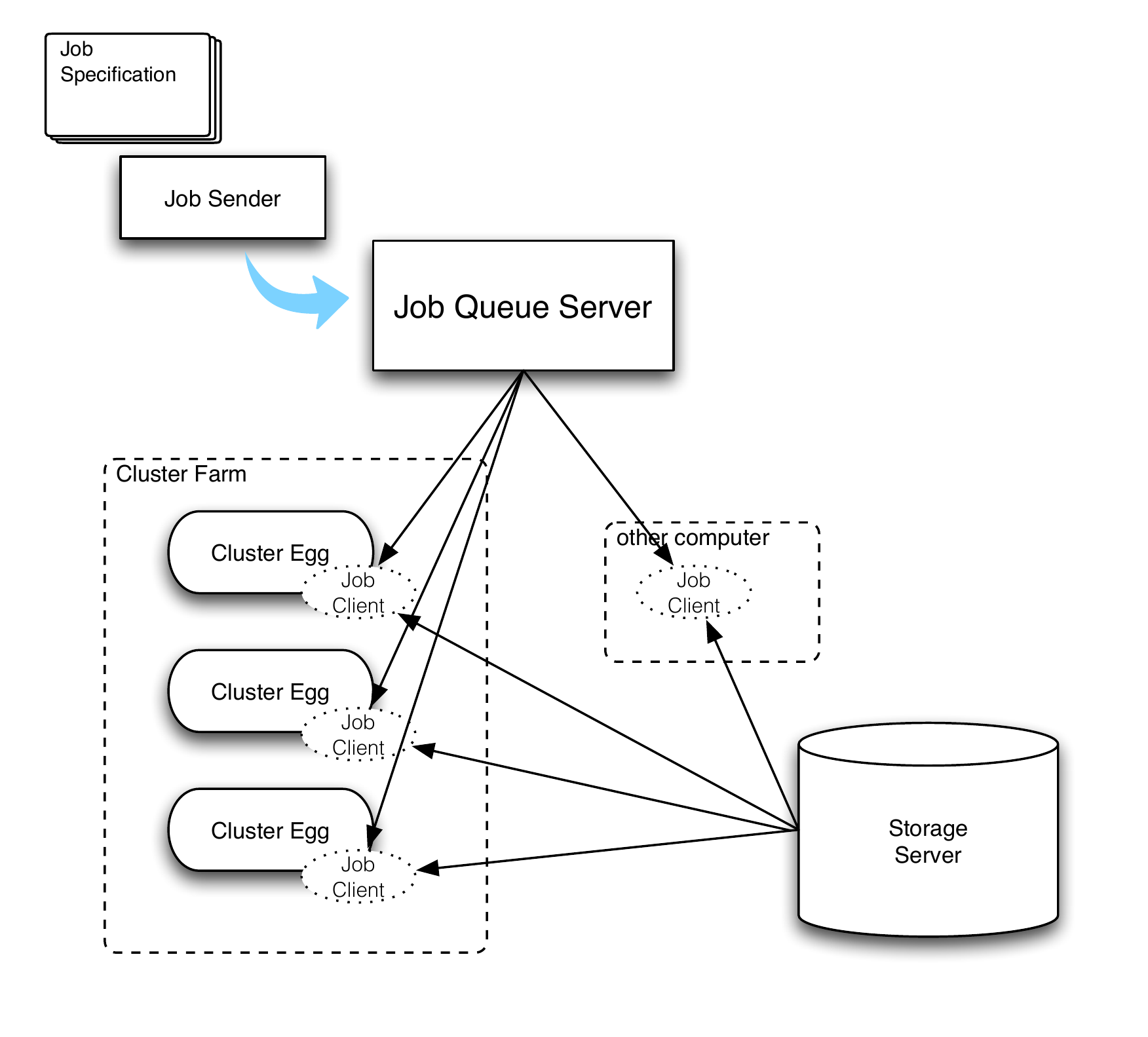}}\qquad
\subfigure[~Detail of client configuration and interaction]{
\includegraphics[width=0.55\textwidth]{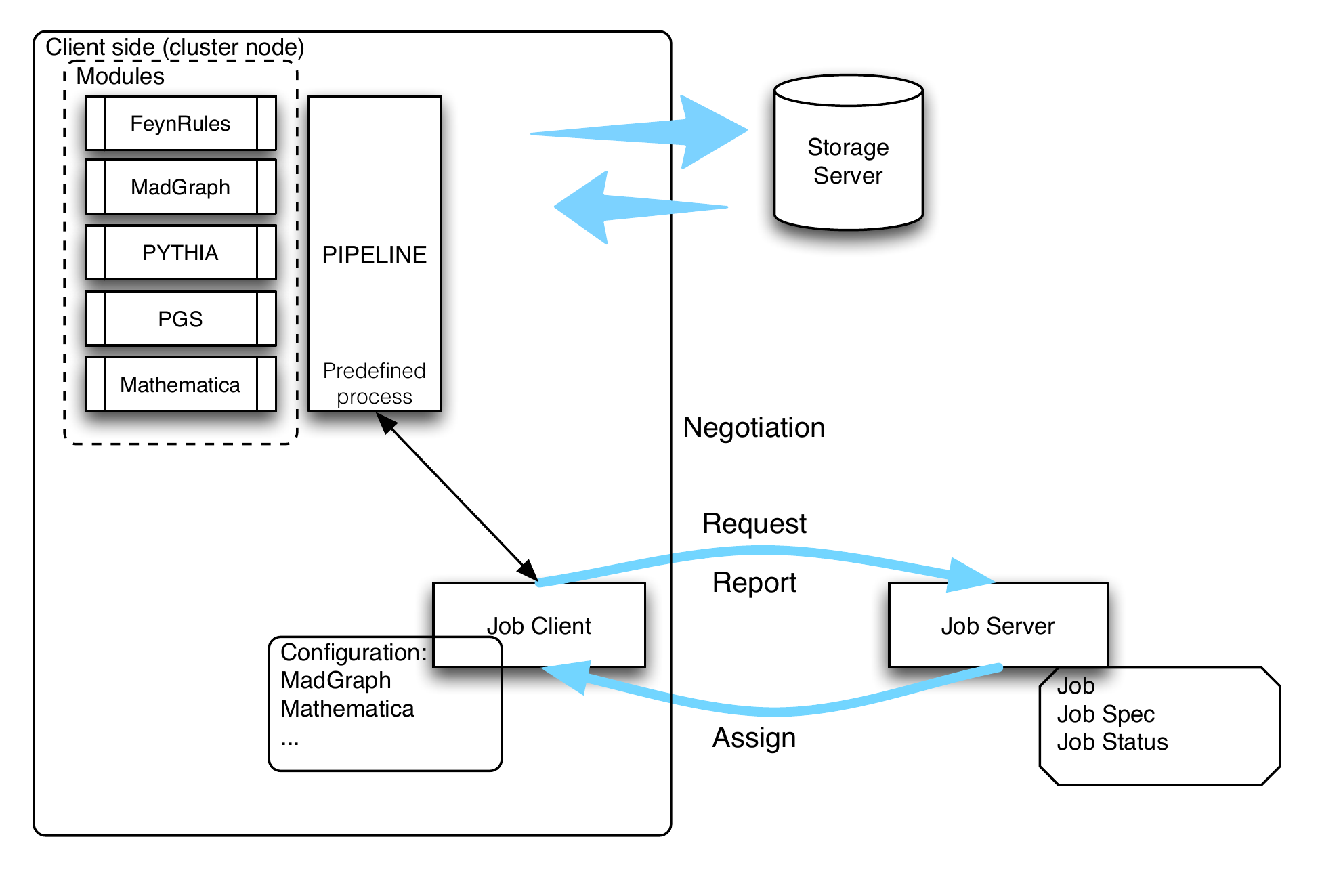}}
\caption{\label{cluster} Cluster job queue system setup and interaction.}
\label{clustersetup}
\end{figure}

Fig.~(\ref{clustersetup}) shows our pipeline setup. We develop the system in 
haskell using Glasgow Haskell Compiler (\texttt{ghc}) 7.0.   The job queue server 
is supposed to be always on and waiting for new jobs or new job requests from 
the client, which can run either in a cluster or on common desktops. If it runs
on a cluster, a bootstrap script called \texttt{clusteregg} automatically 
installs a ready-made setup for a job client with \texttt{MadGraph} and the rest of the needed software.
Since each job client sends its configuration when it requests a new job, 
the job queue server dispatches a new job for which the client is adequate 
(for example, if the job client does not have {\em Mathematica}, then {\em Mathematica} 
jobs are not assigned).  A client also rechecks whether a job is doable 
with its current setup, and finally both parties handshake on the job assignment.
After the negotiation, the job client proceeds with the job according to the job
specification from the server, and the job specification and high energy
physics tools are interfaced with \texttt{pipeline}. After the job is finished, 
a job client sends its results to the storage server and wipes out the 
temporary files. Every step of the job 
status is reported to the job queue server for monitoring purposes.
We will announce details of the \texttt{pipeline} and \texttt{jobqueue} 
systems elsewhere soon.  

High performance computing facilities are now practically mandatory in high energy theory
projects even for understanding the
implication of the current state-of-the-art high energy experiments,
especially in the LHC era. Building a common computing software 
infrastructure adjusted to high energy physics will harness 
our physics community in a very positive way.

\end{document}